\documentclass[fleqn,usenatbib]{mnras}

\usepackage{newtxtext,newtxmath}

\usepackage[T1]{fontenc}
\usepackage{ae,aecompl}

\usepackage{graphicx}	
\usepackage{amsmath}	

\usepackage{soul}
\usepackage{pifont}

\newcommand{\MBDoce}{\textcolor{cMB12}{\bf B12prim}}
\newcommand{\MBInj}{\textcolor{cMB20i}{\bf B20SN}}
\newcommand{\TrMBUp}{\textcolor{cMB12i}{\bf B12primSN}}
\newcommand{\TrMBDown}{\textcolor{cMB13i}{\bf B13primSN}}
\newcommand{\Binj}{B_\text{inj}}
\newcommand{\Einj}{E_\text{mag,inj}}
\newcommand{\Dres}{{\Delta x}_\text{max}}
\newcommand{\emag}{\varepsilon_\text{mag}}
\newcommand{\emagPMF}{\varepsilon_\text{mag,prim}}
\newcommand{\emagSN}{\varepsilon_\text{mag,ast}}

\newcommand{\ikpc}{\text{kpc}^{-1}}
\newcommand{\muG}{\mu \text{G}}
\newcommand{\gram}{\text{g}}
\newcommand{\erg}{\text{erg}}
\newcommand{\cm}{\text{cm}}
\newcommand{\cc}{\cm^3}
\newcommand{\Msun}{\mathrm{M_\odot}}

\newcommand{\cmark}{\ding{51}}%
\newcommand{\xmark}{\ding{55}}%

\title[Primordial vs astrophysical magnetism]{Unraveling the origin of magnetic fields in galaxies}

\author[Martin-Alvarez et al.]{
Sergio Martin-Alvarez$^{1}$\thanks{E-mail: smartin@ast.cam.ac.uk (SMA)},
Harley Katz$^{2}$,
Debora Sijacki$^{1}$,
Julien Devriendt$^{2}$,\newauthor
and Adrianne Slyz$^{2}$
\\
$^{1}$Institute of Astronomy and Kavli Institute for Cosmology, University of Cambridge, Madingley Road, Cambridge CB3 0HA, UK\\
$^{2}$Subdepartment of Astrophysics, University of Oxford, Keble Road, Oxford, OX1 3RH, UK\\
}

\date{Submitted to MNRAS}

\pubyear{2020}

\begin{document}
\label{firstpage}
\pagerange{\pageref{firstpage}--\pageref{lastpage}}
\maketitle

\definecolor{cMB12}{rgb}{0,0.66667,0}
\definecolor{cMB20i}{rgb}{0.66667,0,0}
\definecolor{cMB12i}{rgb}{0.882353,0.686275,0}
\definecolor{cMB13i}{rgb}{1,0.5,0}
\definecolor{dgreen}{rgb}{0,0.8,0.8}

\begin{abstract}
Despite their ubiquity, there are many open questions regarding galactic and cosmic magnetic fields. Specifically, current observational constraints cannot rule out if magnetic fields observed in galaxies were generated in the Early Universe or are of astrophysical nature. Motivated by this we use our magnetic tracers algorithm to investigate whether the signatures of primordial magnetic fields persist in galaxies throughout cosmic time. We simulate a Milky Way-like galaxy down to $z \sim 2 - 1$ in four scenarios: magnetised solely by primordial magnetic fields, magnetised exclusively by SN-injected magnetic fields, and two combined primordial + SN magnetisation cases. We find that once primordial magnetic fields with a comoving strength $B_0 >10^{-12}$ G are considered, they remain the primary source of galaxy magnetisation. Our magnetic tracers show that, even combined with galactic sources of magnetisation, when primordial magnetic fields are strong, they source the large-scale fields in the warm metal-poor phase of the simulated galaxy. In this case, the circumgalactic and intergalactic medium can be used to probe $B_0$ without risk of pollution by magnetic fields originated in the galaxy. Furthermore, whether magnetic fields are primordial or astrophysically-sourced can be inferred by studying local gas metallicity. As a result, we predict that future state-of-the-art observational facilities of magnetic fields in galaxies will have the potential to unravel astrophysical and primordial magnetic components of our Universe. 
\end{abstract}


\section{Introduction}
From planets and stars to clusters of galaxies, magnetic fields pervade our Universe on all scales. Their importance is well recognised in most astrophysical scenarios, be it the process of star formation \citep{Vazquez-Semadeni2011,Federrath2012,Krumholz2019}, the propagation of cosmic rays \citep{Kulsrud1969}, or the their role in active galactic nuclei (AGN) residing in galaxy clusters \citep{Weinberger2017,Bambic2018,Ehlert2018}.

For galaxies in particular, magnetic fields are generally observed at low redshift to have energies comparable or even above the thermal and turbulent energies of the gas \citep[e.g.][]{Beck2007,Beck2015}, corresponding to field strengths of order $\gtrsim \muG$ \citep{Chyzy2011,Basu2013,Mulcahy2014}. Magnetic fields may play an important role in processes shaping galaxies such as star formation \citep{Padoan2011,Zamora-Aviles2018}, the evolution of galactic outflows \citep{Gronnow2018}, the mixing of gas in galaxy haloes \citep{vandeVoort2020}, and regulation of the structure and dynamics of the multi-phase interstellar medium \citep[ISM;][]{Iffrig2017,Kortgen2019}. When their energy is comparable to the kinetic and thermal energies of the gas, magnetic fields can affect the global properties of galaxies \citep{Pillepich2018a,Martin-Alvarez2020} or even reduce the galaxy stellar mass function \citep{Marinacci2016}. As a result, magnetic fields are an important contributor to the process of galaxy formation.

Notwithstanding their importance, the origin of magnetic fields in galaxies and at large scales remains unknown. Amongst the various possibilities to generate these magnetic fields, three primary channels persist \citep{Rees1987}. Two of them are astrophysical in nature: dynamical amplification mechanisms \citep[e.g.][]{Dubois2010,Beck2012,Pakmor2013}, or feedback through magnetised outflows from point-like sources such as stars or AGN \citep[][]{Beck2013a}. Alternatively, magnetic fields could have a primordial origin, with strong fields generated in the early Universe and then preserved and evolved with the baryonic gas \citep[e.g.][]{Ratra1992,Barrow2011}. Cosmological simulations have shown the viability of astrophysical mechanisms to produce $\sim \muG$ magnetic fields through turbulent dynamo amplification \citep{Dolag1999,Pakmor2014,Rieder2017b,Martin-Alvarez2018,Vazza2018} of weak seed fields generated by e.g., a Biermann battery \citep{Attia2021}. Similarly, the injection of magnetic fields by supernova (SN) events can also produce the magnetic fields observed in galaxies \citep{Butsky2017,KMA2019}. Equivalently, AGN could play a similar role for galaxy clusters \citep{Vazza2017}, whereas their role in galaxies is yet to be explored in detail. Finally, if the magnetic fields generated in the early evolution of our Universe are strong enough, they will be amplified to $\muG$ strengths in galaxies during the collapse of density perturbations \citep{Kandus2011} even though additional dynamo amplification might be required for galaxy clusters \citep{Vazza2017}. A variety of theoretical models advocate the production of strong magnetic fields during the early Universe, e.g. during the process of inflation \citep{Ratra1992,Sharma2018}. For comprehensive reviews on the origin of magnetic fields in our Universe see \citet{Widrow2002,Kandus2011,Subramanian2016}. 

Determining the primordial magnetic field of the Universe is an arduous task. In the absence of direct measurements of intergalactic magnetic fields \citep{Taylor2015} permeating cosmic scales, the most reliable prospect is placing upper and lower limits on their strength. However, current constraints allow an exceedingly large range of values that do not adequately constrain the importance of primordial magnetic fields. A well-accepted upper limit for the primordial magnetic field strength\footnote{When discussing the strength of primordial magnetic field or cosmological magnetic fields, their strength can be determined at different scales. Unless indicated otherwise, we will refer generally to magnetic fields with coherence lengths $\lambda \gtrsim 1$ Mpc. Furthermore, the strength of $B_0$ will be provided in comoving units.} is $B_0 < 10^{-9}$G, provided by the absence of B-mode signal induced by primordial magnetic fields in the perturbations of the cosmic microwave background \citep[CMB;][]{PlanckCollaboration2015}. More stringent upper constraints can be extracted from other probes such as the propagation of ultra-high energy cosmic rays (UHECRs) through the intergalactic medium (IGM). Measuring the offset in their arrival direction from their predicted sources, \citet{Bray2018} claim an upper limit $B_0 < 10^{-10}$G. Obtaining a lower limit for $B_0$ is even more problematic, as potential signatures of magnetic fields at large scales rapidly weaken as their strength decreases. Nonetheless, \citet{Neronov2010} proposed a lower limit $B_0 \gtrsim 10^{-16}$G based on the non-detection of secondary GeV emission when observing TeV blazars. This has been contested \cite[e.g. by][]{Broderick2012}, as plasma instabilities could dissipate the energy of particle pairs prior to the GeV emission. More recently, \citet{Broderick2018} also found a non-detection of inverse Compton GeV emission in oblique lines-of-sight to TeV sources. This non-detection places an upper limit inconsistent with (i.e. with a lower value than) the lower limits by \citet{Neronov2010}. Other canonical lower limits tend to be much weaker (e.g. $B_0 > 10^{-27}$ G based on the Harrison mechanism; \citealt{Hutschenreuter2018}), and thus do not elucidate the uncertainty surrounding primordial magnetic fields. This illustrates the need for further study before limits on primordial magnetic fields are well-understood. As a result, it is important to consider alternatives that could aid our understanding of the cosmic magnetic field.  

In the advent of facilities such as SKA, which may directly detect magnetic fields in filaments \citep{Govoni2019} or those permeating the IGM \citep{Gaensler2004,Beck2014}, cosmological simulations are the perfect tool to guide and interpret these observations \citep[e.g.][]{Vazza2017}. However, when studying the origin of magnetic fields, one of the issues is that various channels of magnetisation will operate simultaneously. As a result, simulations must study their co-evolution and devise a method to disentangle magnetic fields of different origins. This can be done using the magnetic tracing method proposed in \citet{KMA2019}. In that paper, we presented the first application of such an algorithm, based on linearly decomposing the total magnetic field in the simulation into multiple extra magnetic fields. A useful decomposition is to then associate each additional {\it tracer} magnetic field with a particular origin or magnetic energy source. In this method the total magnetic field (i.e. the only field allowed to influence the dynamics) and the tracer fields are evolved in time solving the induction equation for each of them separately. As a result, the code allows us to unravel for the first time the {\bf exact} contribution made by each source of magnetism to the magnetisation of a simulation. In this work we apply this method to investigate the co-evolution of primordial and astrophysical magnetic fields in a cosmological zoom-in simulation of a galaxy, disentangling the effects and properties of each type of magnetic field. As upcoming surveys will provide the community with a vast range of observations of magnetic fields in galaxies, we revise whether galaxies and their surroundings can be used to probe and measure the cosmic magnetic field of our Universe.

This work is organized as follows: the numerical methodology to generate and evolve our simulations is described in Section~\ref{s:Methods} as well as the tracer algorithm (Section \ref{ss:Tracers}). Section~\ref{s:Results} explores our main results, subdivided according to the various properties we consider. Due to the numerical approximations and physical models employed in this work, in Section~\ref{s:Caveats} we discuss the implications of our findings and the most important caveats. Finally, we conclude with a summary of our work in Section~\ref{s:Conclusions}.

\section{Numerical methods}
\label{s:Methods}
\subsection{The RAMSES code}
The magnetohydrodynamical (MHD) zoom-in simulations of galaxy formation explored in this work are produced using our own modified version of the public code {\sc ramses} \citep{Teyssier2002}. {\sc ramses} employs an adaptive mesh refinement (AMR) octree grid to solve the evolution of the baryonic gas. This gas is coupled through gravity to the dark matter and stars which we model as collisionless particles. The evolution of the magnetic field in {\sc ramses} is calculated using a constrained transport (CT) method. It models magnetic fields as face-centred quantities, ensuring that the divergence constraint ($\vec{\nabla} \cdot \vec{B} = 0$) is exactly conserved numerically \citep{Teyssier2006, Fromang2006}. We employ our magnetic field tracer algorithm \citep[presented in][]{KMA2019}, which is an extension of the CT solver in {\sc ramses}, to follow the evolution of magnetic fields attending to their origin. Section~\ref{ss:Tracers} provides more details on the configuration used for the magnetic tracers. The magnetic diffusivity $\eta$ in most astrophysical environments is considerably below the numerical diffusivity of {\sc ramses} at the spatial resolutions considered in this study. As a result, we set $\eta = 0$ in the induction equation used to evolve the magnetic field
\begin{equation}
\frac{\partial \vec{B}}{\partial t} = \vec{\nabla} \times \left( \vec{v} \times \vec{B} \right)  +  \eta \vec{\nabla}^2 \vec{B} \,.
\label{eq:Induction}
\end{equation}
Consequently, all diffusive effects in our simulations are of numerical nature.

\subsection{Initial conditions}
\label{ss:ICs}
The setup of our simulation follows the evolution of a Milky Way (MW) like galaxy in the high resolution region of a cosmological zoom-in run ({\sc nut}; \citealt{Powell2011}). The cubic box has a side of 12.5 comoving Mpc (cMpc), containing a zoom sphere of $4.5$~cMpc across. Stellar and dark matter particles in this zoom region have mass resolutions of $m_\text{DM} \simeq 5 \cdot 10^4 \Msun$ and $m_{*} \simeq 5 \cdot 10^3 \Msun$, respectively. We allow refinement to resolve the octree grid down to a minimum physical cell size of $\sim 10$~pc in this region. A grid cell above this minimum size is refined into 8 equal cells when it contains within itself at least 8 dark matter particles or when its total mass (i.e. accounting for baryonic and dark matter) is above $8\, m_\text{DM} \Omega_m / \Omega_b $, where $\Omega_m$ and $\Omega_b$ are the mass density and baryonic mass density parameters, respectively. The studied galaxy is the most massive system forming in the high resolution region, with a halo virial mass $M_\text{vir} (z = 0) \simeq 5 \cdot 10^{11} \Msun$. Cosmological parameters are selected according to the WMAP5 cosmology \citep{Dunkley2009}.

\subsection{Further physics}
\label{ss:Subgrid}
To realistically model galaxy formation, we include in all our simulations the following subgrid physics prescriptions. The process of reionization is modelled by a UV background switched on at $z = 10$ \citep{Haardt1996}. We account for gas metal cooling above and below $10^4$~K 
interpolating {\sc cloudy} cooling tables \citep{Ferland1998} and according \citet{Rosen1995} respectively. We always assume the baryonic gas to be ideal and mono-atomic, with specific heat ratio $\gamma = 5/3$.

To simulate the formation of stars, we use a magneto-thermo-turbulent star formation model, presented in its thermo-turbulent form by \citet{Kimm2017, Trebitsch2017} adopting an extension to account for magnetic fields as in \citet{Martin-Alvarez2020}. In order to form stars, a gas cell must fulfill the following two constraints: 
\begin{itemize}
    \item it must exist on the highest level of refinement \citep{Rasera2006} at the current redshift, and
    \item the gravitational pull on the gas must be higher than the local combination of turbulent, magnetic, and thermal support.
\end{itemize}
In those cells where star formation is allowed, the star formation rate $\dot{\rho}_\text{star}$ follows a simple Schmidt law \citep{Schmidt1959}
\begin{equation}
    \dot{\rho}_\text{star} = \epsilon_\text{ff} \frac{\rho}{t_\text{ff}}\,,
    \label{eq:SchmidtLaw}
\end{equation}
with $\rho$ being the gas density and ${t_\text{ff}}$ the free-fall time. The star formation efficiency $\epsilon_\text{ff}$ is a local value that depends on the magneto-thermodynamical properties of the host and neighbouring gas cells. The efficiency follows the \citet{Padoan2011} model as described by \citet[Section 2.4.5 of their work, MFFPN model]{Federrath2012}. For more details see Appendix B of \citet{Martin-Alvarez2020}.

We model SN events using the mechanical stellar feedback prescription of \citet{Kimm2014, Kimm2015} in which a star particle injects mass, momentum and energy back to its host cell and its neighbours. We assume a Kroupa initial mass function \citep{Kroupa2001}, where each SN event injects a specific energy of $\varepsilon_\text{SN} \sim 10^{51} \erg / 10\,\Msun$. Each explosion returns a fraction of the total exploding mass to the ISM gas $\eta_\text{SN} = 0.213$, of which $\eta_\text{metals} = 0.075$ is returned as metals.

\subsection{Two origins for the galactic magnetic field}
\label{ss:TwoSources}
\begin{figure*}
    \centering
    \includegraphics[width=1.85\columnwidth]{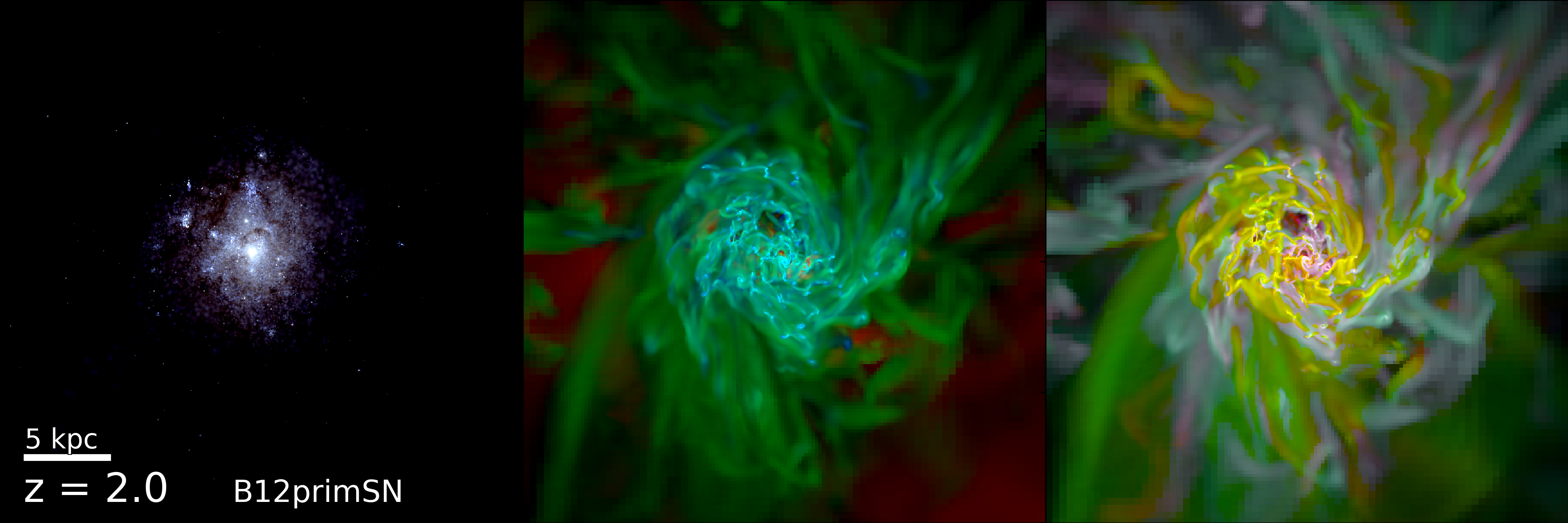}\\   
    \includegraphics[width=1.85\columnwidth]{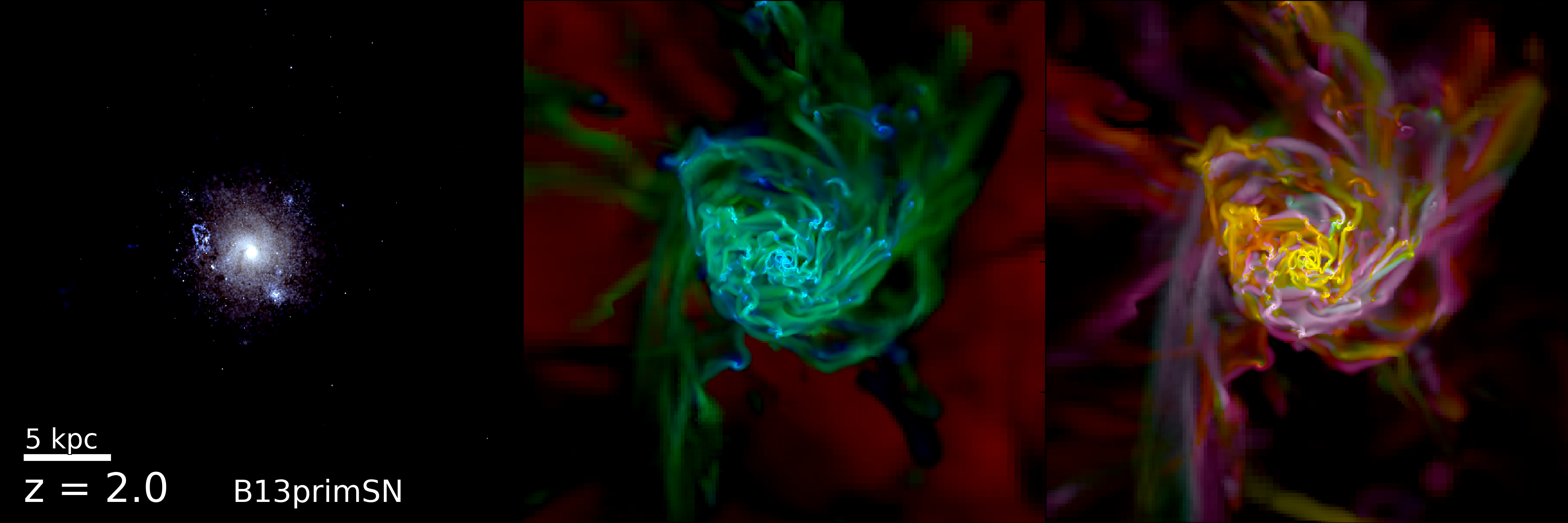}\\
    \includegraphics[width=1.85\columnwidth]{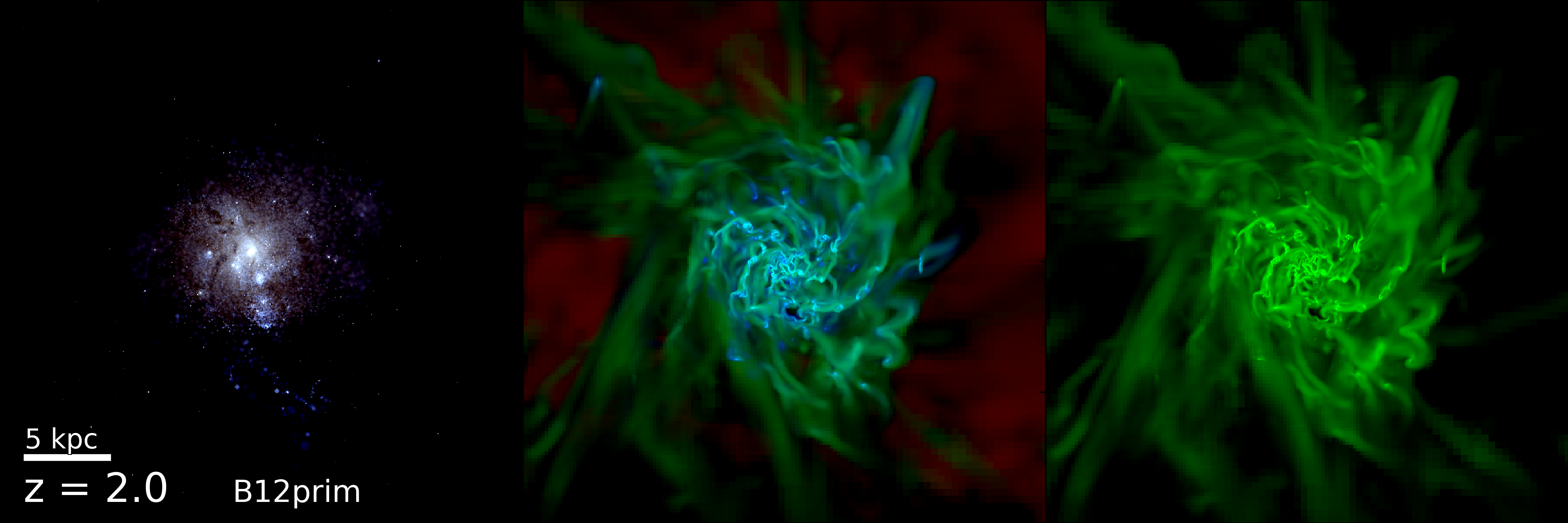}\\
    \includegraphics[width=1.85\columnwidth]{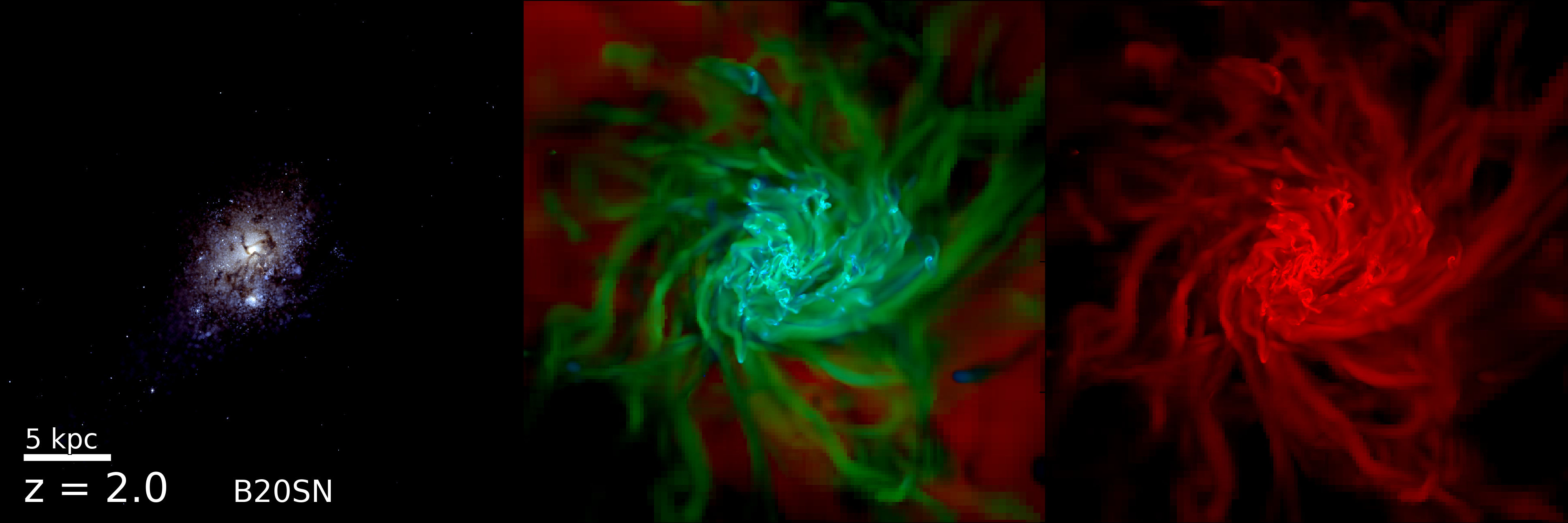}\\
    \caption{Projections of the studied galaxy at $z = 2$. Panels have physical sides and depth of $32$~kpc. Rows correspond from top to bottom to the \TrMBUp, \TrMBDown, \MBDoce, and \MBInj~runs, respectively. {\bf(Left column)} Mock colour-composite optical observation in the SDSS $u$ (blue), $g$ (green) and $r$ (red) filters, in the rest frame of the galaxy. Dust absorption along the line of sight is modelled assuming a $0.4$ dust-to-metal ratio \citep{Kaviraj2017}. {\bf(Central column)} Gas density weighted projections of the gas density (cyan), gas temperature (red), and gas magnetic energy (green). {\bf(Right column)} Gas density weighted projection of the magnetic energy decomposed into primordial (green), SN-injected (red) and cross-sources term (blue; see text for details). The magnetic energy decomposition method is described in Section~\ref{ss:Tracers}. Galaxies formed with different magnetisation channels have similar appearances, and showcase a complex interaction between magnetic fields of distinct origins.}
    \label{fig:GalaxyViews}
\end{figure*}

In this work, we investigate the properties and evolution of two fundamentally different types of magnetic fields in galaxies: magnetic fields with a primordial origin versus magnetic fields of galactic/astrophysical nature. The first will be generated prior to the onset of our simulations, whereas the second will be sourced throughout the galaxy formation process. Due to the absence of source terms in the induction equation (Equation~\ref{eq:Induction}), we can model the two sources of interest by two distinct procedures. We will model primordial magnetic fields as an ab-initio magnetic field. More specifically, our simulations will have a uniform and homogeneous magnetic field (divergenceless by construction) oriented along the z-axis of the simulated box, with a constant comoving strength $B_0$. The second source of magnetic fields will be modelled by magnetising SN ejecta. We insert $6$ small-scale closed loops of magnetic field around each SN event adjusted to produce a magnetic energy $\Einj \sim 0.01 E_\text{SN}$. This value is comparable with those used by other authors: e.g. $\sim0.03 E_\text{SN}$ by \citet{Beck2013a}, $\sim1/3\;\times\;0.01 E_\text{SN}$ by \citet{Butsky2017}, $\sim0.01 - 0.1 E_\text{SN}$ by \citet{Vazza2017}. The vast majority of our SN events take place at the highest level of refinement, which implies that the typical size of each injected loop is $\sim$10 physical pc. We find our choice to typically inject realistic magnetic fields of $\gtrsim 10^{-5}$ G \citep[e.g.][]{Parizot2006} when injected over these $\sim$10 pc scales. We provide more details on the magnetic injection method in Appendix~\ref{ap:MagInjection}. 

Due to limitations in spatial resolution, the temporal rate of turbulent amplification in galaxy formation simulations such as those studied here is low compared with that expected in the real ISM. While similar simulations with magnetic fields in the kinematic regime are able to produce relatively fast amplification (e.g. departing from $B_0 \sim 10^{-20}$ G seeds; \citealt{Rieder2016,Martin-Alvarez2018}), this is not the case for the magnetic fields studied in this work. Here our galaxies rapidly reach fields across their volume with strengths $>\muG$ shortly after their formation ($z \sim 13$), either through primordial or SN-injection seeding. Therefore, we expect minor turbulent dynamo activity, which will in turn only produce a limited growth of the magnetic energy \citep{Rieder2017a}. Consequently, the two sources of magnetic field described above are the main drivers of the galactic magnetic energy, whereas the turbulent dynamo would only be considered an important contender when contributing to the final magnetic energy budget if the magnetic field in the galaxy was weaker. Note that other dynamical processes such as e.g. disk shear (after disk formation at $z \sim 4$) will also influence the evolution of the two traced magnetic fields. We discuss this in Section \ref{s:Caveats}. Our selection of $\Einj$ is motivated by observations and previous work. $B_0$ is selected instead with the specific aim of directly obtaining realistic magnetisations in the simulated galaxy. We show in Section~\ref{ss:SameEnergy} that our choices for the $B_0$ and $\Einj$ yield both primordial and SN-injected magnetic fields of comparable strength in all runs, pointing towards saturated magnetisations. Similarly, the typical magnetic fields in the simulated galaxy compare well with those deduced by observations. Therefore, the ab-initio magnetic field $B_0 \sim 10^{-12}$ G will be a good representation of a primordial magnetic field that provides realistic magnetisation for similar {\sc ramses} cosmological simulations of spiral galaxies. In contrast, the SN-injected magnetic field will depict a star-sourced magnetic field, but will also serve as a rough approximation for other galactic channels of magnetisation: e.g. a small-scale turbulent dynamo produced magnetic field.

\begin{table*}
\caption{Summary of all the simulations studied. Columns show for each run the maximum spatial resolution ($\Dres$), the stellar feedback prescription, initial primordial magnetic field with comoving strength $B_0$, the magnetic energy injected by each SN event $\Einj$, and whether the simulation contains magnetic tracers. The final column provides further commentary on the case studied by each simulation.}
\centering
\label{table:setups}
\begin{tabular}{l l l l l l l}
\hline
Simulation & $\Dres$ & Stellar feedback & $B_0$ (G) & $\Einj$ & Tracers & Further details \\
\hline
\MBDoce & 10 pc & Mechanical & $3 \cdot 10^{-12}$ & \xmark & \xmark & Reference for primordial magnetic field-only scenario\\
\MBInj & 10 pc & Mechanical & $3 \cdot 10^{-20}$ & $0.01 E_\text{SN}$ & \xmark & Reference for astrophysical sources-only scenario\\
\TrMBUp & 10 pc & Mechanical & $3 \cdot 10^{-12}$ & $0.01 E_\text{SN}$ & \cmark & Astrophysical vs. primordial (primordial dominated)\\
\TrMBDown & 10 pc & Mechanical & $3 \cdot 10^{-13}$ & $0.01 E_\text{SN}$ & \cmark & Astrophysical vs. primordial (astro. dominated)\\
\hline
\end{tabular}
\end{table*}

While there is uncertainty about whether primordial magnetic fields are prominent in our Universe, we expect astrophysical magnetisation to be relevant in most scenarios. As a result, we require an understanding of the effects of each of these sourcing mechanisms separately, as well as of their combined role. Through four different simulations, we compare all these scenarios for the two sources of magnetisation. Each simulation is named {\it BX} where $X$ determines the strength of $B_0 \sim 3 \cdot 10^{-X}$ G. If the primordial magnetic field modelled is dynamically important, the simulation has the suffix {\it prim} added to its name. Similarly, if SN-injection is included, the name contains the suffix {\it SN}. Simulations with both a $B_0 > 10^{-15}$ G (i.e. {\it prim}) and SN injection include our magnetic tracers. The four simulated galaxies are shown in Figure~\ref{fig:GalaxyViews} and correspond to:
\begin{itemize}
    \item \MBDoce: a purely primordial magnetisation scenario with a strong primordial magnetic field ($B_0 \sim 3 \cdot 10^{-12}$~G, $\Einj = 0$). The magnetisation of the galaxy is dominated by the choice of $B_0$.
    \item \MBInj: an exclusively astrophysical origin for magnetic fields ($B_0 \sim 3 \cdot 10^{-20}$~G, $\Einj = 0.01 E_\text{SN}$). This run represents the case in which the primordial magnetic field is extremely weak and galaxy magnetisation takes place in-situ.
    \item \TrMBUp: a combination of a prominent primordial magnetic field and astrophysical sourcing ($B_0 \sim 3 \cdot 10^{-12}$~G, $\Einj = 0.01 E_\text{SN}$). This run combines the two different magnetisation scenarios with our tracing algorithm to explore the co-evolution of the two fields with a strong primordial magnetisation. 
    \item \TrMBDown: as a final addition to our suite, we include a simulation similar to \TrMBUp~but with a slightly weaker primordial magnetic field ($B_0 \sim 3 \cdot 10^{-13}$~G, $\Einj = 0.01 E_\text{SN}$). This simulation will allow us to place a lower bound for the importance of primordial magnetic fields in the magnetisation of the studied galaxy.
\end{itemize}

\begin{figure}
    \centering
    \includegraphics[width=\columnwidth]{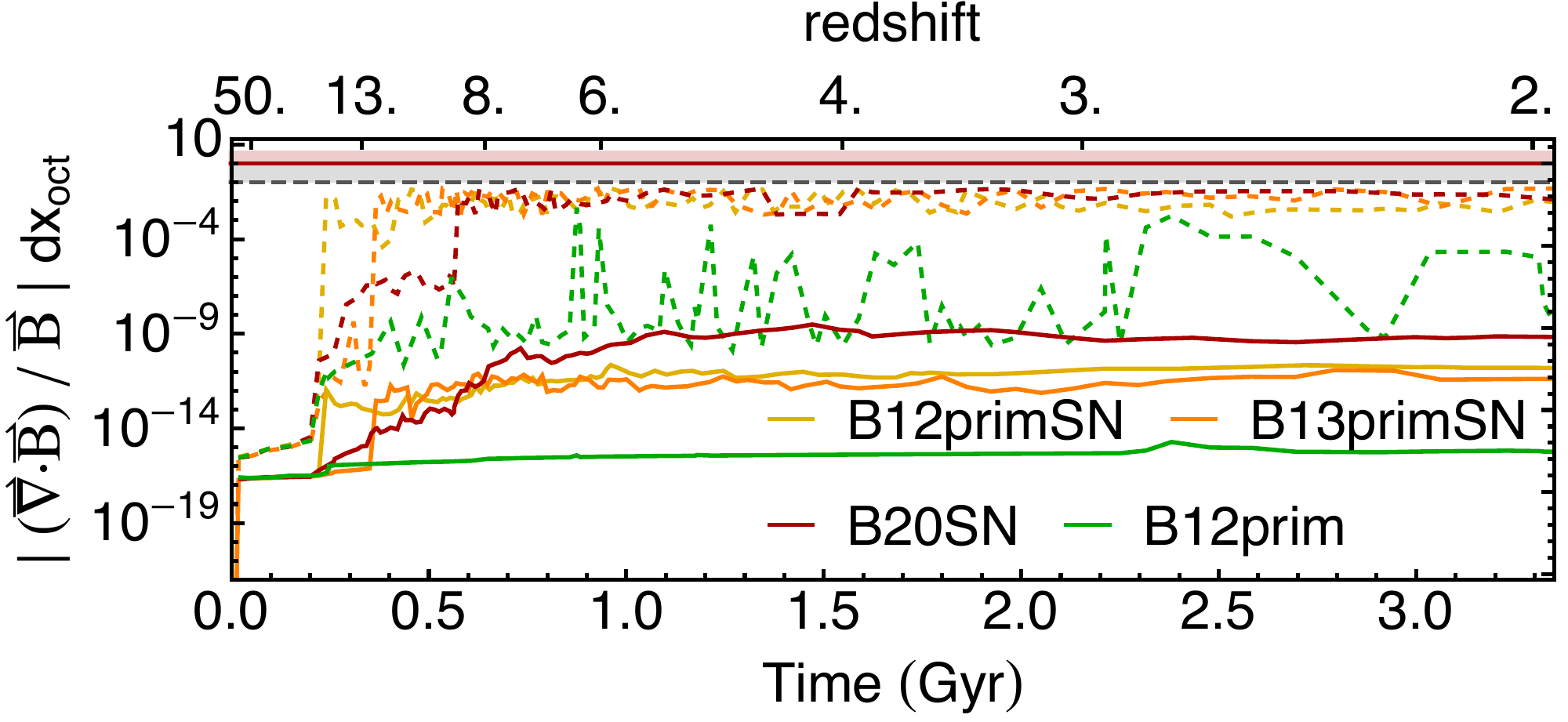}\\
    \caption{Time evolution of the ratio between magnetic field divergence produced in the oct length $|\vec{\nabla} \cdot \vec{B}| \cdot d\text{x}_\text{oct}$ to its total magnetic field $|\vec{B}|$, for the whole volume of the studied simulations. We measure divergences and total magnetic fields in octs (i.e. groups of 8 cells) to reduce the importance of {\it X} and {\it O} points where $|\vec{B}| \sim 0$. Solid (dashed) lines show the average (maximum) of the ratio. Red and grey lines show the $|(\vec{\nabla} \cdot \vec{B}) / \vec{B}| \cdot d\text{x}_\text{oct} = $1 and 0.1 ratios. While runs with injection produce higher relative divergences, maxima remain below $|\vec{B}|$ and the average divergence is markedly negligible.}
    \label{fig:Divergence}
\end{figure}

Due to the computational cost, we evolve all the runs to $z = 2$ and only continue \TrMBUp~to $z = 1$. The divergenceless behaviour of all runs is shown in Fig.~\ref{fig:Divergence}. It depicts the ratio of divergence to magnetic field, which informs on the relative importance of numerical errors, but not on their absolute values. The average of the ratio between the local magnetic divergence and magnetic fields remain significantly below the percent level, with maximum values corresponding to O- and X-points. We note that our injection mechanism produces O-points by construction and yields higher divergences, but even their maxima remain smaller than the magnetic field $|\vec{B}|$ at all times. All simulations are summarised in Table~\ref{table:setups}.

\subsection{Disentangling the magnetic field sources}
\label{ss:Tracers}
All our simulations use the MHD equations to follow directly the evolution of the total magnetic field $\vec{B}_\text{tot}$. On top of this, we employ the method we presented in \citet{KMA2019} to decompose the total magnetic field into two separate fields. These two additional fields are sourced by either $B_0$ (primordial) or $\Einj$ (astrophysical). As a result, they always add up to the total magnetic field $\vec{B}_\text{tot} = \vec{B}_\text{prim} + \vec{B}_\text{ast}$. Each of these fields is evolved independently through the induction equation and equally fulfils the solenoidal constraint. Neither of the two tracer fields interacts with the gas dynamics and can in this way be understood as a {\it colour} field.

The first magnetic {\it colour} will correspond to the primordial magnetic field, and is simply set by requiring $\vec{B}_\text{prim} = \vec{B}_\text{tot}$ at the beginning of the simulation. The second magnetic {\it colour} will be the astrophysical magnetic field, set to $\vec{B}_\text{ast} = 0$ initially and injected (sourced) by magnetised feedback as the simulation progresses. Whenever a SN event takes place with $\Einj \neq 0$, both $\vec{B}_\text{tot}$ and $\vec{B}_\text{ast}$ are modified (as described in Appendix~\ref{ap:MagInjection}). This separation into magnetic colours is illustrated in the right column of Fig.~\ref{fig:GalaxyViews}, where the colour green represents the magnetic energy associated with $\vec{B}_\text{prim}$ and the red colour corresponds to the magnetic energy found for $\vec{B}_\text{ast}$.

When separating the magnetic field into two magnetic tracers, the total magnetic energy density can be decomposed into three terms
\begin{equation}
\epsilon_\text{mag,tot} = \frac{\vec{B}^2_\text{tot}}{8 \pi} = \frac{1}{8 \pi} \left(\vec{B}^2_\text{prim} + \vec{B}_\text{ast}^2 + 2 \vec{B}_\text{ast} \cdot  \vec{B}_\text{prim}\right)\,, 
    \label{eq:CrossTerm}
\end{equation}
where the last term on the right hand side is the product of the interaction between the two magnetic fields. We will refer to this term as the cross-term magnetic energy. We note that the misalignment of $\vec{B}_\text{prim}$ and $\vec{B}_\text{ast}$ will allow this term to become negative. This term is shown in the rightmost column of Fig.~\ref{fig:GalaxyViews} as the blue colour. We discuss the meaning of the other colours in the rightmost column in Section \ref{ss:TracersEmag}. Throughout the paper, we will associate the green colour with a primordial magnetic field, the red colour with SN-injected magnetic fields, and the blue colour with the energy resulting from the interaction of the two tracer fields.

\section{Results}
\label{s:Results}
\subsection{Realistic magnetic fields in the galaxy: different origins, similar energy}
\label{ss:SameEnergy}
\begin{figure}
    \centering
    \includegraphics[width=\columnwidth]{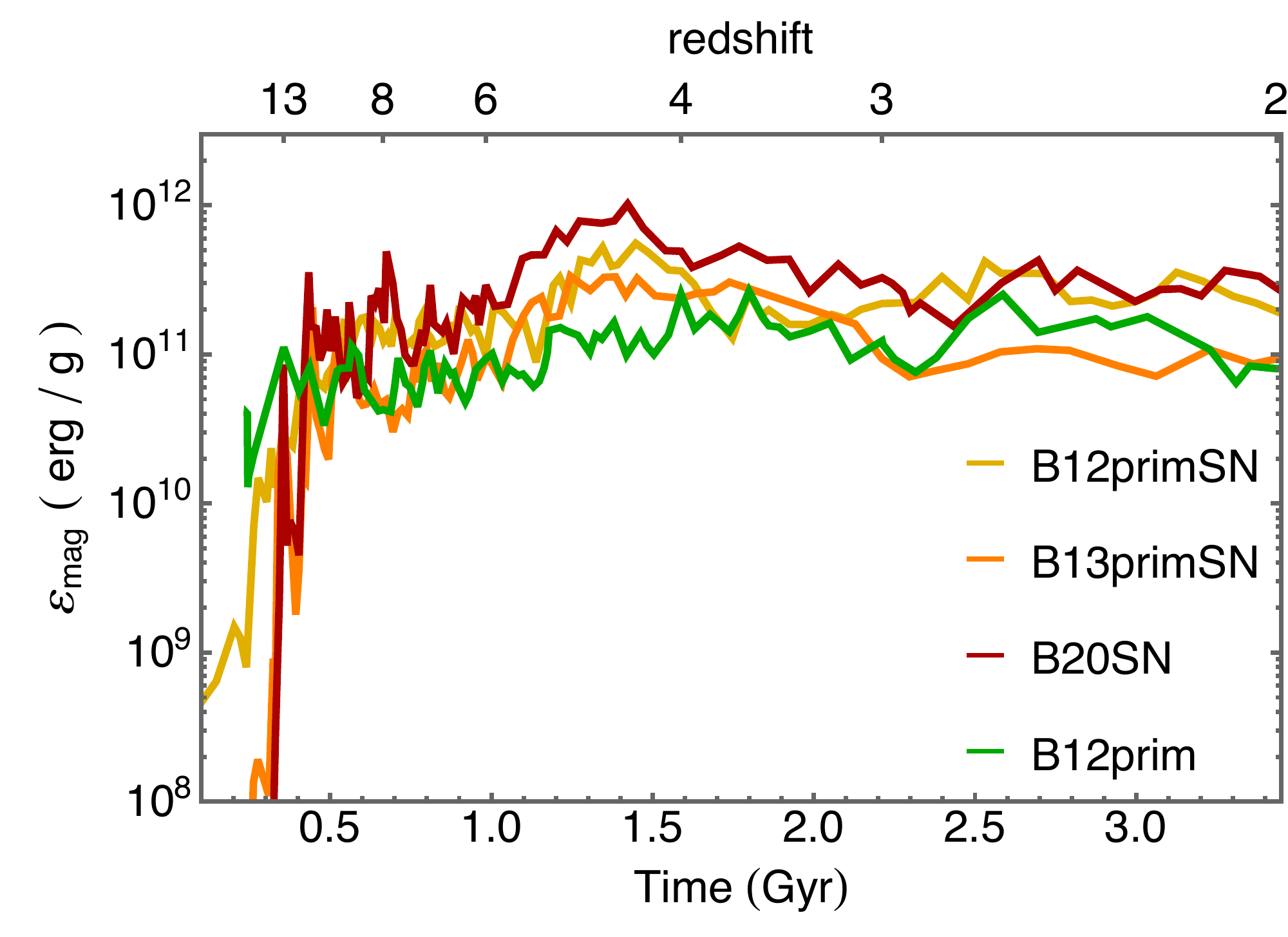}\\
    \caption{Comparison of the specific magnetic energy $\emag$ in the galactic region for all simulations. As all runs have comparable levels of magnetisation, differences in the properties and distribution of the traced magnetic energies are caused by their distinct origins.}
    \label{fig:sEmCompare}
\end{figure}

\begin{figure}
    \centering
    \includegraphics[width=\columnwidth]{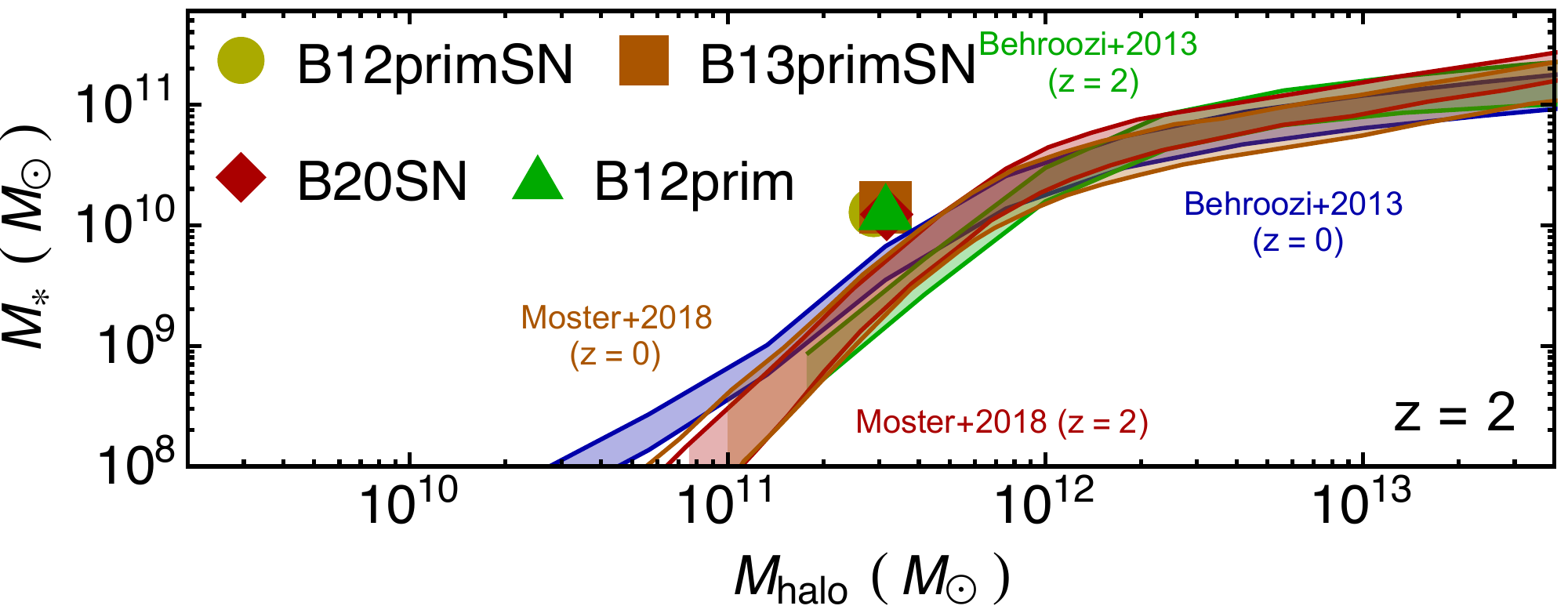}\\
    \includegraphics[width=\columnwidth]{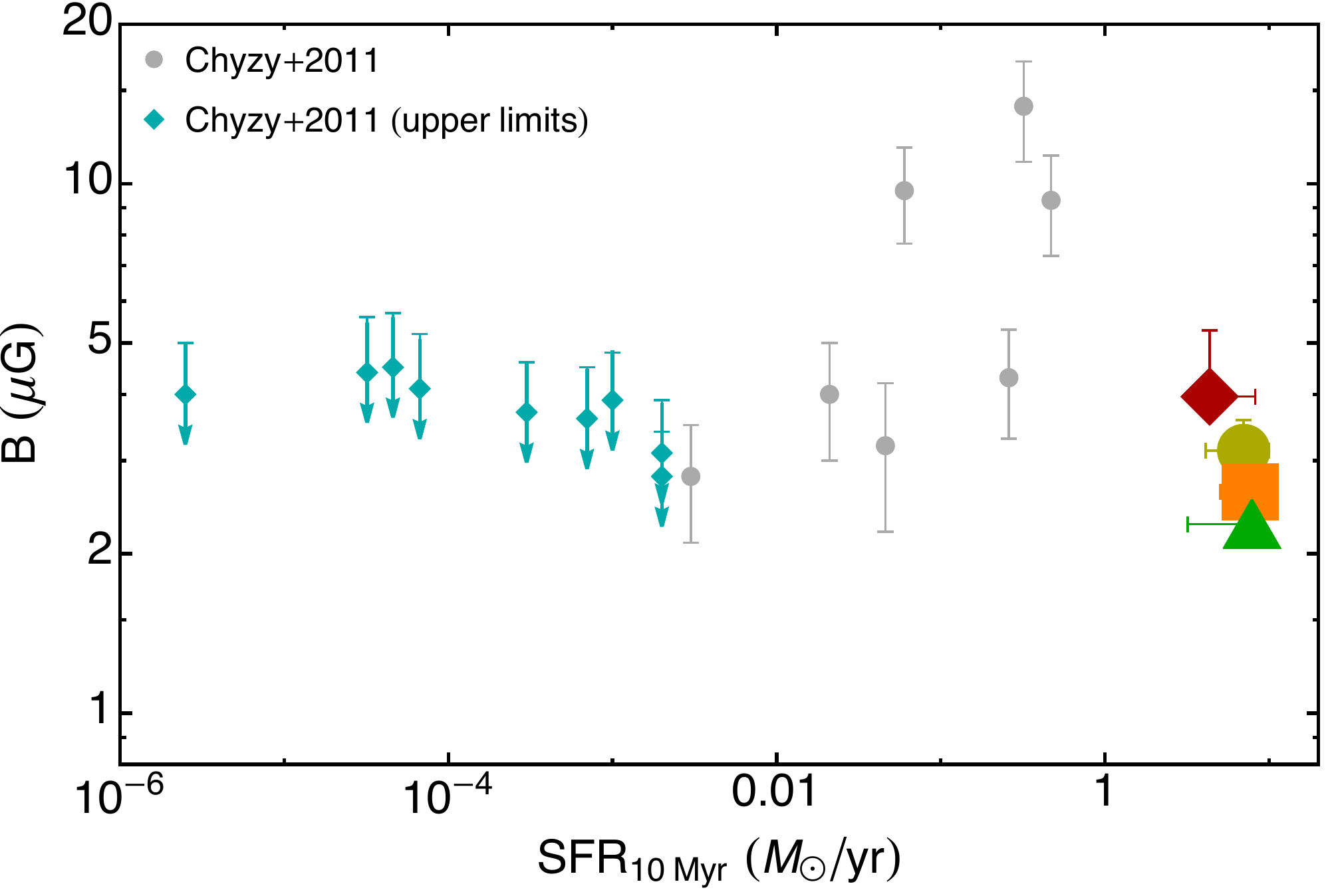}\\
    \caption{{\bf(Top panel)} stellar mass vs halo mass relation for the galaxy in all our runs at $z = 2$. All simulations display approximately the same halo and stellar masses, only marginally above the relations by \citet{Behroozi2013,Moster2018} for $z = 2$ and $z = 0$. Note that our simulations do not boost stellar feedback to calibrate this relation. {\bf(Bottom panel)} average magnetic field in the galaxy vs SFR (see text for details). Our galaxies are star forming and have magnetic field on the typical order of $B \sim \muG$. Comparison values are for Local Group systems \citep{Chyzy2011}, with upper limits shown as cyan arrows.}
    \label{fig:HMSM}
\end{figure}

In Fig.~\ref{fig:sEmCompare} we show the specific magnetic energy $\emag = E_\text{mag} / m_\text{gas}$ in the {\it galactic region}. We define the {\it galactic region} as a sphere centred on the galaxy with radius $0.2\;r_\text{DM}$, where $r_\text{DM}$ is the virial radius of the dark matter halo. The centre of the region is determined by applying our version of the {\sc halomaker} software \citep{Tweed2009} to the baryonic mass, and recursively computing the central position through the shrinking spheres method \citep{Power2003}. Fig.~\ref{fig:sEmCompare} shows that the galaxy has the same approximate magnetisation in all four simulations. \MBDoce~has the lowest $\emag$ due to our particular choice of $B_0$, whereas \MBInj~has a slightly higher specific magnetic energy, due to a more efficient injection in the absence of pre-existing magnetic fields (see Appendix~\ref{ap:MagInjection}). Nonetheless, the attained level of magnetisation does not allow us to distinguish between the origin of the magnetic field across these simulations. The global properties of galaxies are predicted to undergo only minor changes due to magnetic fields, unless greater magnetisations are explored \citep{Su2017,Pakmor2017,Marinacci2015,Martin-Alvarez2020}. For value of $B_0$ used by \MBDoce, \citet{Martin-Alvarez2020} showed that the impact of magnetic fields on the properties of the galaxy is modest. As \TrMBUp, \TrMBDown, and \MBInj~all have similar magnetisations to \MBDoce, we expect the general properties of all our galaxies to be similar. This is apparent in the mock observations and gas projections in Fig.~\ref{fig:GalaxyViews}. We also validate our galaxies against observations using the stellar mass-halo mass relation and their typical magnetic fields (see Fig.~\ref{fig:HMSM} and associated discussion).

Fig.~\ref{fig:GalaxyViews} shows projections for the galaxy in each of the studied runs. The left column shows a composite colour mock observation in rest-frame SDSS [$u$, $g$, $r$] filters. We model the obscuration by dust purely as absorption, approximating the dust mass in a given cell as $0.4$ of the metal mass \citep{Kaviraj2017}. The central column presents a gas density-weighted projection for the gas density (cyan), gas temperature (red), and magnetic energy density (green). These two panels depict a gas-rich galaxy with an extended star forming disk at approximately the beginning of Cosmic Noon - the peak of star formation of the Universe at $z \sim 2$. The appearance of the galaxy in the mock observations and gas density projections is considerably similar across runs, indicating that the different initial and injected magnetic fields portrayed do not cause significant differences between the systems. This is expected, as we have shown that all four galaxies have comparable magnetic energy budgets (Fig.~\ref{fig:sEmCompare}).

In Fig.~\ref{fig:HMSM}, we compare the simulated galaxy with two observational relations. In the top panel, we show the stellar mass vs halo mass relation. All runs have approximately the same median halo and stellar masses at $z \sim 2$. They are found in relatively good agreement with the relations by \citet{Behroozi2013} and \citet{Moster2018} at $z = 2$ and $z = 0$, where the shaded bands show $2 \sigma$ dispersion. To avoid an excess of stellar mass, simulations typically calibrate the intensity of stellar feedback \citep[e.g.][]{Crain2015,Rosdahl2018}. This aims to replace typically unaccounted for physics such as magnetic fields, cosmic rays, and radiation. While our galaxies have an excess of stellar mass compared to the empirical relations (factor of $\sim 2 - 5$), we opt not to do a calibration of the SN feedback in order to explore the topic at hand using the standard physical values attributed to $\varepsilon_\text{SN}$ and $E_\text{SN}$ (see further discussion in Section \ref{s:Caveats}). The bottom panel shows the average magnetic field vs star formation rate (SFR). The SFR$_\text{10 Myr}$ is measured as the mass of stars in the galaxy with an age lower than $10$~Myr in a given snapshot - the mass of young stars, divided by a $10$~Myr time interval. Data points and error bars correspond to the time-weighted median and upper and lower quartiles of all snapshots in the interval $2.5 > z > 2.0$, combining typically $\sim 10$ snapshots per simulation. The average magnetic field in our galaxy is in good agreement with the order of magnitude found in observations ($B \sim \muG$).

\subsection{Unraveling the origin of magnetic fields}
\label{ss:TracersEmag}

\begin{figure*}
    \centering
    \includegraphics[width=\columnwidth]{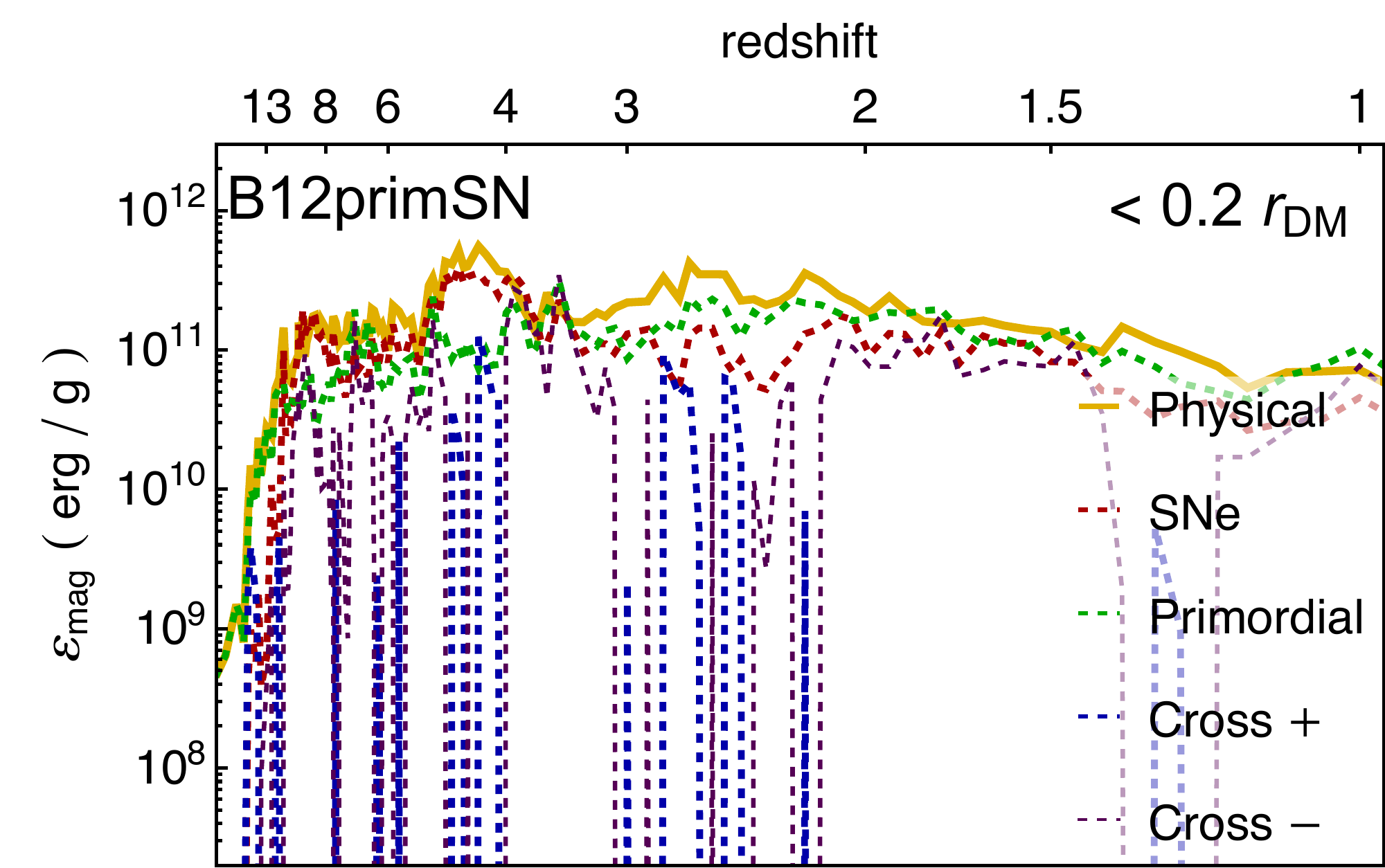}%
    \includegraphics[width=\columnwidth]{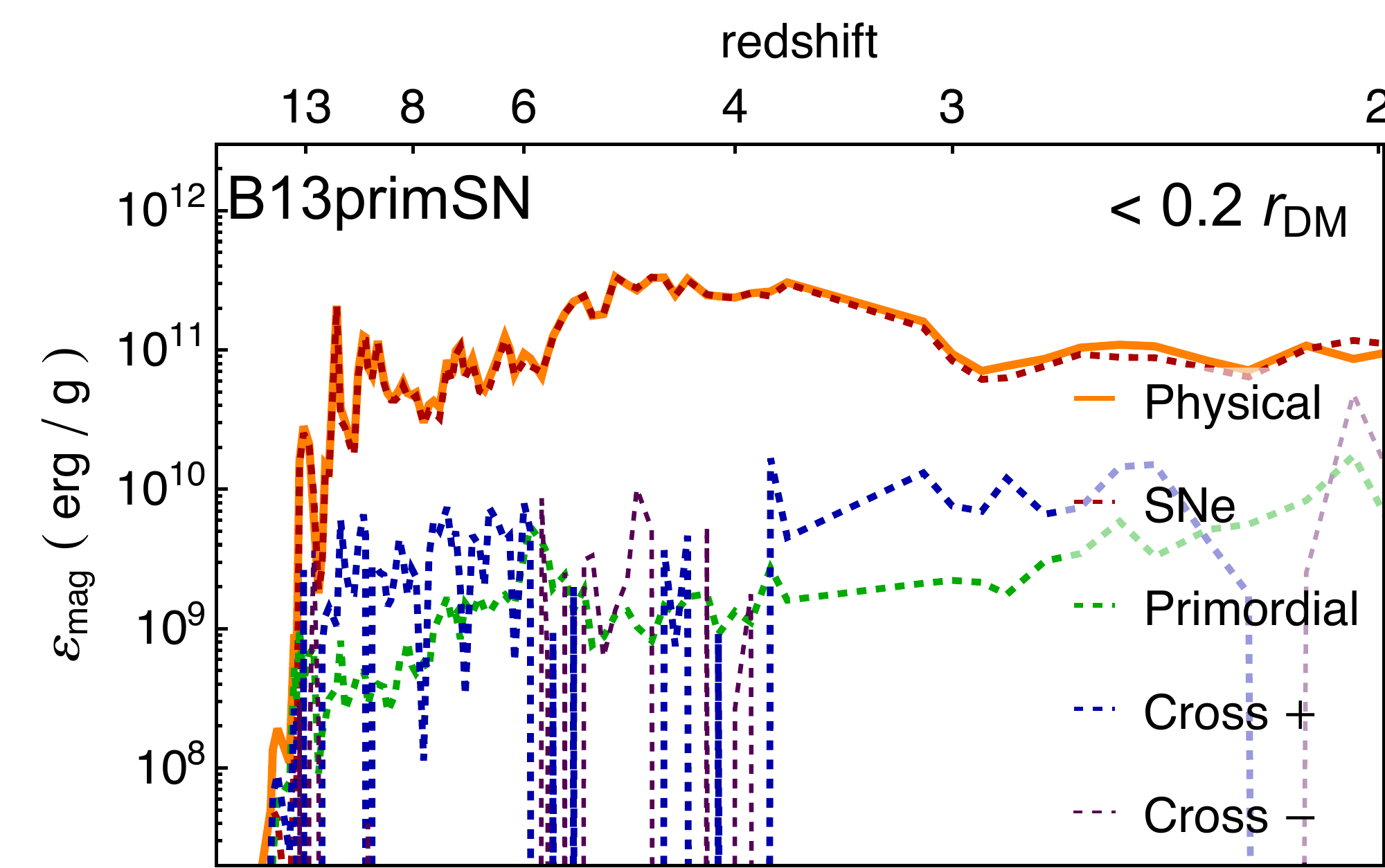}\\
    \includegraphics[width=\columnwidth]{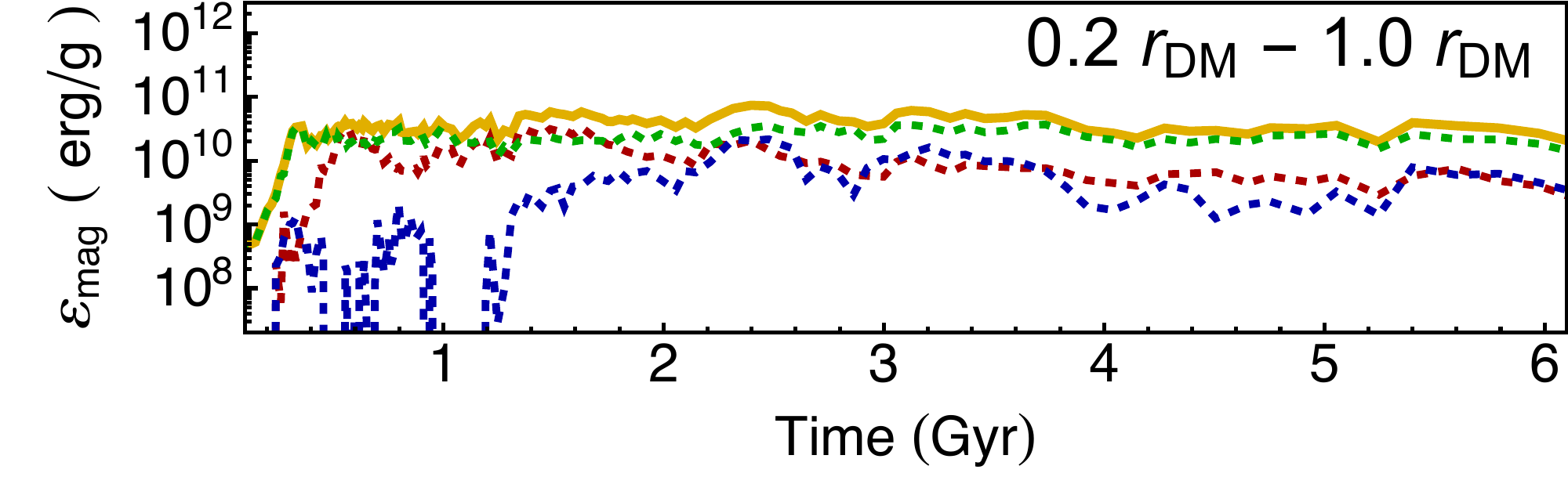}%
    \includegraphics[width=\columnwidth]{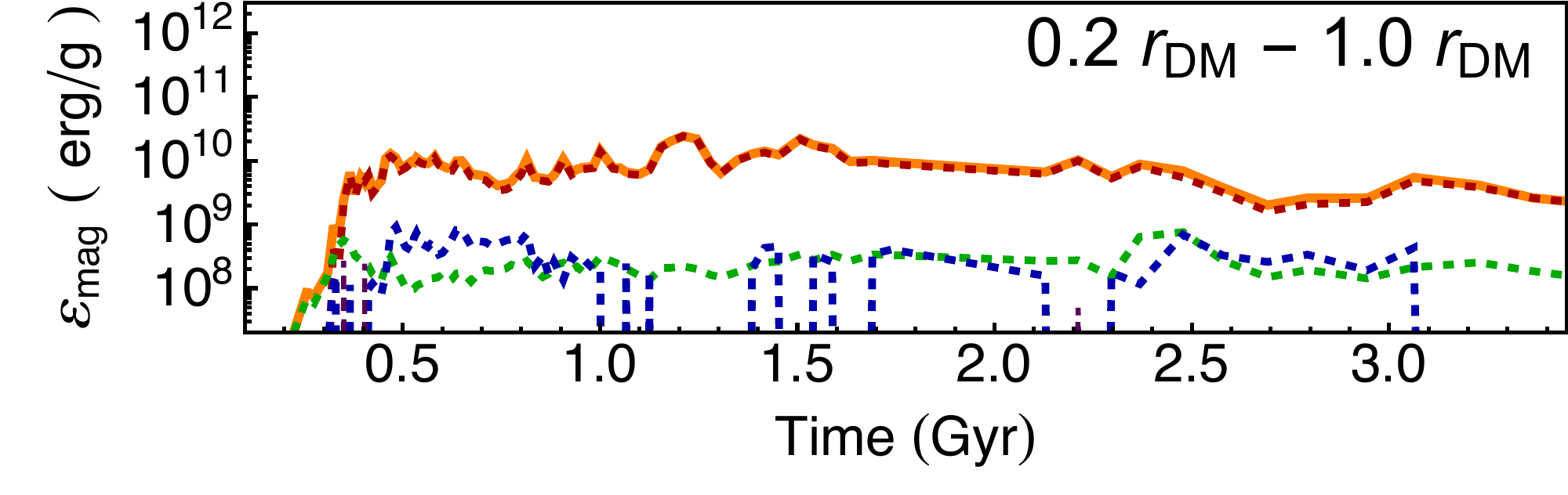}\\
    \caption{Solid lines depict the specific magnetic energy in \TrMBUp~(left column) and \TrMBDown~(right column) decomposed into its different contributions: primordial (green dashed line), SN-injected (red dashed line), and cross-term. The cross-term is further subdivided into a constructive (positive energy; blue dashed line) and destructive (negative energy; purple dashed line) interaction between the two tracer fields. {\bf (Upper panel)} galactic region ($r < 0.2\;r_\text{DM}$, see text). {\bf (Lower panel)} galactic halo ($0.2\;r_\text{DM} < r < 1.0\;r_\text{DM}$). \TrMBUp~illustrates the prevalence of primordial magnetic fields with strengths $B_0 > 10^{-12}$~G in galaxies down to $z \sim 1$. \TrMBDown~is dominated by the magnetic energy injected by SN events. While the primordial energy is not totally negligible, primordial magnetic fields become unimportant for $B_0 < 10^{-12}$ G, even in the halo of the galaxy.}
    \label{fig:sEmTracersUp}
\end{figure*}

The right-most column of Fig.~\ref{fig:GalaxyViews} displays the three forms of traced magnetic energy density: SN-injected $\epsilon_\text{mag,ast}$ (red), primordial $\epsilon_\text{mag,prim}$ (green), and cross-term $\epsilon_\text{mag,cross}$ (blue). In these projections, whenever the cross-term is negative, it is removed in equal parts from $\epsilon_\text{mag,ast}$ and $\epsilon_\text{mag,prim}$. Due to the colour composition employed, yellow tones correspond to regions featuring high values of both $\epsilon_\text{mag,prim}$ and $\epsilon_\text{mag,ast}$. Conversely, cyan and violet tones are respectively a mixture of $\epsilon_\text{mag,prim}$ and $\epsilon_\text{mag,cross}$, or $\epsilon_\text{mag,ast}$ and $\epsilon_\text{mag,cross}$. While a number of interesting properties can be appreciated in the two runs with tracers, perhaps the most striking feature is that $\epsilon_\text{mag,ast}$ and $\epsilon_\text{mag,prim}$ cannot be clearly differentiated in any region of the galaxy. All of the galaxy is instead a combination of the three different magnetic energies, both for the case of \TrMBUp~(primordial dominated) and \TrMBDown~(SN-injection dominated). However, Fig.~\ref{fig:GalaxyViews} indicates that there is a radial transition to one of the two tracers dominating as the distance from the centre increases. The outskirts of \TrMBUp~are primordially dominated whereas the SN-injected energy dominates in the outer regions of \TrMBDown. The primordially-dominated run also appears to have a more radially extended magnetisation than the SN-injected case. This could be expected, as these primordial magnetic fields should be ubiquitous at all scales (\citealt{Vazza2017} find a similar result at larger scales). We will review in more detail the radial distribution of the different types of magnetic energy in Section~\ref{ss:Profiles}. These projections showcase an important conclusion recurrent through this work: primordial magnetic fields with strengths $B_0 > 10^{-12}$ G are non-negligible in the magnetisation and evolution of galaxies. They dominate the magnetic energy budget in these systems at least up to $z \sim 1$, past the peak of star formation of the Universe, and potentially until the present day. $B_0 > 10^{-12}$ G was also found in \citet{Martin-Alvarez2020} to be the approximate magnetic field above which noticeable effects on the global properties of galaxies due to the presence of magnetic fields manifest, with larger effects when studying even stronger $B_0$.

To review more quantitatively how magnetic energy sources compare in \TrMBUp~and \TrMBDown, we show in Fig.~\ref{fig:sEmTracersUp} their corresponding specific magnetic energies, separated into magnetic energy tracers. The upper panels show these for the galactic region ($r < 0.2\;r_\text{DM}$), while the bottom panels show the same quantities for the remainder of the halo ($0.2\;r_\text{DM} < r < 1.0\;r_\text{DM}$). Both runs have a total (solid lines) specific magnetic energy $\emag \sim 10^{11} \erg / \gram$ (and magnetic energy density $\epsilon_\text{mag} \sim 10^{-14} \erg / \cc$) in the galactic region. Fig.~\ref{fig:sEmTracersUp} confirms that when decomposed into magnetic tracers (dashed lines), \TrMBUp~is dominated by the primordial magnetic field (green dashed lines) whereas \TrMBDown~is dominated by the SN-injected magnetic field (red dashed lines). The discussion of the cross-term energy (dashed blue line for the positive/constructive fraction and dashed purple line for the negative/destructive fraction) is deferred until the end of this Section. While for \TrMBUp, the two tracer energies contain a significant fraction of the total magnetic energy (at $z < 4$, both energies are comparable $\emagPMF \sim 2\;\emagSN$), $\emagPMF$ is secondary for \TrMBDown. 

\begin{figure*}
    \centering
    \includegraphics[width=\columnwidth]{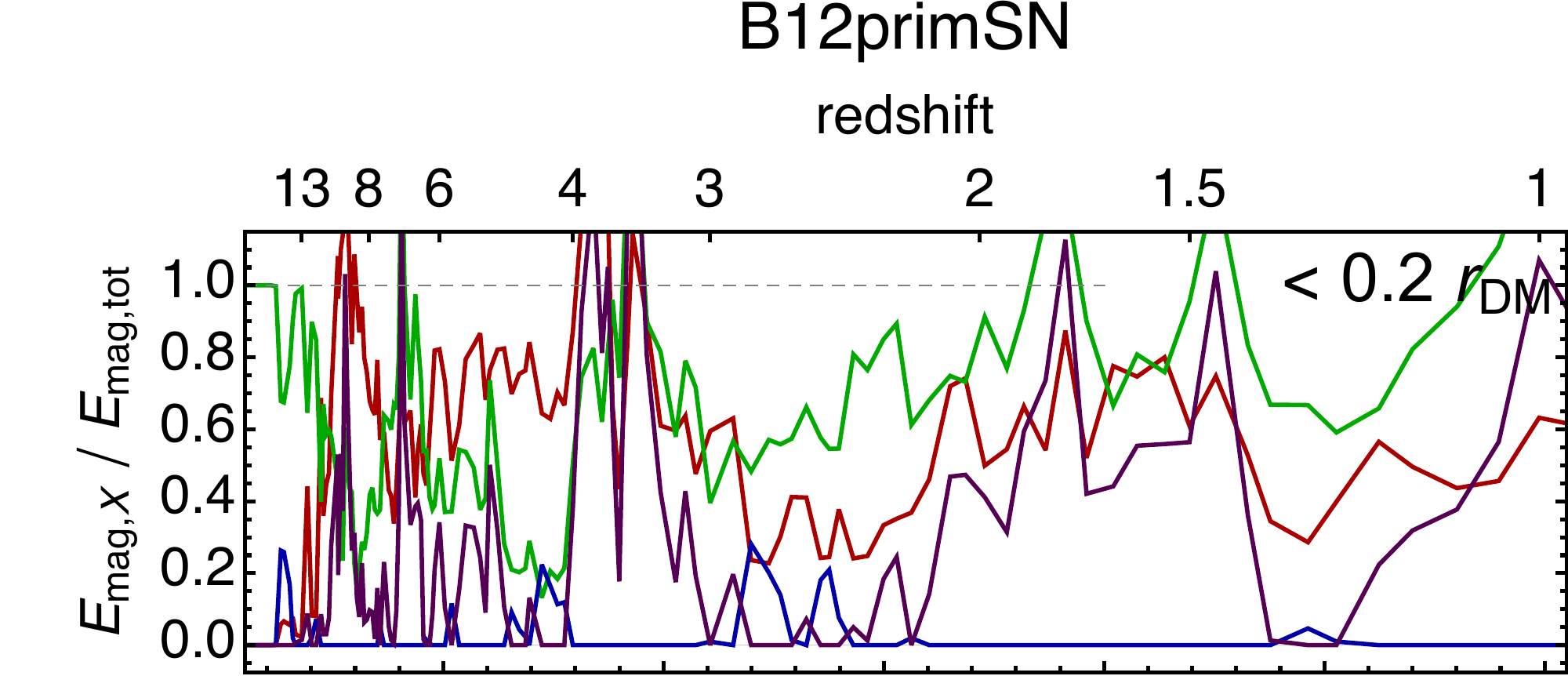}%
    \includegraphics[width=\columnwidth]{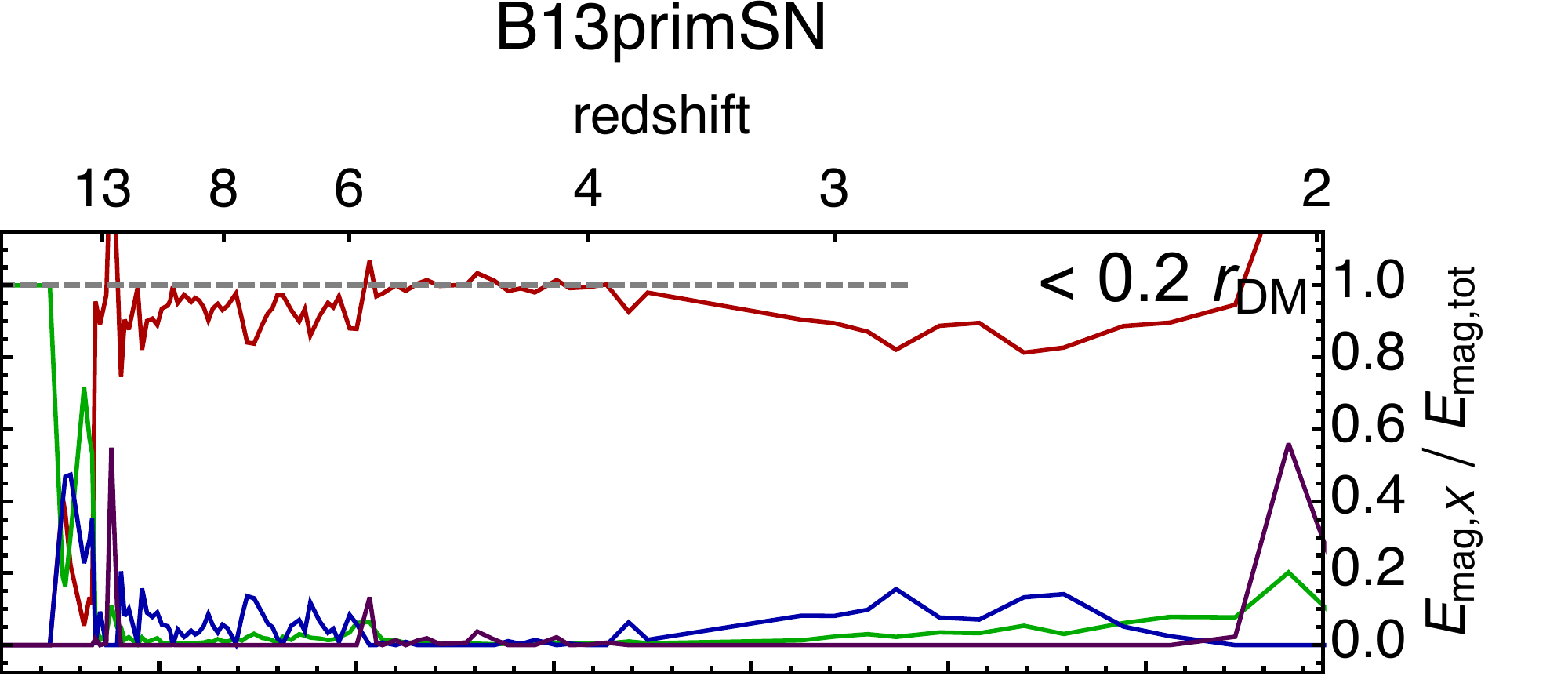}\\
    \includegraphics[width=\columnwidth]{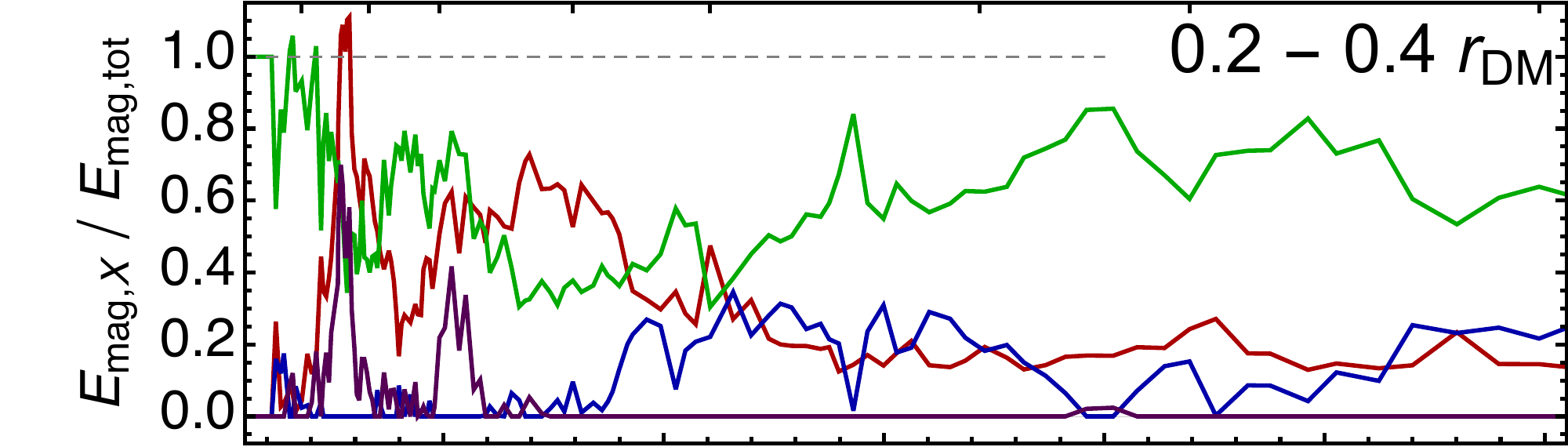}%
    \includegraphics[width=\columnwidth]{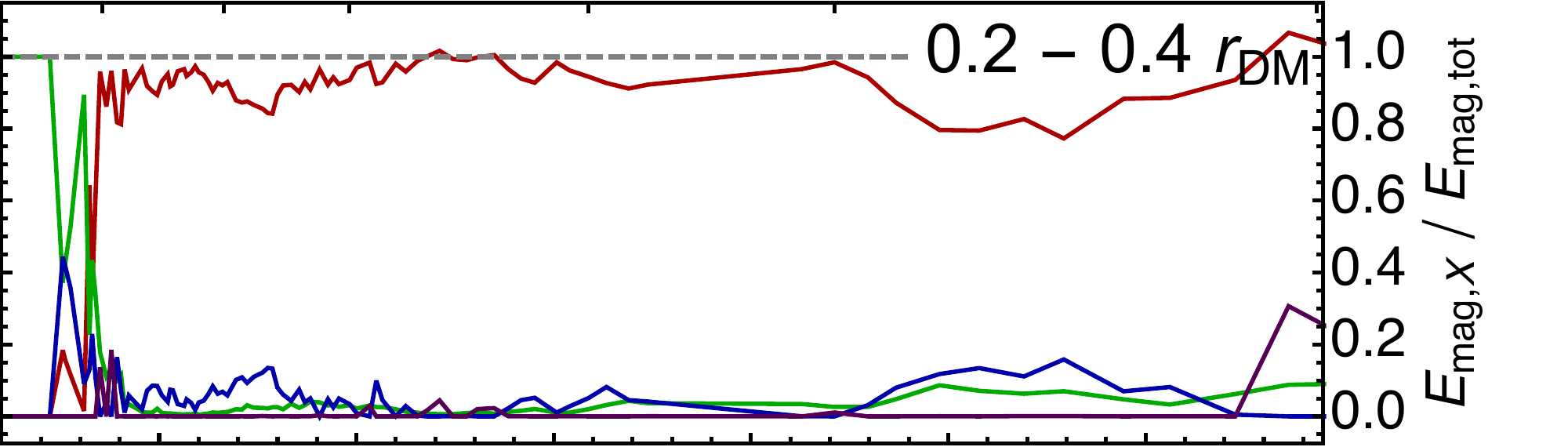}\\
    \includegraphics[width=\columnwidth]{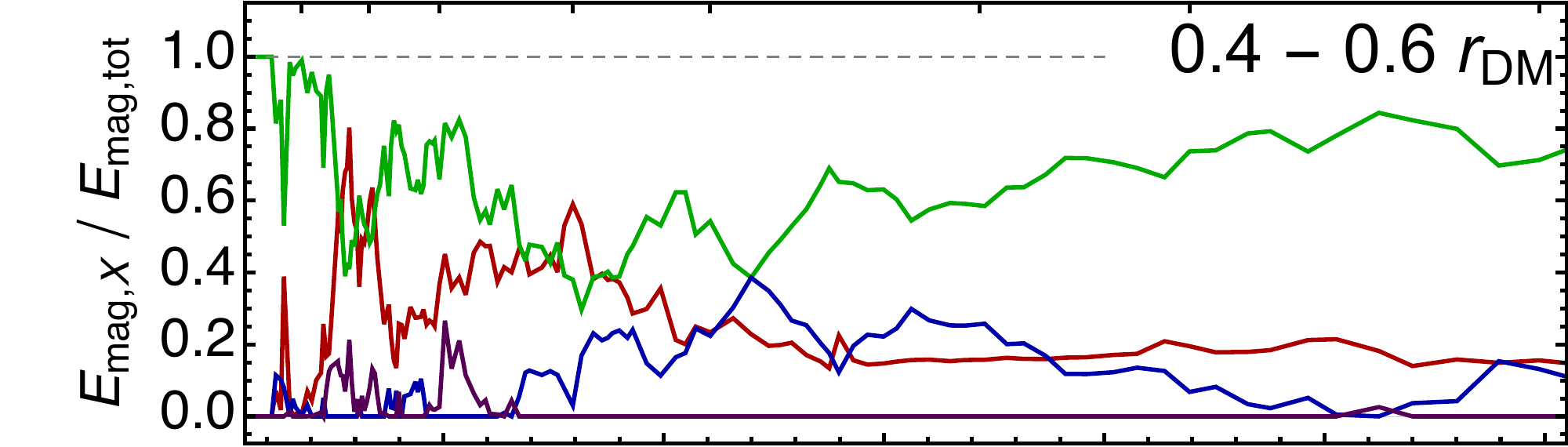}%
    \includegraphics[width=\columnwidth]{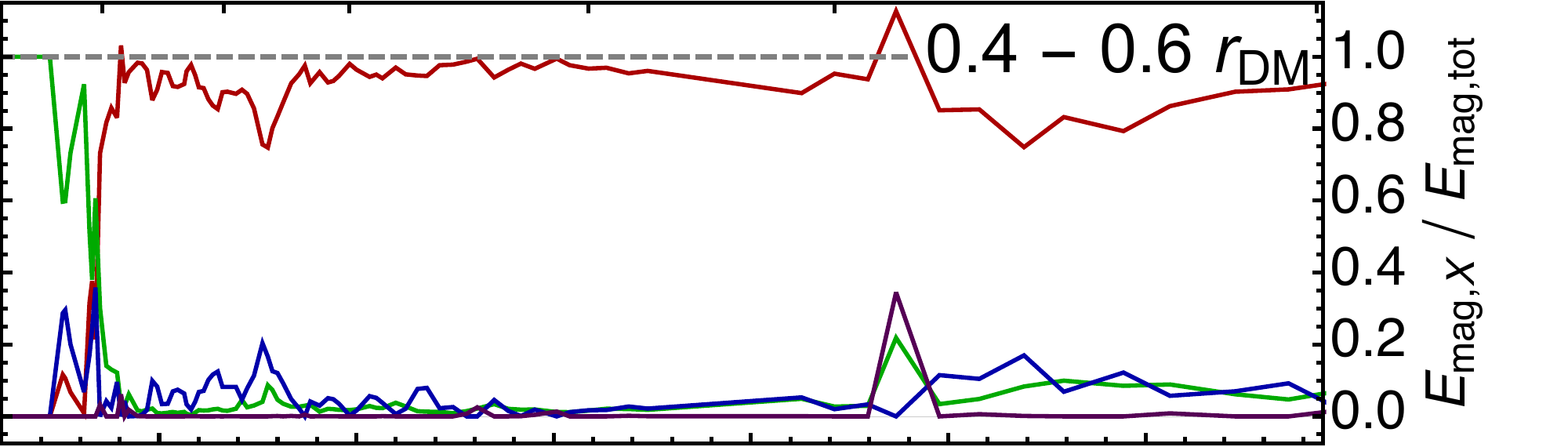}\\
    \includegraphics[width=\columnwidth]{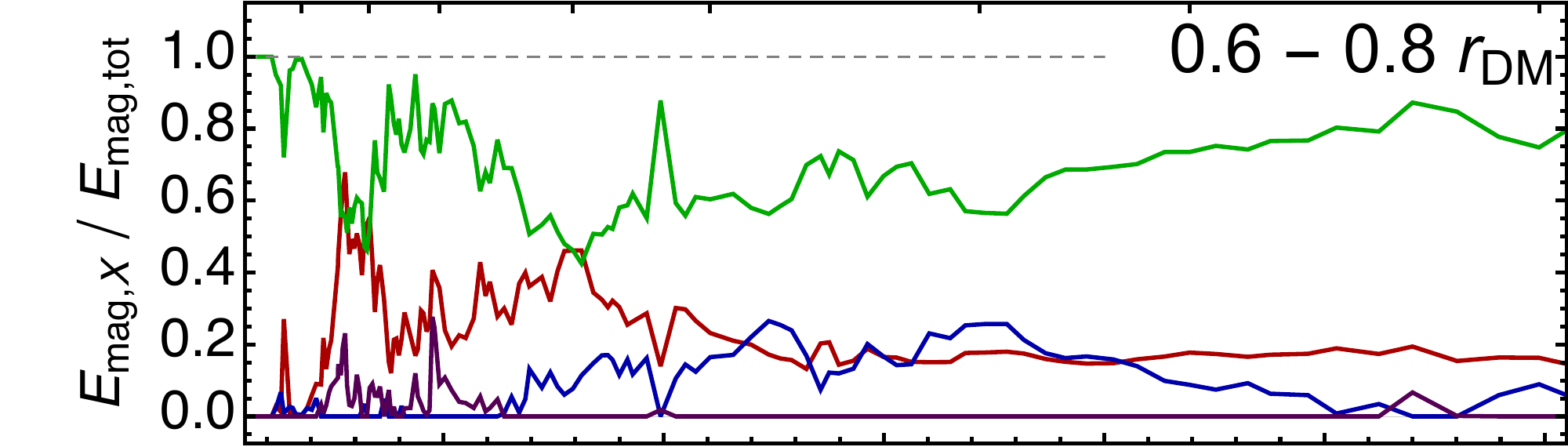}%
    \includegraphics[width=\columnwidth]{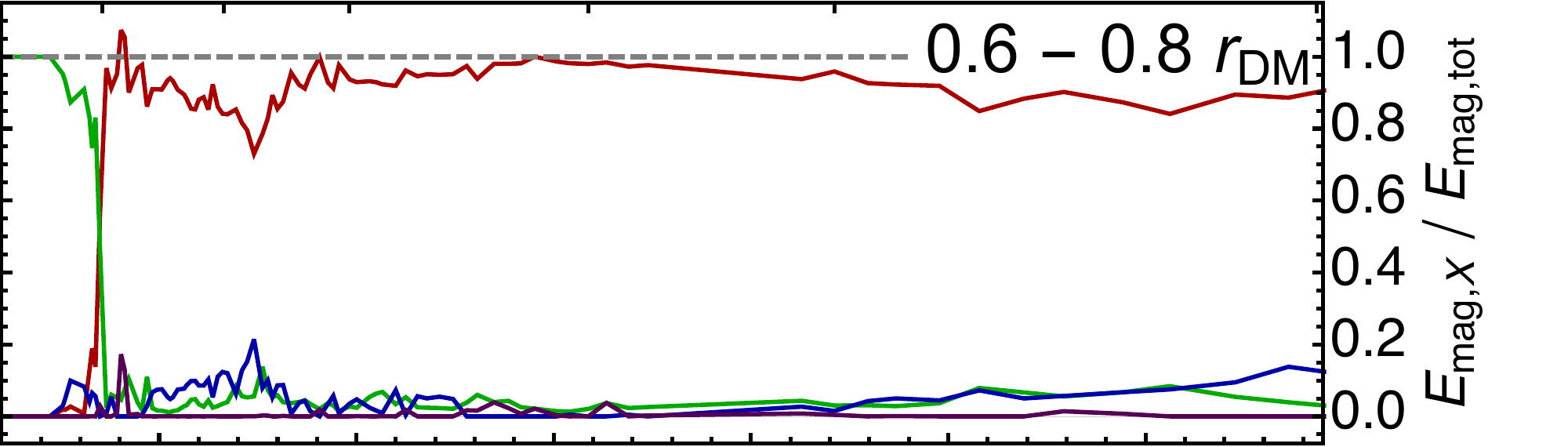}\\
    \includegraphics[width=\columnwidth]{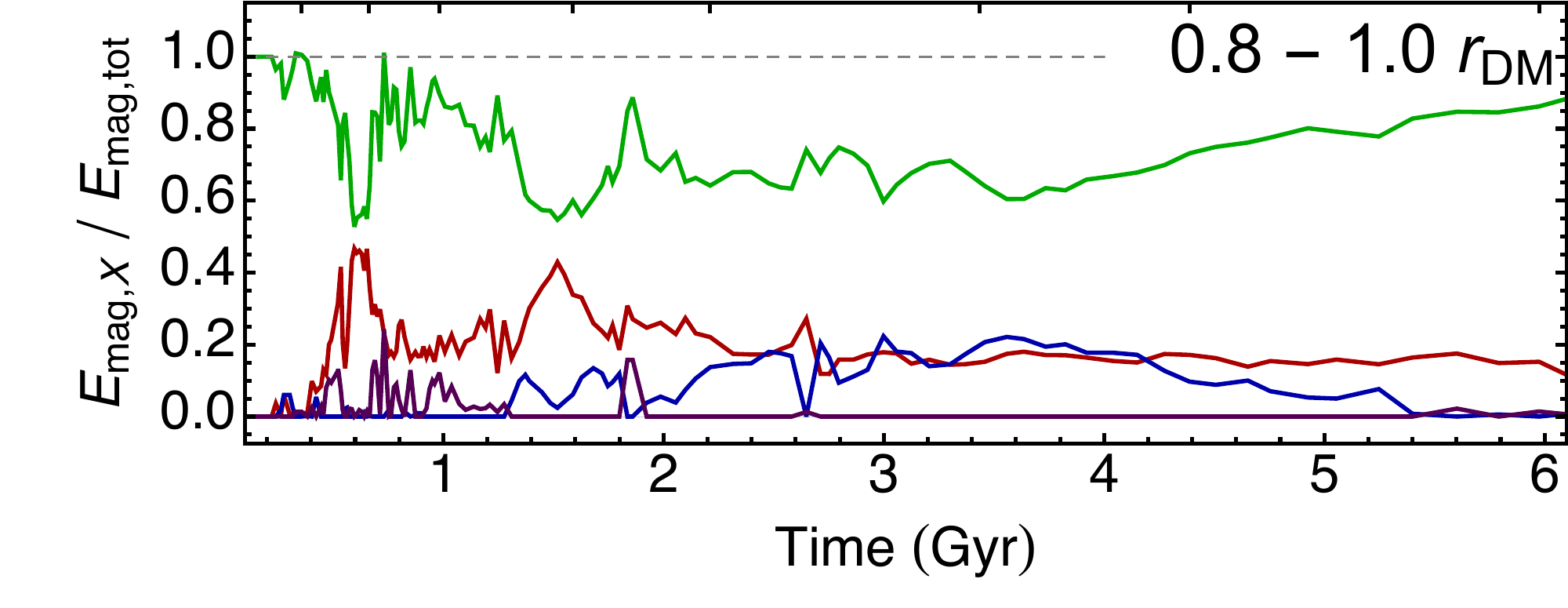}%
    \includegraphics[width=\columnwidth]{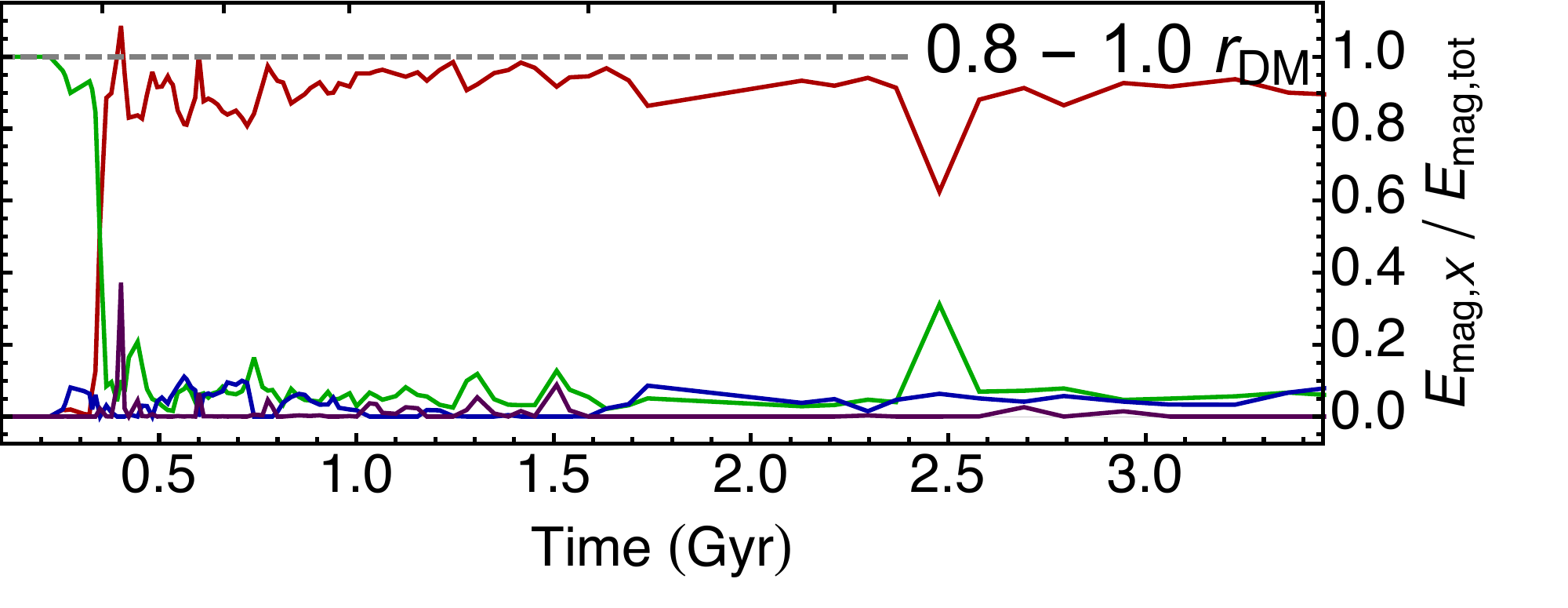}\\
    \caption{Magnetic tracers fractional contribution to the total magnetic energy as a function of time for different halo sections. From top to bottom, panels show the $r < 0.2\;r_\text{DM}$, $0.2\;r_\text{DM} < r < 0.4\;r_\text{DM}$, $0.4\;r_\text{DM} < r < 0.6\;r_\text{DM}$, $0.6\;r_\text{DM} < r < 0.8\;r_\text{DM}$, $0.8\;r_\text{DM} < r < 1.0\;r_\text{DM}$ regions. Left and right columns show the \TrMBUp~and \TrMBDown~simulations, respectively. Coloured lines correspond to the primordial (green), SN-injected (red), positive cross-term (blue), and negative cross-term (purple) magnetic energies. SN-injected fields are important for the \TrMBUp~galactic region, while the rest of its halo is dominated by primordial magnetic energy. Contrarily, shortly after the onset of star formation, \TrMBDown~is always dominated by the energy injected by SN events.}
    \label{fig:sEmSlicesUp}
\end{figure*}

To facilitate the analysis of the relative importance of each tracer field, we show the ratio of each traced magnetic energy to the total magnetic energy in Fig.~\ref{fig:sEmSlicesUp}. Primordial, SN-injected, positive cross-term and negative cross-term are once again shown by green, red, blue, and purple coloured lines, respectively. Each row shows a radial slice of the dark matter halo. Left and right columns show these measurements for \TrMBUp~and \TrMBDown, respectively.

Similar ratios for each traced magnetic energy are observed in the halos: \TrMBUp~is primordially dominated (after $z = 4$, $\emagSN\sim20\%$), whereas the magnetic energy in the halo of \TrMBDown~is dominated by the outflows from the galaxy. Understanding the magnitude of astrophysically-generated magnetic fields expelled to the CGM (circumgalactic medium), the distances they reach into the IGM, as well as their volume fraction and relative importance with respect to primordial magnetic fields is paramount for observations desiring to constrain the cosmic magnetic field using measurements around galaxies as a proxy \citep[e.g.][]{Neronov2010,Broderick2018}. If primordial magnetic fields are strong (e.g., $B_0 > 10^{-12}$ G), it will be unlikely that observations are polluted by astrophysical magnetic fields. In \TrMBUp, $\emagSN$ drops below $5\%\; \emag$ at 
$\sim 60$~kpc, with $r_\text{DM} \sim 90$~kpc. In contrast, the primordial magnetic energy in \TrMBDown~only becomes $\emagPMF \simeq \emagSN$ outside the halo ($r \gtrsim 100$~kpc). Note that the indicated distances are only lower limits, as additional physics, such as cosmic rays, could potentially enhance the driving of galactic winds \citep{Salem2014b,Dashyan2020} and even increase their magnetisation \citep{Ruszkowski2017}. Furthermore, we do not account for AGN in our simulations, but these are expected to be active at some point during the evolution of typical MW-like galaxies \citep[][]{Heywood2019InflationEvent}. Similarly, AGN-driven winds may be considerable even in dwarf galaxies \citep[see e.g.][]{Koudmani2019, Koudmani2020}, with AGN capable of polluting the primordial magnetic field around galaxies with a range of masses. We discuss this issue further in Section~\ref{s:Caveats}, and defer the inclusion of AGN to future work. If the studied $B_0$ is high, the importance of primordial magnetic fields in the halo at high redshift continually increases in time as additional pristine and magnetised gas is accreted by the system. After $z \sim 3$, all slices of the \TrMBUp~halo remain dominated by primordial magnetic energy. Furthermore, we find that $\emag \gtrsim 10^{10} \erg/\gram$ in the halo of a galaxy indicates the existence of strong primordial magnetic fields (according to the caveats discussed in Section \ref{s:Caveats}).

\begin{figure*}
    \centering
    \includegraphics[width=2\columnwidth]{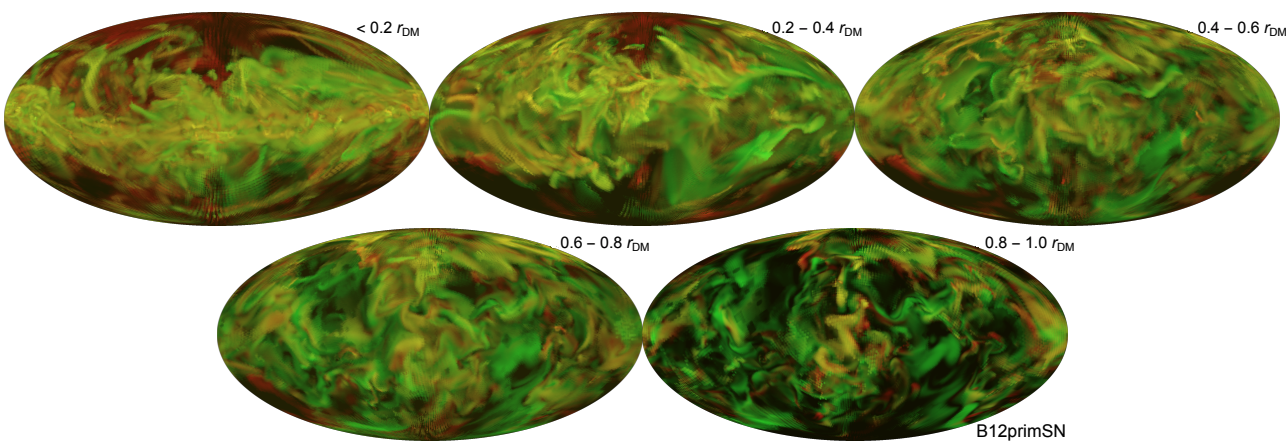}\\
    \includegraphics[width=0.667\columnwidth]{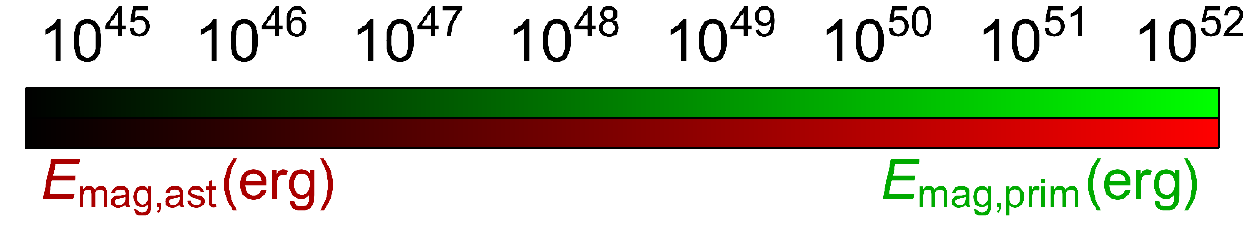}\\
    \includegraphics[width=2\columnwidth]{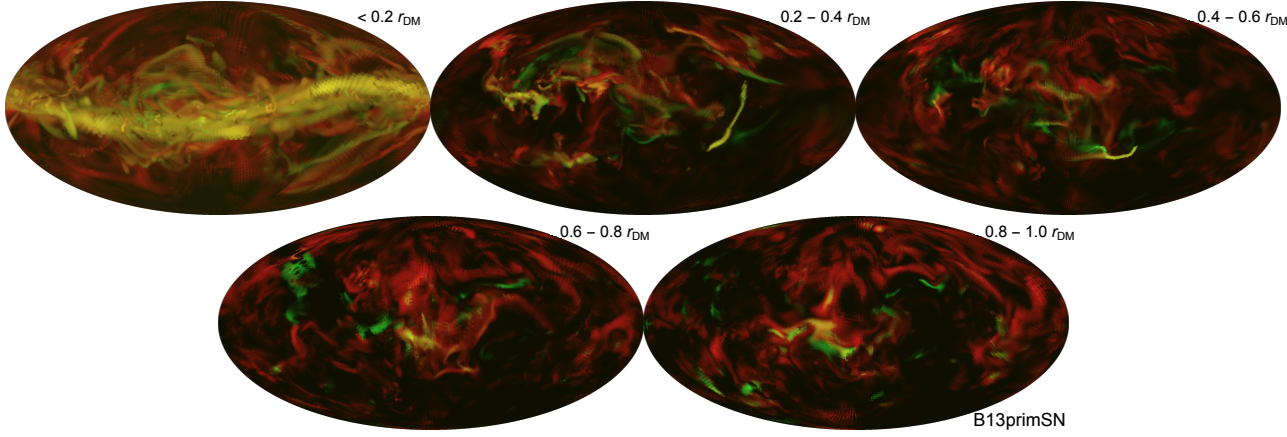}%
    \caption{All-sky Mollweide projections of various halo radii intervals centred on the \TrMBUp~(top) and \TrMBDown~(bottom) galaxies. The projection z-axis is aligned with the angular momentum of the galaxy. From top left to bottom right, each image shows the $r < 0.2\;r_\text{DM}$, $0.2\;r_\text{DM} < r < 0.4\;r_\text{DM}$, $0.4\;r_\text{DM} < r < 0.6\;r_\text{DM}$, $0.6\;r_\text{DM} < r < 0.8\;r_\text{DM}$, $0.8\;r_\text{DM} < r < 1.0\;r_\text{DM}$. Colours represent the primordial (green) and SN-injected (red) magnetic energies. Blue is omitted for the sake of clarity. Close to the \TrMBUp~galaxy SN-injected fields dominate the outflows. A higher mixture of both sources is found at larger radii, where primordial magnetic field dominate the magnetic energy. While secondary, SN-injected magnetic fields extend to some regions of the halo outer edges. \TrMBDown~is dominated by SN-injected fields in the entire volume, but in a patchy distribution.}
    \label{fig:Mollweide}
\end{figure*}

In the presence of a disk, galactic outflows will preferentially emerge perpendicularly to its plane. Thus, it is interesting to review the distribution of the different types of magnetic energy across the sky. All-sky Mollweide projections of the simulated galaxies at $z = 2$ are shown in Fig.~\ref{fig:Mollweide}. For each projection, we place an observer at the galactic centre and divide the sky using a HEALPix decomposition \citep{Gorski2005}. Each pixel projects cells with an angular distance from the pixel centre $< 0.5 \Delta \theta$, with angular resolution $\Delta\theta\sim0.46^\text{o}$. The map is shown in Mollweide coordinates, where the equatorial poles are aligned with the galactic angular momentum. Fig.~\ref{fig:Mollweide} is an RGB colour composite for the SN-injected (red) and primordial (green) magnetic energies - blue colour is omitted for clarity. In agreement with our previous results, primordial magnetic fields dominate \TrMBUp~at all radii. The top left map shows the galactic region (the inner $200$~pc are ignored to avoid projection effects), where the galactic disk is evidenced by the amber band along the equator. The colour indicates that in the disk, both SN-injected and primordial magnetic fields are comparable. The top left and the top middle ($0.2\;r_\text{DM} < r < 0.4\;r_\text{DM}$) maps have SN-injected magnetic energy prevailing at latitudes $> 50^\text{o}$ above and below the equator. However, in the top right ($0.4\;r_\text{DM} < r < 0.6\;r_\text{DM}$), bottom left  ($0.6\;r_\text{DM} < r < 0.8\;r_\text{DM}$), and bottom right ($0.8\;r_\text{DM} < r < 1.0\;r_\text{DM}$) panels the distribution is more homogeneous and randomised. While primordial magnetic fields are more important, certain smaller regions exist that are dominated by SN-injected magnetic fields. Due to this, and the complex morphology of the magnetic field, some degree of primordial magnetic field pollution is expected in haloes, unless $B_0 \gg 10^{-12}$ G. These more extreme primordial magnetic fields have been shown to have other potential signatures of their presence including modified properties of galaxies or reionisation histories \citep[e.g.][]{Marinacci2016,Safarzadeh2019,Martin-Alvarez2020,Sanati2020,KMA2021}. \TrMBDown~is globally dominated by the SN tracer magnetic field, and portrays a more patchy magnetisation at the outskirts of the halo. There are some filaments that are magnetised by the primordial component. Interestingly, as we move from the inner to the outer region of the halo, the coherence length of magnetic energy structures appears to increase, most likely correlated with the coherence of bulk velocity flows, as well as the initialisation of the primordial magnetic field as a uniform field \citep{Marinacci2015}. We will explore in Sections~\ref{ss:Phases} and \ref{ss:Metal} whether the origin of the magnetic fields in the halo can be distinguished by examining gas properties such as temperature or metallicity. 

Up until this point, we have not discussed the cross-term magnetic energy resulting from the interaction between the two types of magnetic fields considered. Figs.~\ref{fig:sEmTracersUp} and \ref{fig:sEmSlicesUp} show that the cross-term magnetic energy is typically well below the other two types of energies. The proportionality of the cross-term with both tracers in Equation~\ref{eq:CrossTerm} suppresses the importance of this cross energy when one of the tracers comprises the majority of the magnetic field. As a result, the cross-term energy is unimportant in the halos ($r > 0.2\;r_\text{DM}$) of \TrMBUp~and \TrMBDown, and equally so for the \TrMBDown~galaxy. In contrast, the galactic region of \TrMBUp~features both significant $\vec{B}_\text{ast}$ and $\vec{B}_\text{prim}$, which leads to a non-negligible contribution of the cross-term. Our maximisation of the injected energy per loop (described in Appendix \ref{ap:MagInjection}) leads to the injection mechanism predominantly injecting positive cross-term energy. Interestingly, the negative cross-term is above the positive fraction in the galactic region of \TrMBUp~throughout most of the simulation (Fig.~\ref{fig:sEmSlicesUp}). We attribute this to the behaviour of the magnetic tracers during magnetic reconnection. If two reconnecting magnetic lines are associated with different tracers, the total magnetic field cancels. However the traced magnetic fields superimpose in anti-alignment, producing a negative contribution to the cross-term. The turbulent nature of the ISM causes significant magnetic reconnection, which in turn increases the negative cross-term. In contrast, in the simulated halo of \TrMBUp, the positive cross-term energy is prevalent due to bulk motions mostly driven by galactic outflows and the low amount of turbulence. Nonetheless, the cross-term energy is seldom comparable to the SN-injected energy and mostly unimportant with respect to the primordial budget. The high fraction of cross-term magnetic energy in \TrMBUp~indicates a substantial interaction between the two traced magnetic fields, thus both being important for the magnetic effects taking place in the galaxy.

We have shown in this Section that primordial magnetic fields are prevalent in galaxies and their haloes when their comoving strength is $B_0 > 10^{-12}$~G. We now proceed to examine how the energy is distributed spatially across the galaxy.

\subsection{The spatial separation of astrophysical and primordial magnetic fields}
\label{ss:Profiles}
\begin{figure*}
    \centering
    \includegraphics[width=\columnwidth]{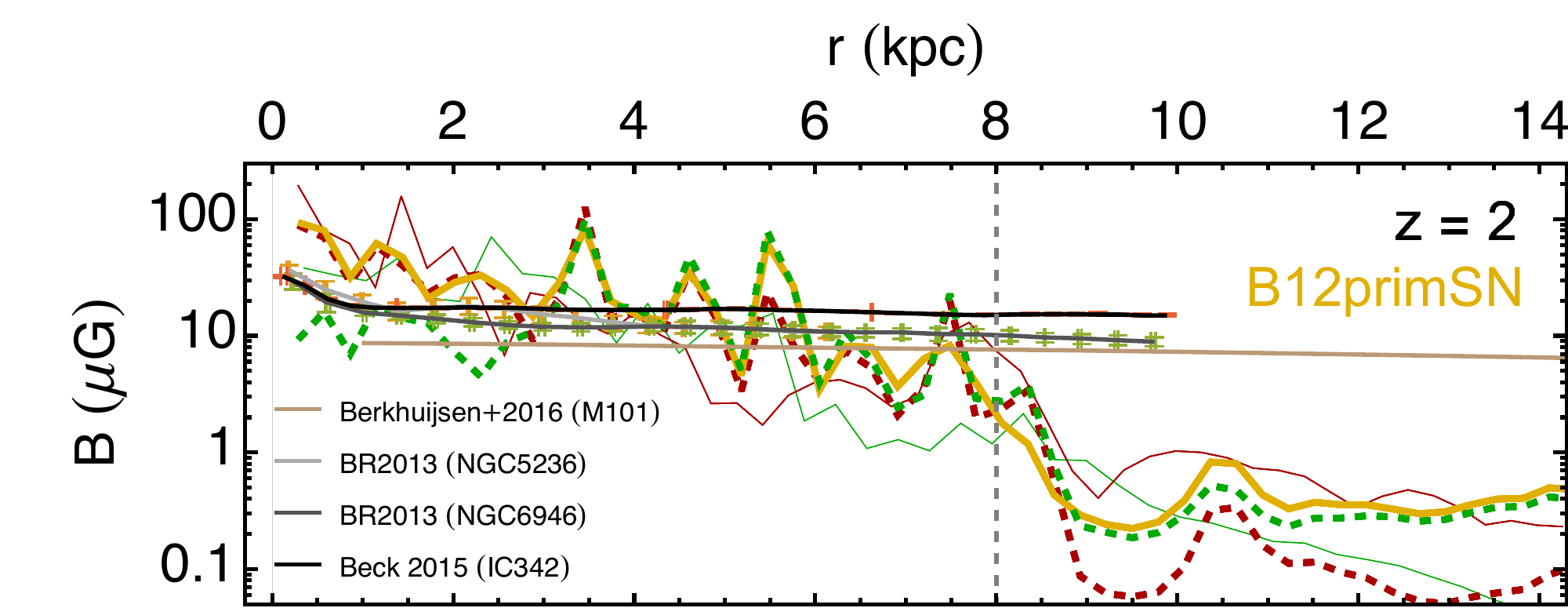}%
    \includegraphics[width=\columnwidth]{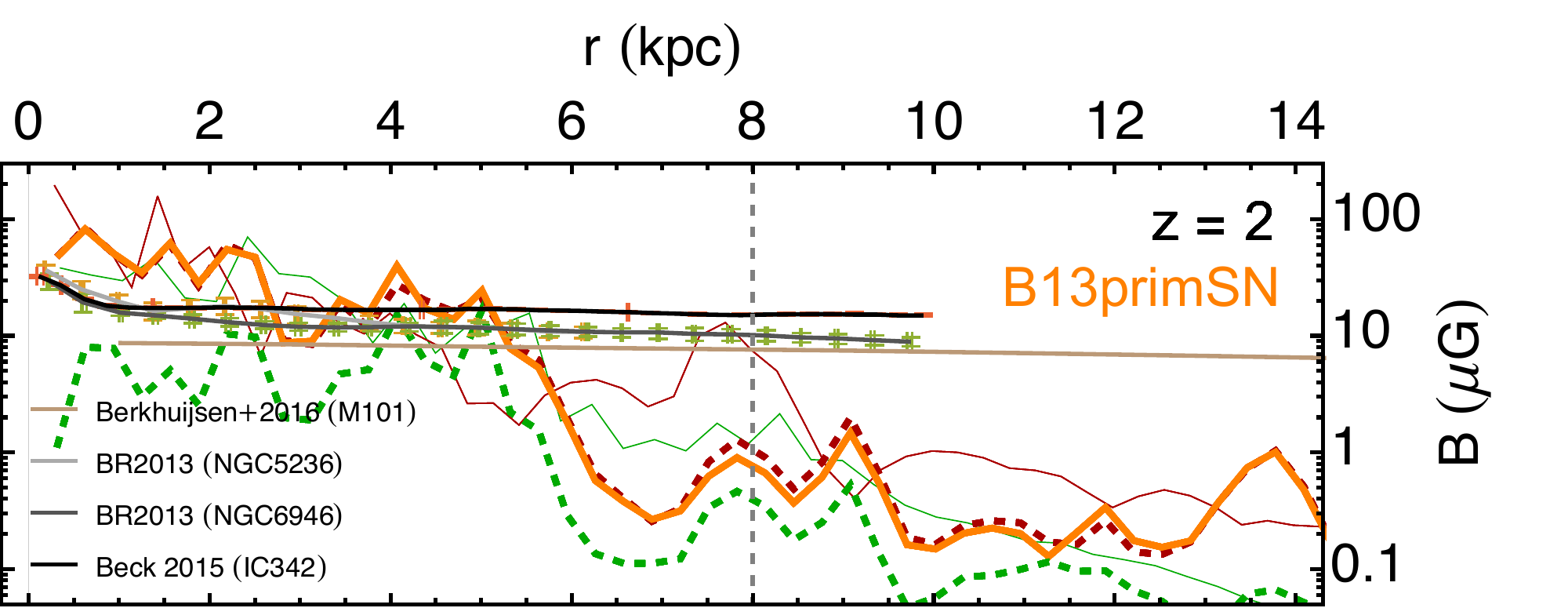}\\
    \includegraphics[width=\columnwidth]{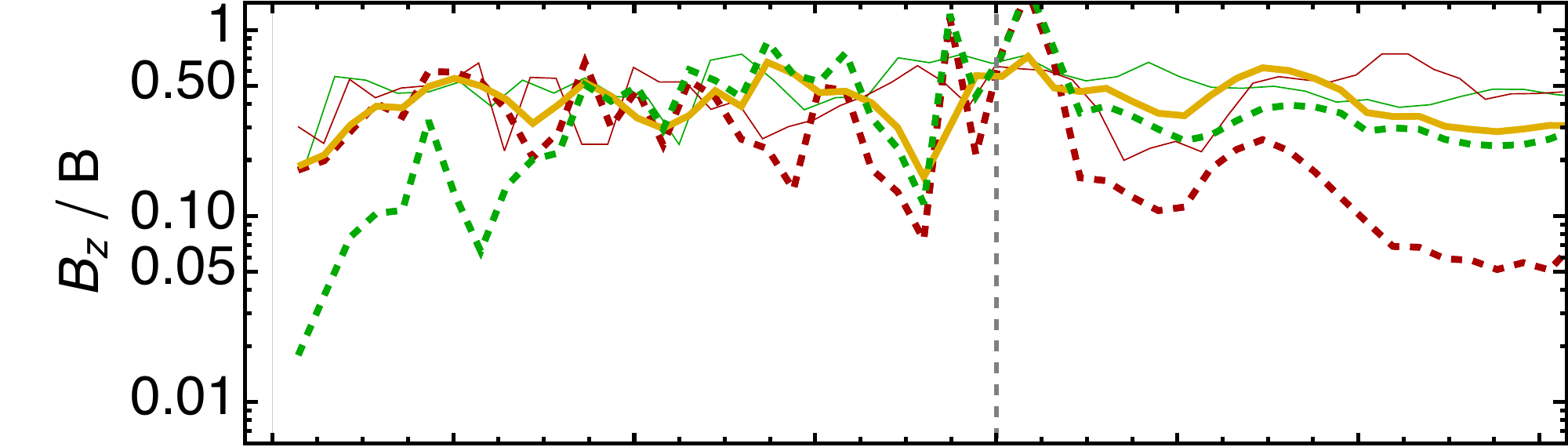}%
    \includegraphics[width=\columnwidth]{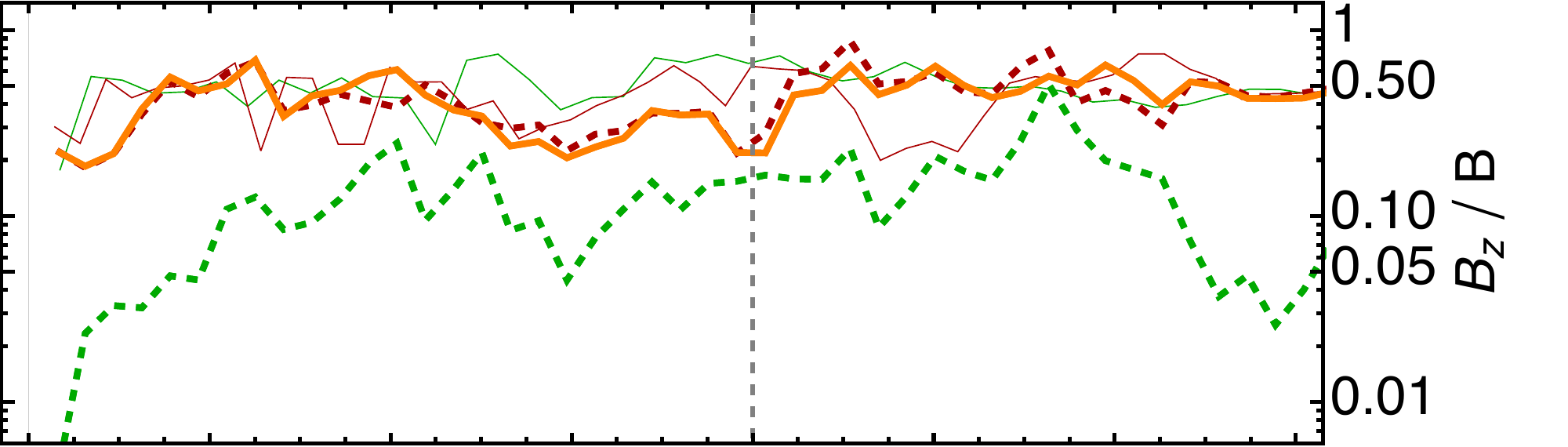}\\
    \includegraphics[width=\columnwidth]{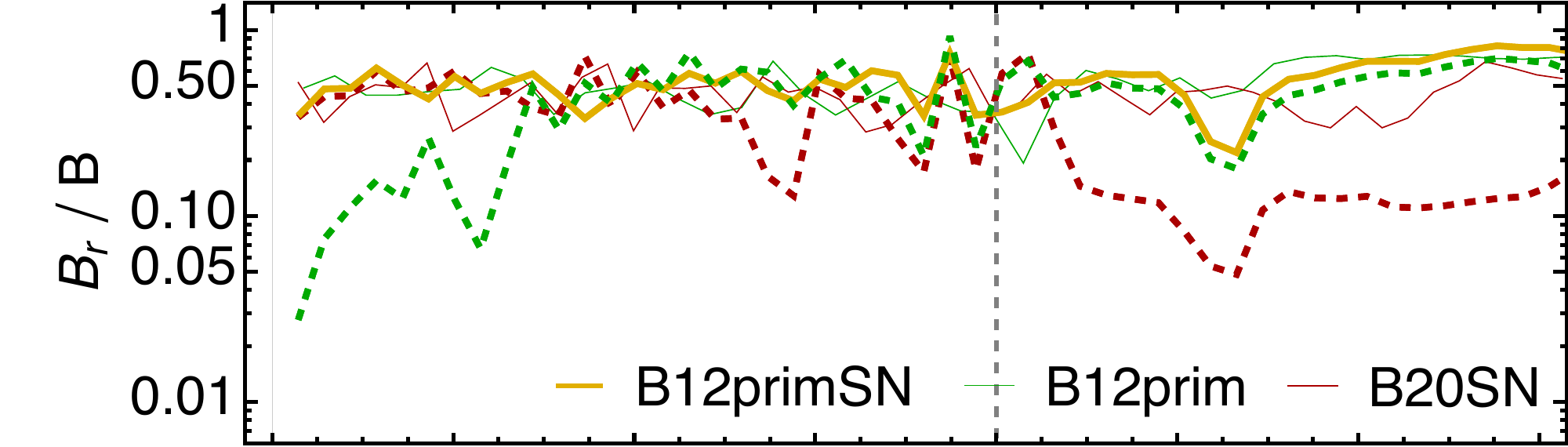}%
    \includegraphics[width=\columnwidth]{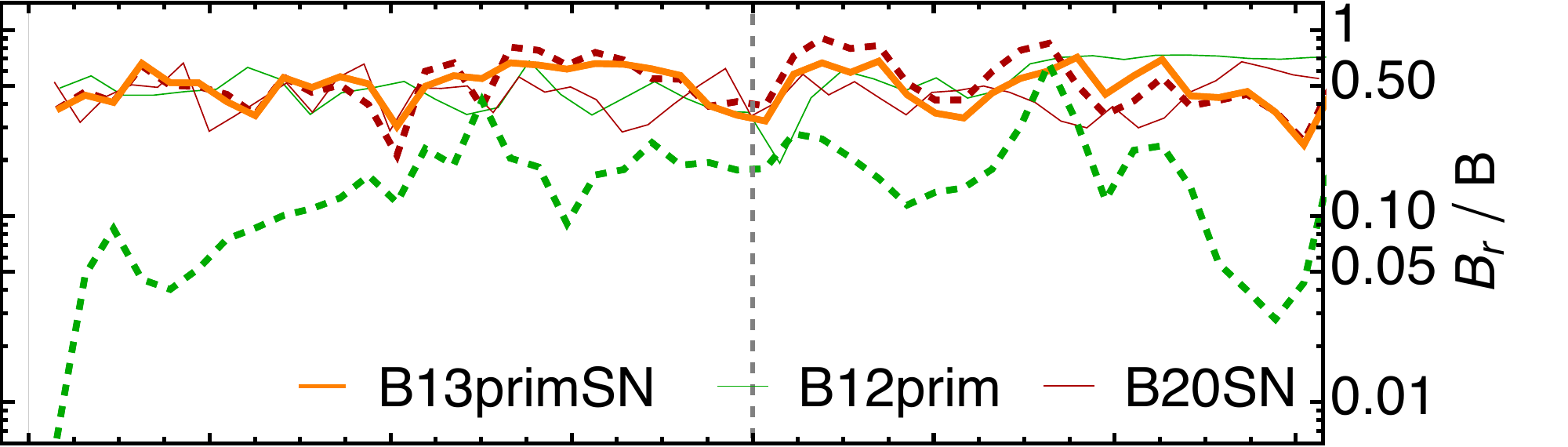}\\
    \includegraphics[width=\columnwidth]{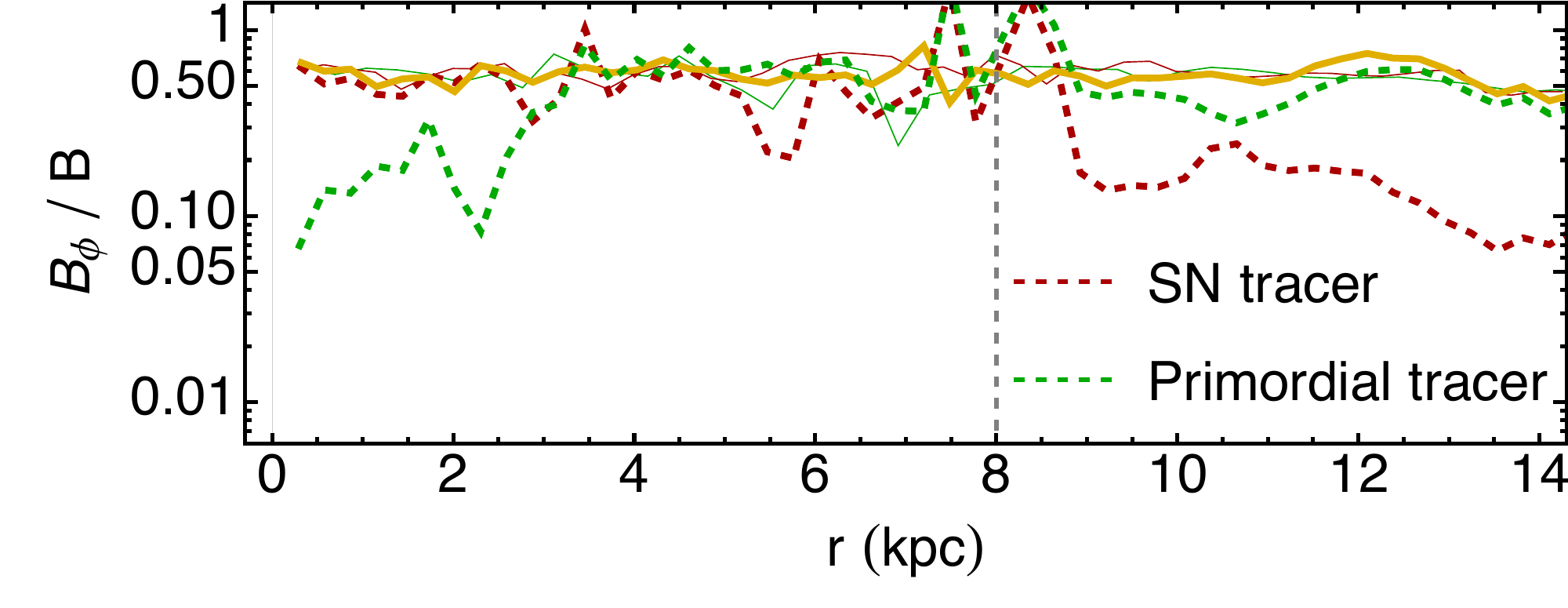}%
    \includegraphics[width=\columnwidth]{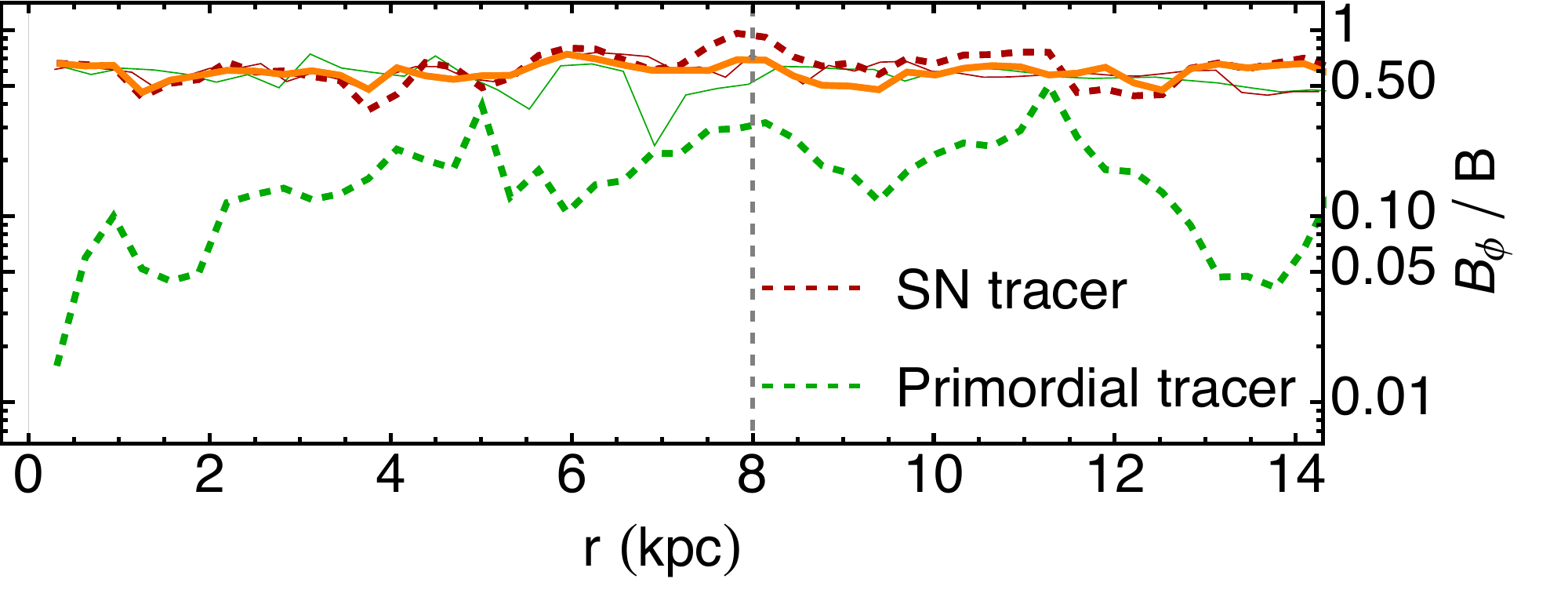}\\
    \caption{Radial profiles of the magnetic field in the galactic region of the \TrMBUp~(left column) and \TrMBDown~(right column) runs at $z = 2$. Top row shows the absolute magnetic field ($B$). Remaining rows show from top to bottom the ratio of the axial ($B_z$), radial ($B_r$), and toroidal ($B_\phi$) magnetic field components to $B$. We add for comparison the same profiles for runs \MBDoce~(thin green solid line) and \MBInj~(thin red solid line). The vertical dashed line demarcates the approximate extension of the gas disk, i.e. $8$~kpc. Observational measures of $B (r)$ at $z = 0$ are included for comparison for M101 \citep[black line;][]{Berkhuijsen2016}, NGC5235 and NGC6949 \citep[gray lines;][]{Basu2013}, and IC342 \citep[brown line;][]{Beck2015}. When primordial magnetic fields are important ($B_0 > 10^{-12}$ G), the central region of the galaxy (i.e. $r \lesssim 2$ kpc) remains dominated by astrophysical fields, but the rest of the galaxy is magnetised by the primordial field.}
    \label{fig:Profiles}
\end{figure*}

Due to their intrinsic properties, each channel of magnetisation is more important in different regions of the galaxy. Magnetic fields produced by stars will concentrate in the extended gas disk and the centre of the galaxy, where most star formation occurs. Conversely, primordial magnetic fields are initially amplified by compressional processes when pristine gas is accreted onto the cosmic web, and eventually fed to the outskirts of galaxies. As a result, the relative importance of primordial magnetisation will be higher at large radii. To quantitatively explore the spatial distribution of the magnetic field in the galaxy, we show in Fig.~\ref{fig:Profiles} the total and traced magnetic fields mass-weighted radial profiles\footnote{We find that employing a volume-weighted measurement yields unchanged results, but reduces the typical values of $B$ by $\lesssim 1$ dex.} at $z = 2$. The profiles are computed using cylindrical coordinates for thin cylinders of $r = 0.2 r_\text{DM}$ and half-thickness $h = 300$~pc, where the z-coordinate is aligned with the angular momentum of the galactic baryonic mass. The left and right columns show the profiles for the \TrMBUp~and \TrMBDown~runs, respectively. Different rows correspond from top to bottom to the total magnetic field ($B$) and the axial to total ($B_z / B$), radial to total ($B_r / B$), and toroidal to total ($B_\phi / B$) magnetic field ratios. Each panel presents the combined (i.e. the sum of both tracers) magnetic field (thick solid line), and its decomposition into SN-injected (red dashed) and primordial (green dashed) magnetic fields. For further comparison, we include the profiles for the \MBDoce~(thin green solid line) and \MBInj~(thin red solid line) simulations, which only consider one source of magnetic field. Finally, we include the profiles calculated by \citet{Berkhuijsen2016} for M101, \citet{Beck2015} for IC342, and \citet{Basu2013} for NGC5236 and NGC6949, which portray galaxies with size and morphology comparable to our system.

We focus on $B\,(r)$ first, for which we find various interesting features. All profiles display a break at about $\sim 8$~kpc, corresponding to the approximate gas disk radial extension. Within $r = 8$~kpc, all profiles are relatively flat, with $B \sim 10 \muG$. However, SN-injected magnetic fields (\MBInj~and the corresponding tracer in \TrMBUp~and \TrMBDown) progressively increase in their strength towards the centre of the galaxy. On the contrary, magnetic fields of primordial nature (\MBDoce~and the primordial magnetic field tracer in \TrMBUp~and \TrMBDown) have a shallower horizontal profile with a more constant $B\,(r)$ at all radii. As a result, even though the magnetic field in the majority of the \TrMBUp~disk is dominated by the primordial component, the inner $\sim 2$~kpc of the disk is dominated by an astrophysical magnetic field. In fact, we expect astrophysical magnetic fields to dominate in the centre of most galaxies, even more so in the presence of AGN. In \TrMBUp, the difference in the slopes of each magnetic tracer profile leads to the dominance of primordial magnetic fields at $r \gtrsim 4$~kpc. Finally, for weaker primordial magnetic fields (i.e. \TrMBDown) $B\,(r)$ is always dominated by fields generated in the galaxy. Our profiles are in good agreement with the observations shown in Fig.~\ref{fig:Profiles}. We note that the observed galaxies are found at $z \sim 0$, whereas our system is studied at higher redshift. This comparison assumes that the behaviour we find between $z \sim 4$ and $z = 1$ is conserved down to $z = 0$. While there is evidence supporting the existence of comparable $\gtrsim \muG$ magnetic fields in galaxies at redshifts as high as $z \sim 4$ \citep{Bernet2008}, their radial distribution might have some redshift evolution. Similarly, M101 extends to larger radii than our system \citep[][report a break of the profile at approximately 16 kpc]{Berkhuijsen2016}. At the centre of the observed galaxies, the profiles portray weaker magnetic fields than our simulations. However, this is likely due to a lack of resolution. For example, \citet{Aitken1998} estimate magnetic fields as high as $B \sim 2$ mG at the centre of the MW. These strong magnetic fields would be in better agreement with the astrophysically-amplified magnetic fields in our simulations. The profiles have a similar behaviour to those obtained by \citet{Pakmor2017}. 

We decompose the magnetic field into a cylindrical component, finding a similar behaviour across the four simulations. $B_\phi$ is approximately constant at all radii. In contrast, $B_z$ slightly decreases at small and large radii. $B_r$ increases at large distances due to bulk flow accretion onto the disk. When reviewing the behaviour of the two tracer fields we find that their interrelation is, as expected, mostly dictated by the modulus of each tracer field (i.e. top panels). However, we note that the aforementioned increase of $B_r$ at larger distances has a larger reflection on the primordial magnetic field tracer field. 

The apparent equipartition of the magnetic tracers across the three coordinates is likely caused by the turbulent, small-scale magnetic field, for which we expect all coordinates to be equidistributed. Dominance of the strength of the magnetic field by its turbulent component is also found in $z \sim 0$ observations \citep[e.g.][]{Beck2015}. We have shown that primordial magnetic fields with $B_0 > 10^{-12}$~G dominate at intermediate radii of galactic disks and become even more important at the outskirts. As a result, it is now interesting to pin down at which scales each tracer dominates. One possibility would be to examine whether this behaviour is still in place when separating turbulent and large-scale magnetic fields \citep[e.g. as done by][]{Ntormousi2020}. We leave this for future work and opt instead to explore the magnetic energy spectra, which provides a better overview of the relevant scales for magnetic fields. This will help us answer whether strong primordial magnetic fields could be responsible for the large scale magnetic fields that we observe in galaxies \citep{Nixon2018}.

\subsection{The prevalence of primordial magnetic field at large scales in galaxies}
\label{ss:Spectra}

\begin{figure*}
    \centering
    \includegraphics[width=\columnwidth]{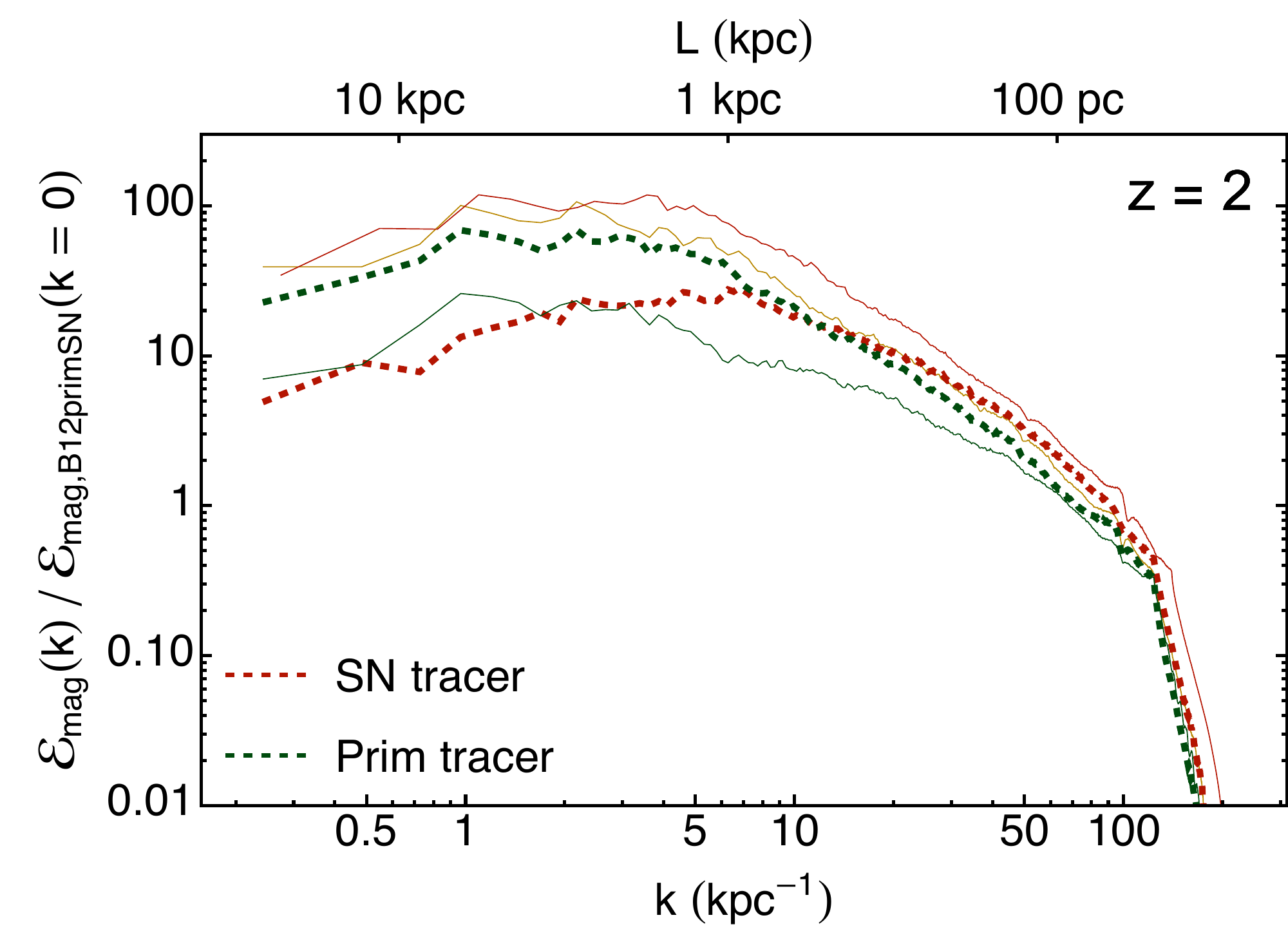}%
    \includegraphics[width=\columnwidth]{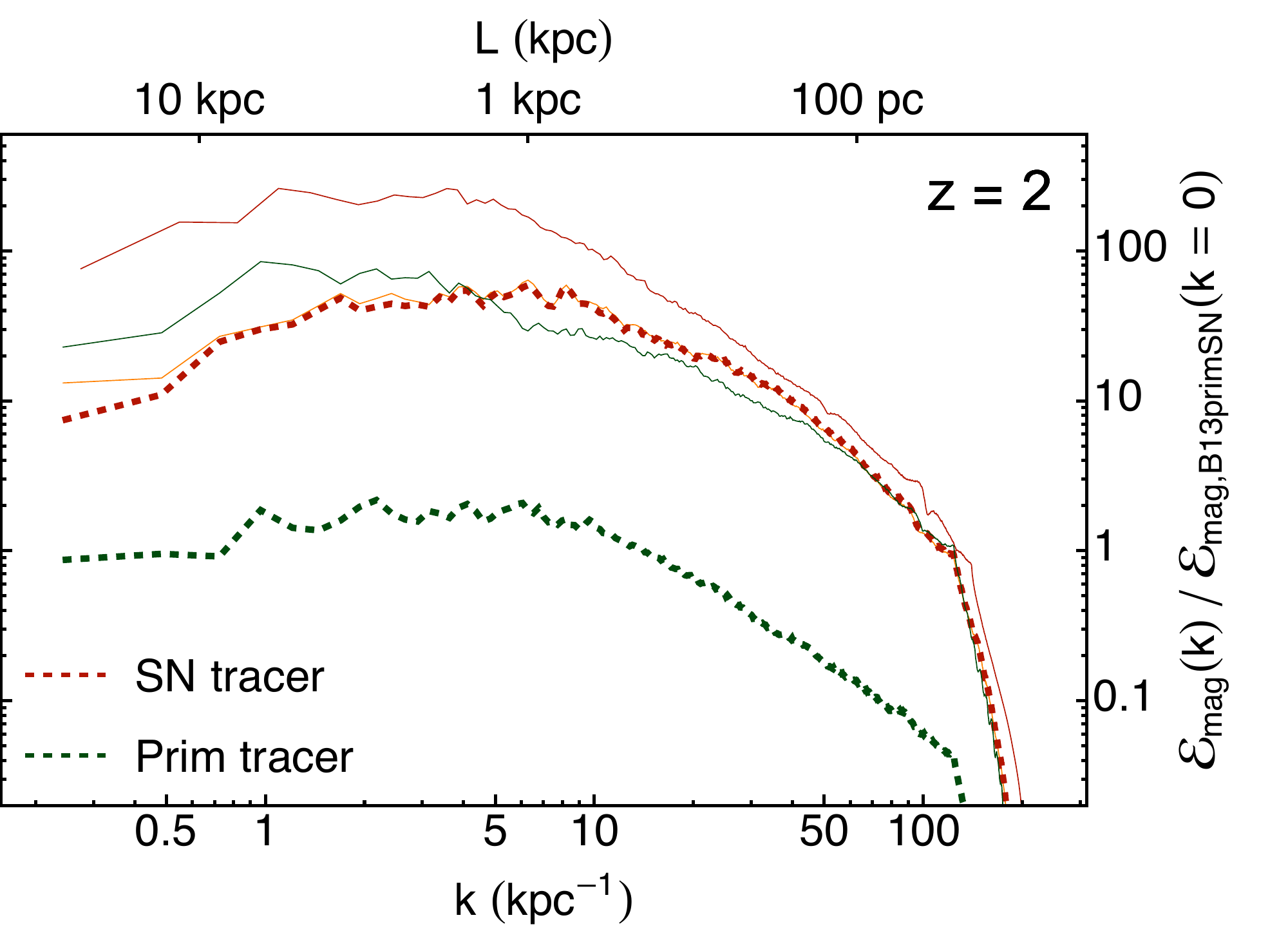}\\
    \caption{Magnetic energy spectra of the galaxy at $z = 2$. The FFT is computed using $26$ physical kpc per side cubic boxes. Left and right panels show the \TrMBUp~(solid amber line) and \TrMBDown~(solid orange line) runs, respectively. We include the spectra for the \MBDoce~(green solid line) and \MBInj~(red solid line) runs for comparison. Dashed lines show the spectra for the primordial (green dashed line) and SN-injected (red dashed line) tracer magnetic field in each run. On large scales ($> 1$~kpc), primordial magnetic fields contain more power than their astrophysical counterpart and have an almost flat spectrum. On the other hand SN-injected fields have a more pronounced decay towards lower k. For strong values of $B_0$ (above $10^{-12}$ G), we predict large-scale galactic magnetic fields to be the remnant of magnetic fields originated in the early Universe.}
    \label{fig:Spectra}
\end{figure*}

Due to their vectorial nature, magnetic fields distribute their energy differently across the span of spatial scales. To unfold this information, it is useful to study the magnetic energy spectra in Fourier space. We compute our spectra through the Fast Fourier Transform\footnote{We acknowledge the use of the FFTW library (\href{http://www.fftw.org/}{http://www.fftw.org/}).} (FFT) of 1024 cells-per-side cubic boxes centred at the position of the galaxy, onto which the entire AMR structure is interpolated. We fix the box size to $L_\text{FFT} =$~26~kpc, assuming periodic boundary conditions in our FFT computation\footnote{Appendix B in \citet{Martin-Alvarez2018} discusses how periodicity affects these type of spectra in FFT transformations.}. To reach higher resolutions, we decide not to 0-pad our FFT, but we note that this would yield a more pronounced decay of power towards the lower $k$ values. Finally, to aid comparison between the different runs, we normalise the spectra to the $k = 0$ value of the total magnetic field in the main run of each panel. These runs are \TrMBUp~(amber solid line) and \TrMBDown~(orange solid line) for the left and right panels of Fig.~\ref{fig:Spectra}. Green and red solid lines show the spectra for the \MBDoce~and \MBInj~runs. Finally, the dashed lines in each panel are the spectra resulting from the FFT applied to each magnetic tracer in the corresponding run.

The energy spectra of the magnetic tracers reveal striking differences according to the source of the magnetisation. Magnetic fields of primordial nature have more power at scales larger than the galaxy ($k \lesssim 1\;\ikpc$) due to the environmental magnetisation and higher field at large radii. For the SN-injected field, we observe a more pronounced decay towards smaller $k$ due to the magnetic field being produced within the galaxy. At intermediate scales, spanning from the size of the galaxy to the thickness of the disk ($1\; \ikpc \lesssim k < 20\; \ikpc$), we find a relatively flat spectrum for primordial fields, accumulating power at $k \sim 2 - 3\;\ikpc$. This suggests that primordial magnetic fields dominate and peak at the large-scales of \TrMBUp. In contrast, the SN-injected fields have a more pronounced inverse-cascade that continues to grow up to the smallest scales of the intermediate range ($k \sim 20\; \ikpc$). SN-injected fields peak at a scale of approximately $300$~pc, roughly the thickness of the disk and slightly above the scale at which SN forcing is most efficient. We expect the magnetisation of hot gas in large SN bubbles to be fully governed by this magnetisation channel. This is shown and further discussed in Section~\ref{ss:Phases}. Finally, at the smallest scales ($k > 20\; \ikpc$), both fields decay due to the limited resolution and the convolution with the disk of the galaxy (see Appendix B, \citealt{Martin-Alvarez2018}). However, we find a steeper decline of the primordial magnetic energy spectrum towards larger $k$ than of the SN-injected one. This is in agreement with an injection scale of approximately the size of the disk for the primordial fields whereas the SN-injected ones are generated at the smallest scales in our simulation. Finally, when focusing on our primary simulation, \TrMBUp, we find an interesting change of regime at the scale of the disk, $k \sim 10 - 20\; \ikpc$: the large-scale magnetic fields of the galaxy are dominated by a primordial origin whereas the small-scale field has an astrophysical one. Our spectra suggest that magnetic fields produced by SN are not as efficiently re-organised into large-scale magnetic fields as those of primordial origin. As a result, if the magnetic field of the Universe is $B_0 > 10^{-12}$ G, we predict galactic-scale magnetic fields to be the remnants of magnetic fields originated in the early Universe. We discuss other considerations that might be important for this statement in Section \ref{s:Caveats}. Finally, the overall behaviour observed for the tracers is also confirmed by our simulations with only one magnetisation channel. That is: \MBInj~resembles the SN-injected tracer, whereas \MBDoce~is similar to the primordial magnetic field-tracer.

\subsection{The distribution of magnetic fields across ISM phases}
\label{ss:Phases}

\begin{figure*}
    \centering
    \includegraphics[width=0.67\columnwidth]{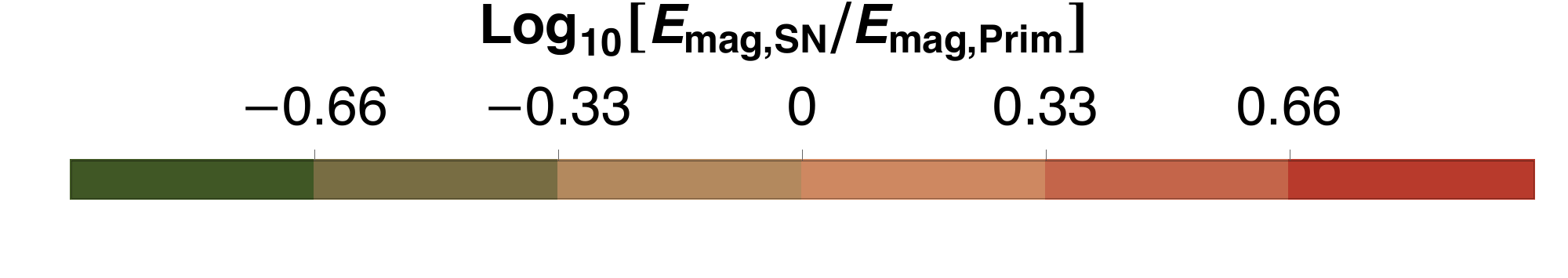}%
    \includegraphics[width=0.67\columnwidth]{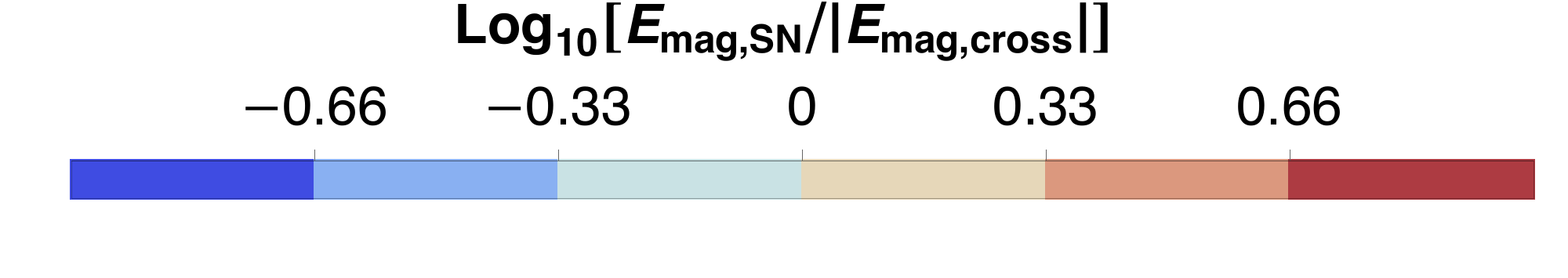}%
    \includegraphics[width=0.67\columnwidth]{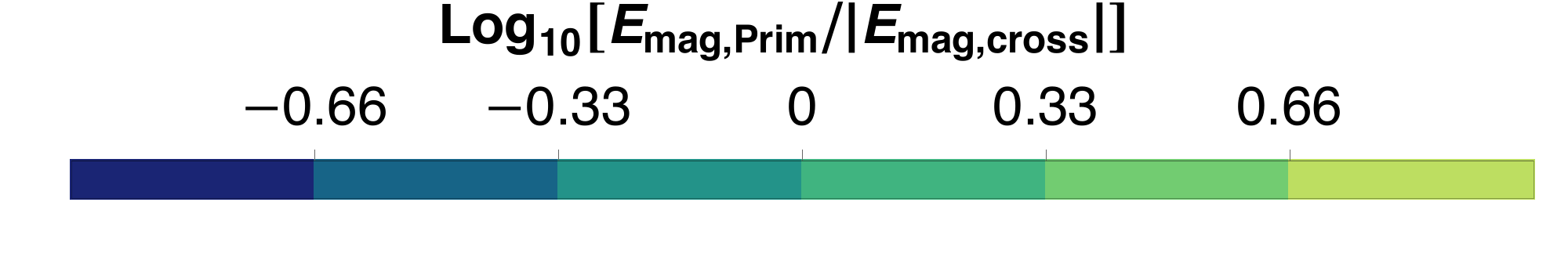}\\
    \includegraphics[width=0.67\columnwidth]{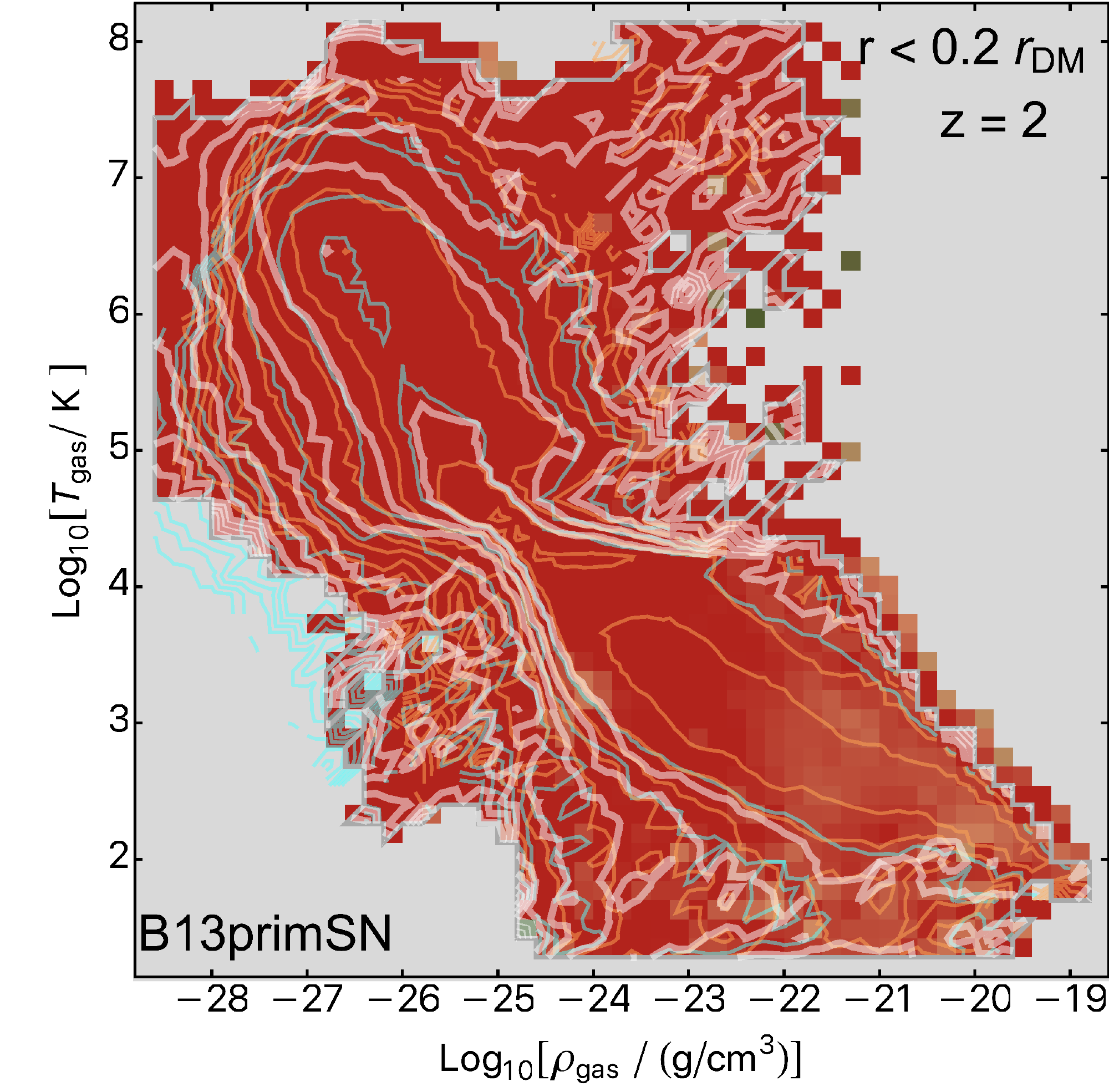}%
    \includegraphics[width=0.67\columnwidth]{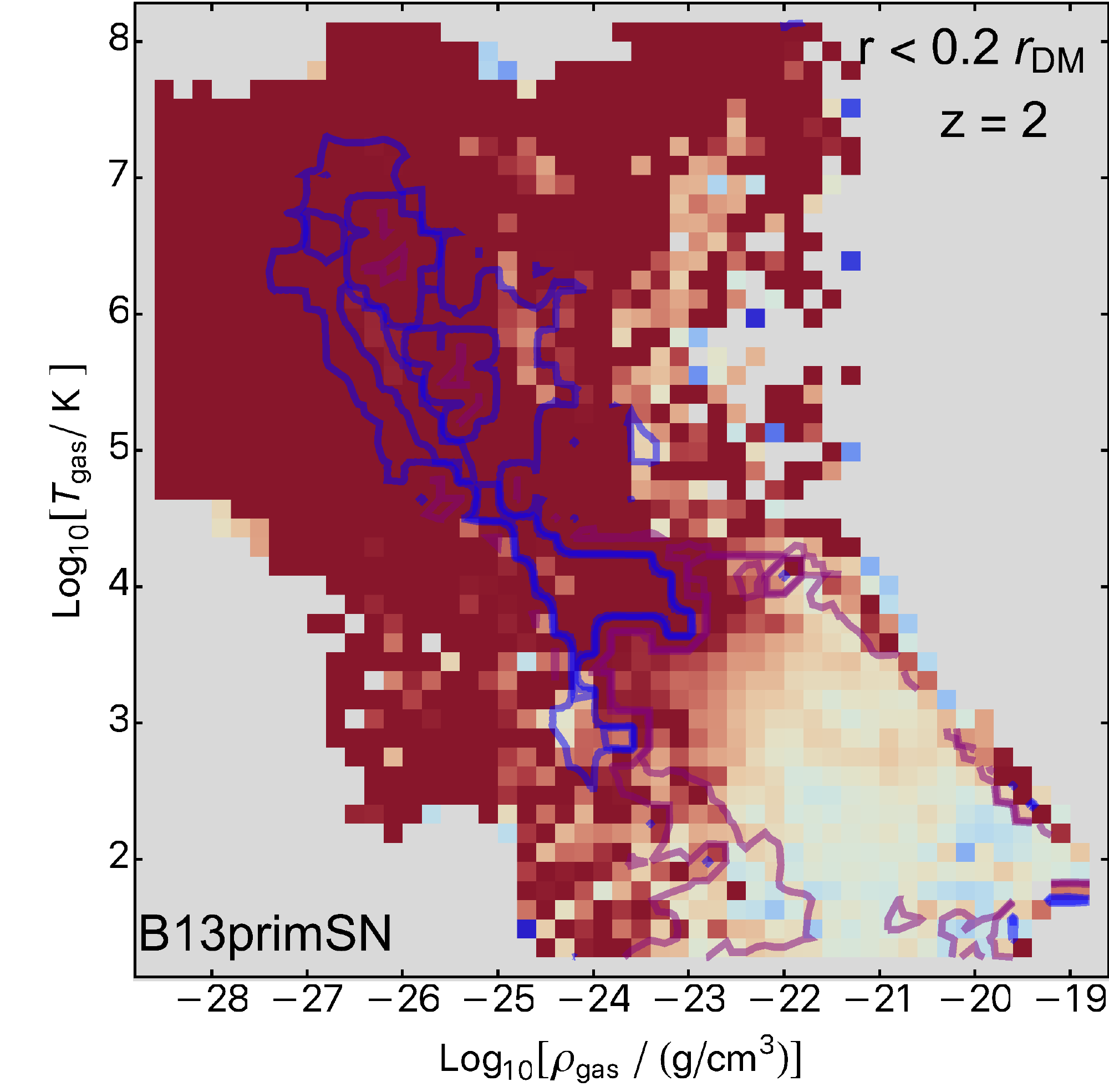}%
    \includegraphics[width=0.67\columnwidth]{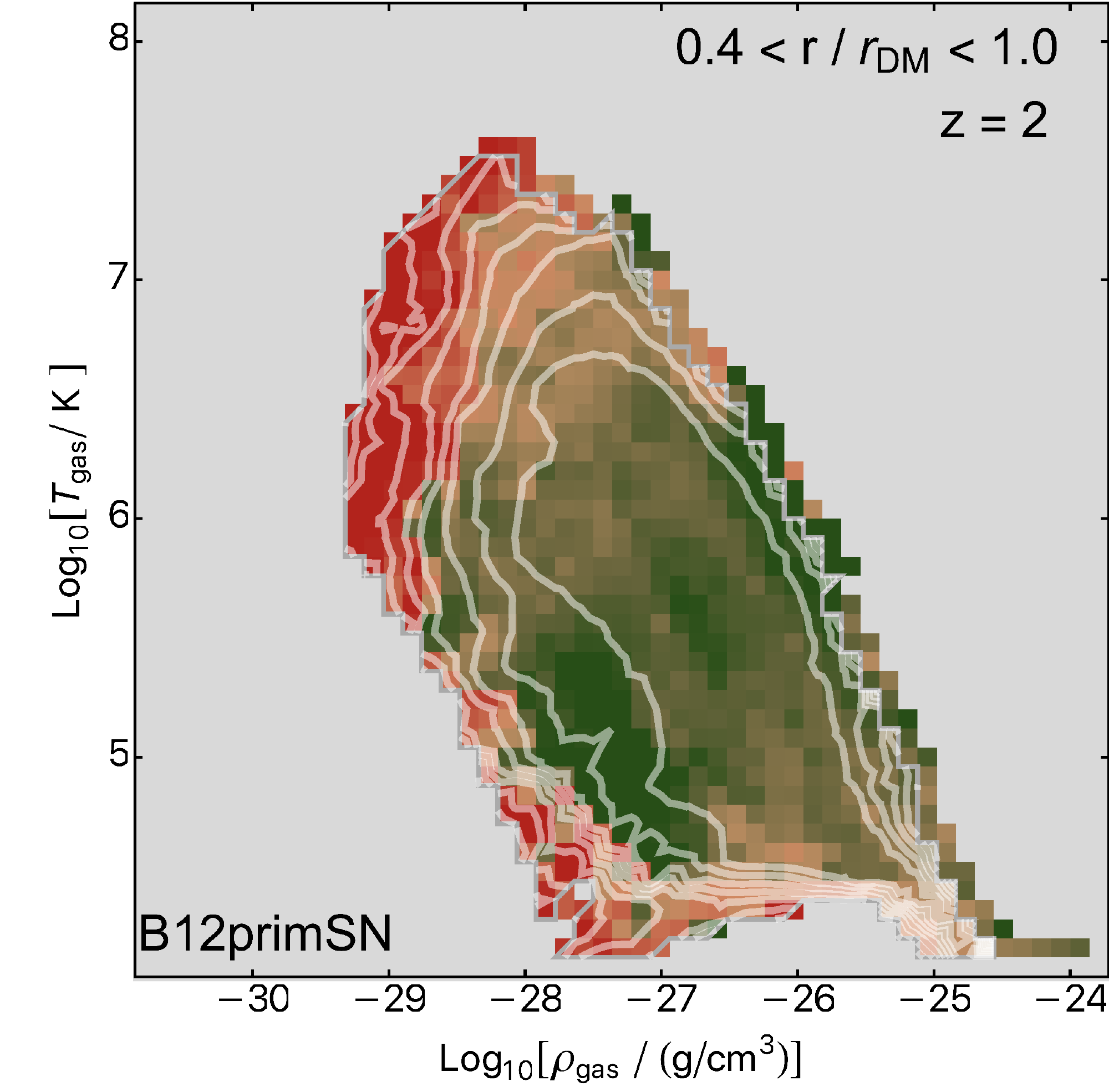}\\
    \includegraphics[width=0.67\columnwidth]{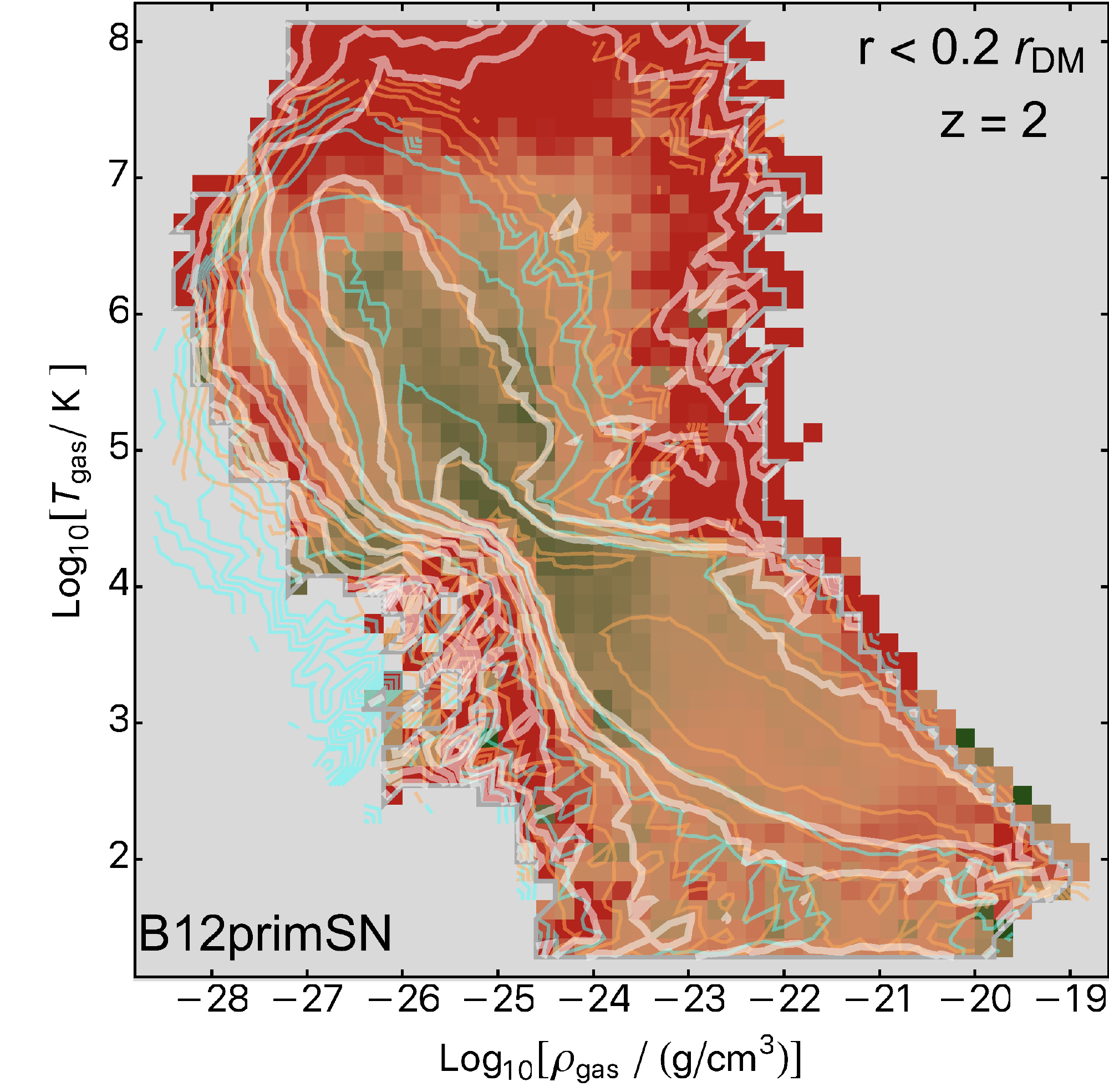}%
    \includegraphics[width=0.67\columnwidth]{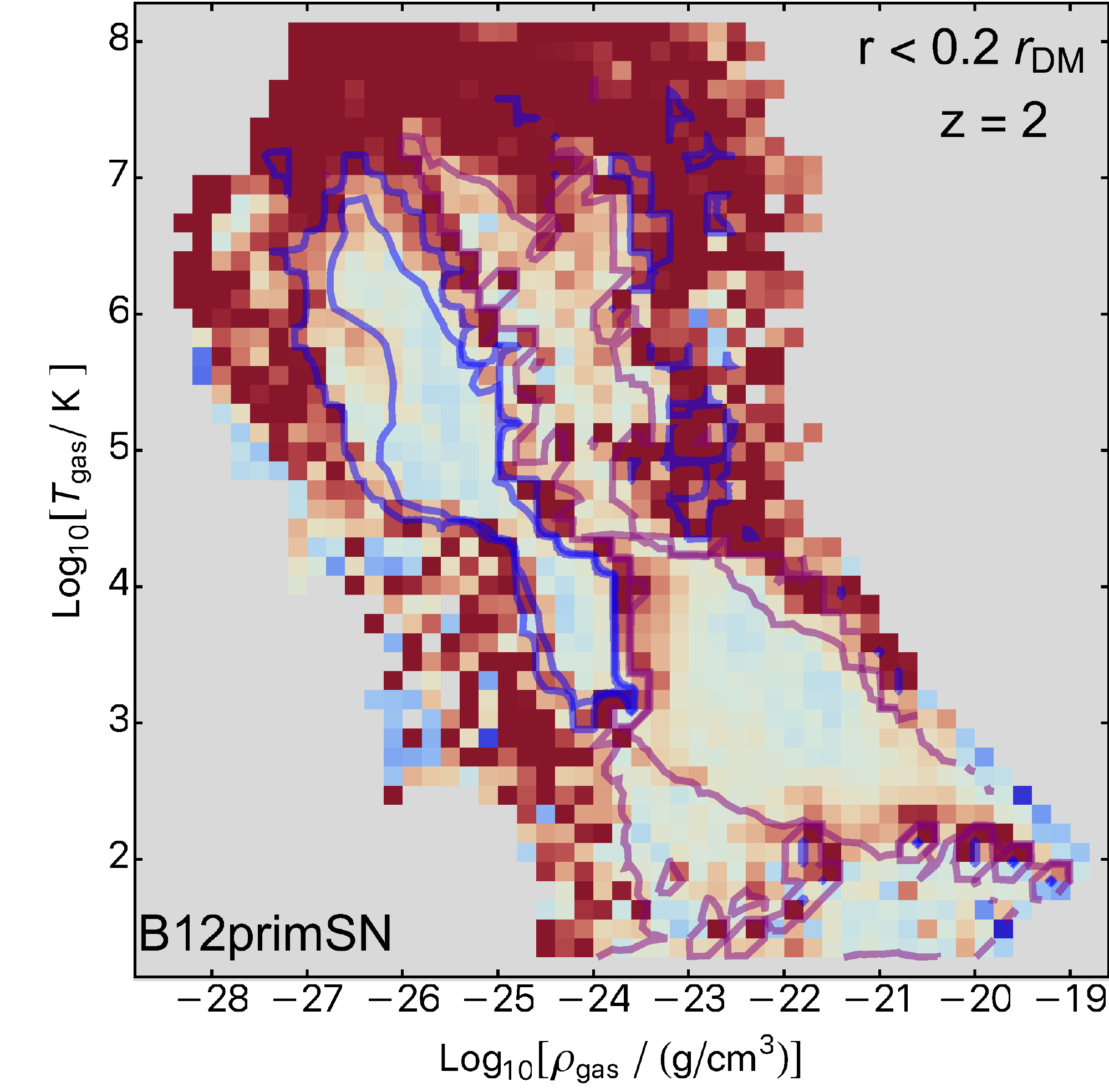}%
    \includegraphics[width=0.67\columnwidth]{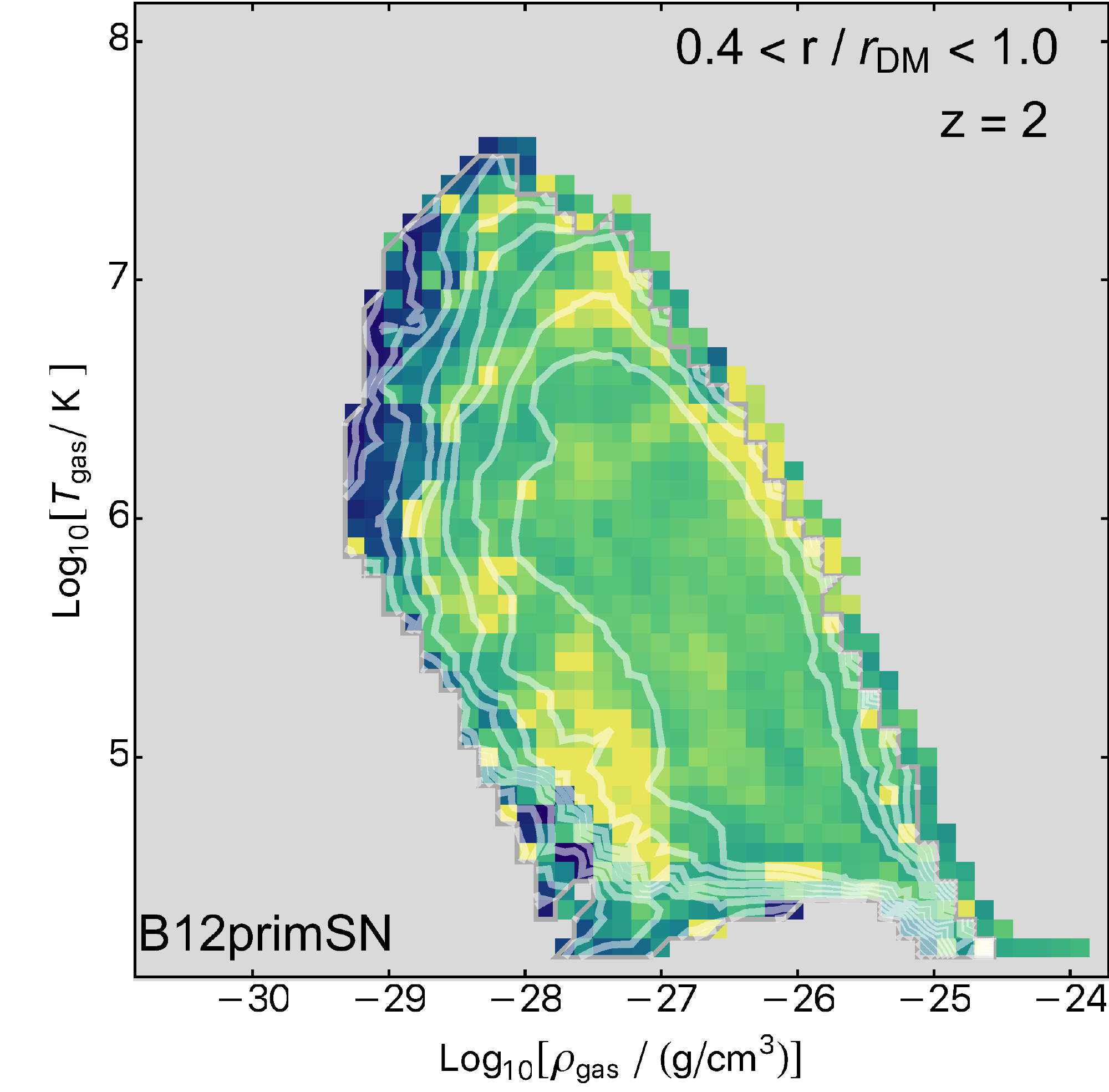}\\
    \caption{Phase diagrams coloured as a function of the ratio between different magnetic energy tracers at $z = 2$. {\bf(Left column)} SN-injected to primordial magnetic energies ratio for the galactic region in the \TrMBDown~(top) and \TrMBUp~(bottom) runs. Contours show the total magnetic energy in each corresponding run (white). We include the contours for the magnetic energy in \MBInj~(orange) and \MBDoce~(cyan) for comparison. {\bf(Centre column)} SN-injected to cross term energy ratio in the galactic region of the \TrMBDown~(top) and \TrMBUp~(bottom) simulations. Contours now represent the distribution of the positive (blue) and negative (purple) parts of the cross-term energy. {\bf(Right column})) Ratio of the SN to primordial (top) and primordial to cross term (bottom) energies in the halo ($0.4\;r_\text{DM} < r < 1.0\;r_\text{DM}$) of \TrMBUp. White contours show the total magnetic energy distribution. The phase space of \TrMBDown~is mostly dominated by the SN-injected energy. \TrMBUp~has distinct distributions for each traced magnetic energy across the phases of the galaxy, with the primordial energy dominating the warm and diffuse medium. The halo of \TrMBUp~is equally dominated by primordial energy.}
    \label{fig:ThermoPhases}
\end{figure*}

In the same manner our two traced magnetic fields have distinct scales of injection into the galaxy, the thermodynamical properties of the magnetised gas are different for each channel. Pristine gas is mostly accreted through cold flows whereas SN-injected magnetic fields are produced in hot SN bubbles. We explore the distribution of the traced magnetic energies across the temperature - density phase space in Fig.~\ref{fig:ThermoPhases}. Plot colours showcase the ratio between two types of tracer energies, which are determined by the colour scale. We include contours depicting the absolute energy distribution. The first column in Fig.~\ref{fig:ThermoPhases} corresponds to the galactic region in \TrMBDown~(top) and \TrMBUp~(bottom). They are coloured according the ratio of SN-injected (red colour) to primordial (green colour) magnetic energy ($E_\text{mag,SN} / E_\text{mag,prim}$). Contours show the distribution of the total magnetic energy (white, $E_\text{mag}$). For comparison, we also include the total magnetic energy contours for \MBInj~(orange) and \MBDoce~(green). The distribution of total magnetic energy is comparable across runs, with a more substantial concentration in the warm phase ($T \sim 10^4$ K, $\rho_\text{gas} \sim 10
^{-24} \gram/\cm^3$) when primordial magnetic fields are present, and higher magnetic energy in the hot gas ($T \gtrsim 10^6$ K) when injection is included. For the weaker $B_0$ case (\TrMBDown), the magnetic energy is dominated by the SN-injected energy across the entire phase space. \TrMBUp~is more intriguing: in the cold and dense phase of the galaxy ($T < 10^4$ K, $\rho_\text{gas} > 10
^{-23} \gram/\cm^3$) SN-injected magnetic fields prevail. However, at intermediate densities ($\rho_\text{gas} \sim 10
^{-24} \gram/\cm^3$) primordial magnetic fields dominate. Altogether with Section~\ref{ss:Spectra}, this suggests that the large-scale magnetic fields permeating the warm phase of the galaxy are the remnants of primordial magnetism for $B_0 > 10^{-12}$~G. As expected, the hottest gas is produced in SN events, and therefore magnetised by the astrophysical magnetic energy tracer. The behaviour portrayed by these two panels is in place from approximately $z \sim 6$ onward. 

The central column shows the two same plots, now coloured by the ratio of SN-injected to absolute cross-term energy ($E_\text{mag,prim} / | E_\text{mag,cross}|$). Contours show the distribution of the positive (blue lines) and negative (purple lines) parts of the cross term energy. The cross-term energy is generally unimportant, except in the cold and dense phase. Here, both galaxies feature a non-negligible contribution from the negative cross-term energy. This supports that magnetic reconnection is large in the galaxy, with the majority of the interaction between the two tracers taking place in the phase space boundary between the warm-cold phases. However, we note that the cross-term energy does not fully dominate the energy budget in any region of the phase space. 

Finally, we want to understand how magnetic fields are distributed in the halo of the galaxy, in particular for $B_0 > 10^{-12}$~G, as future observations may probe the CGM and IGM to determine the magnetic field of our Universe. In the right-most column of Fig.~\ref{fig:ThermoPhases} we show the ratio of SN-injected to primordial (top; $E_\text{mag,SN} / E_\text{mag,prim}$) and primordial to absolute cross (bottom; $E_\text{mag,prim} / | E_\text{mag,cross}|$) magnetic energies for the outer region of the halo ($0.4\;r_\text{DM} < r < 1.0\;r_\text{DM}$) in the \TrMBUp~simulation. Once again white contours show the distribution of the total magnetic energy. While the energy budgets discussed in Section~\ref{ss:TracersEmag} show that SN-generated magnetic fields are secondary in the halo of the galaxy, Fig.~\ref{fig:ThermoPhases} indicates that SN-generated magnetic fields are still present at the highest temperatures. 

\subsection{The magnetic field - metallicity correlation}
\label{ss:Metal}
\begin{figure*}
    \centering
    \includegraphics[width=0.67\columnwidth]{Images/PhaseDiag/PhaseCBar}%
    \includegraphics[width=0.67\columnwidth]{Images/PhaseDiag/CBarCross2}\\    \includegraphics[width=0.67\columnwidth]{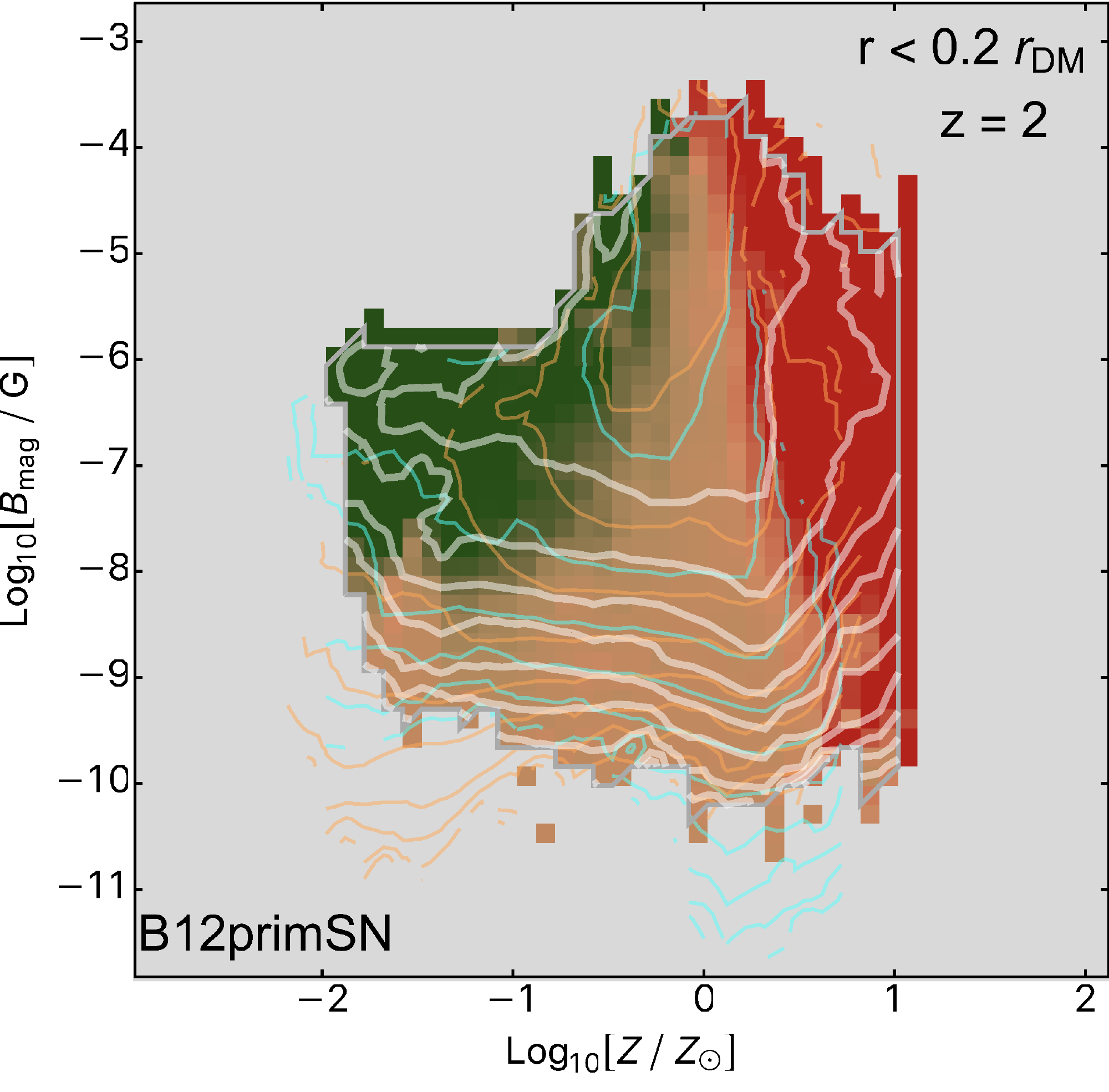}%
    \includegraphics[width=0.67\columnwidth]{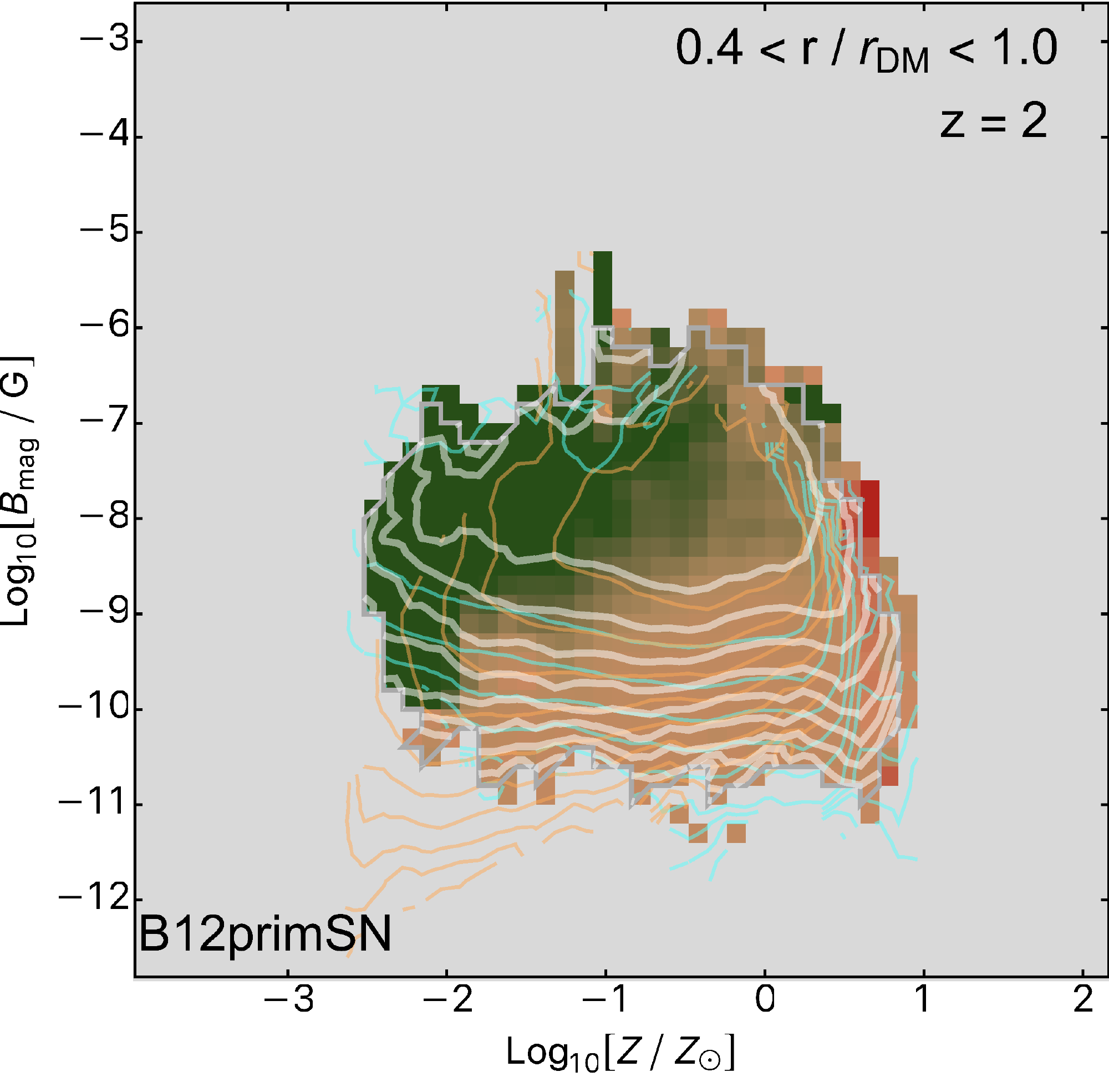}%
    \includegraphics[width=0.67\columnwidth]{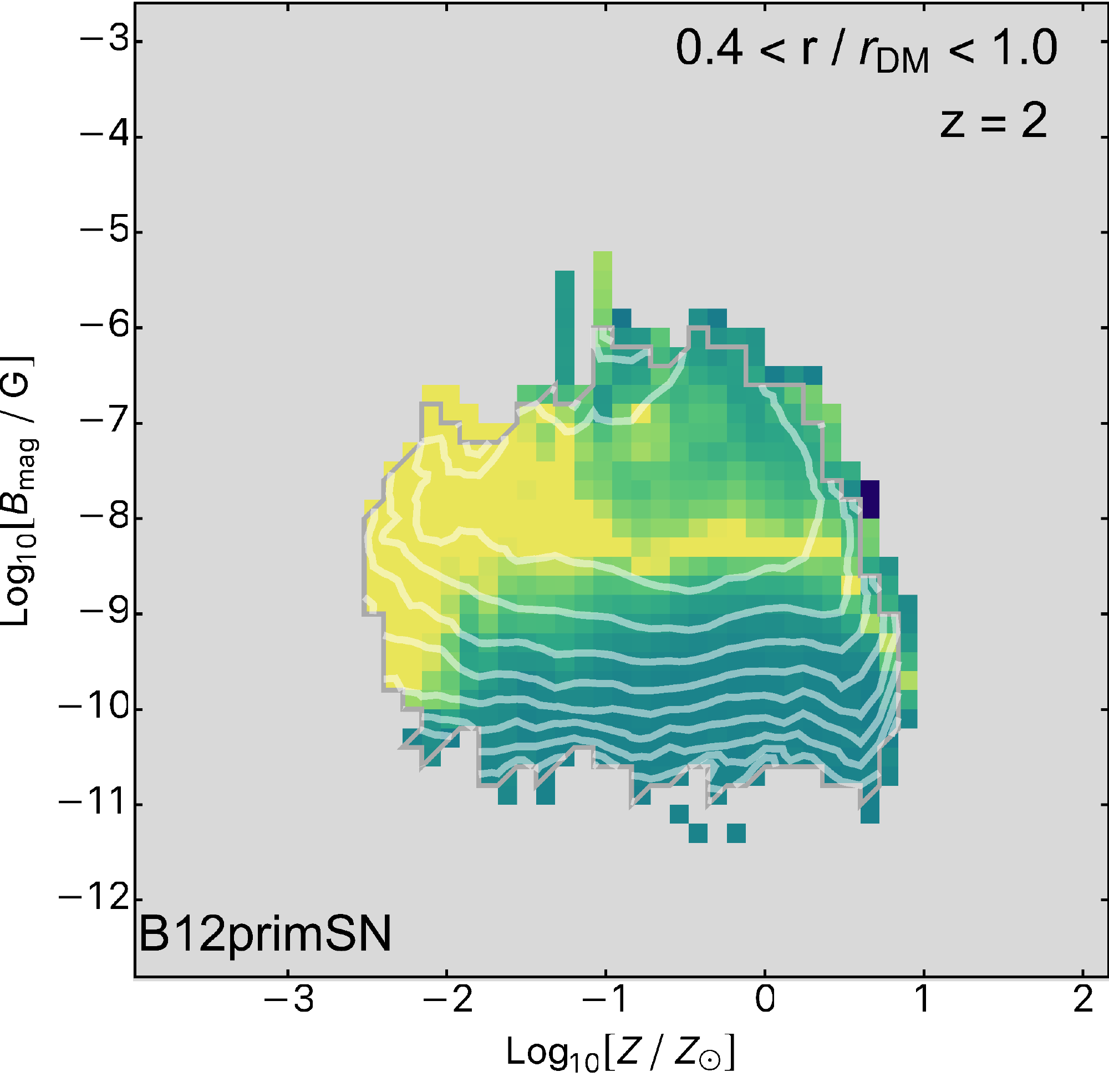}\\
    \includegraphics[width=0.67\columnwidth]{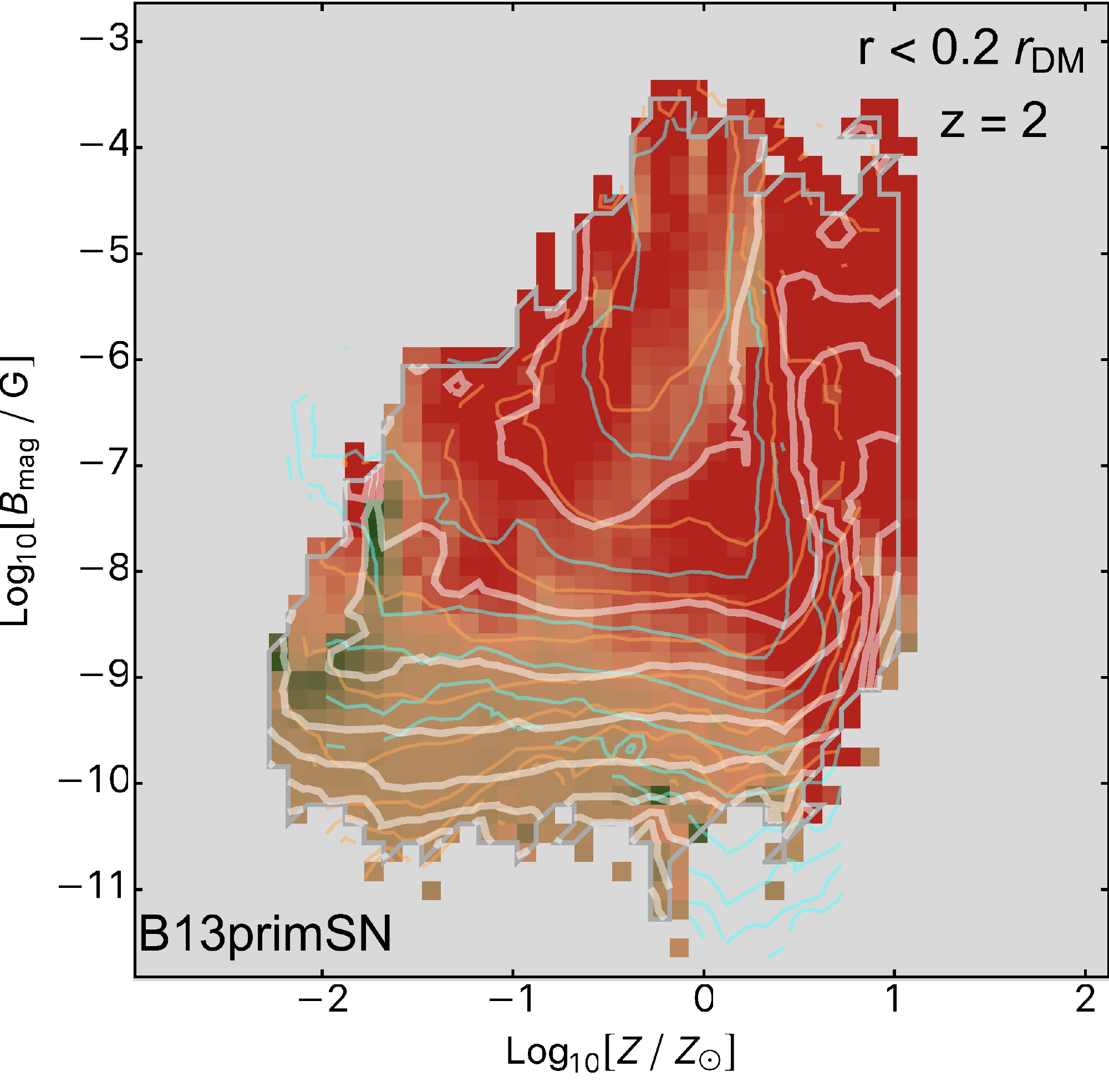}%
    \includegraphics[width=0.67\columnwidth]{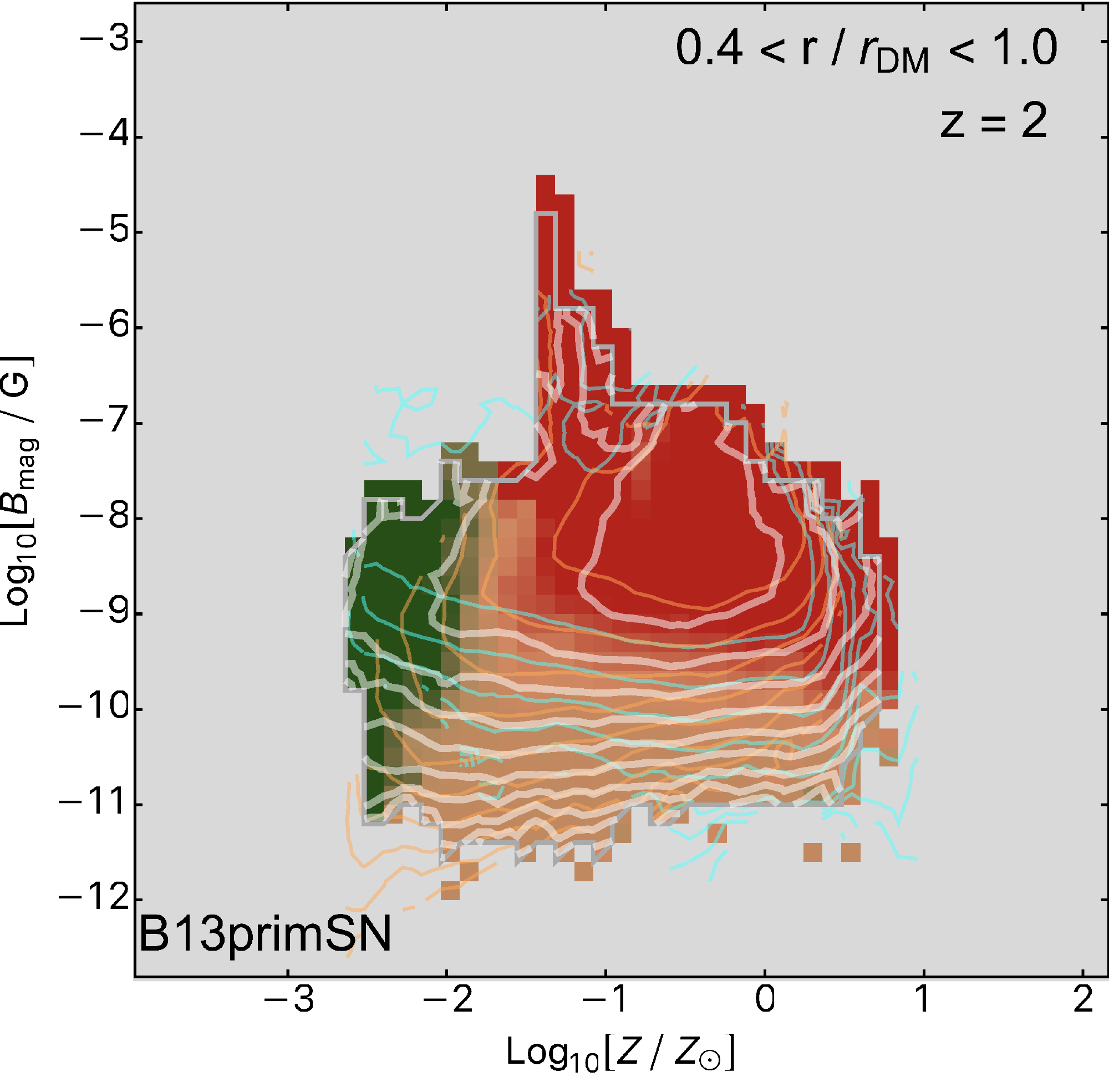}%
    \includegraphics[width=0.67\columnwidth]{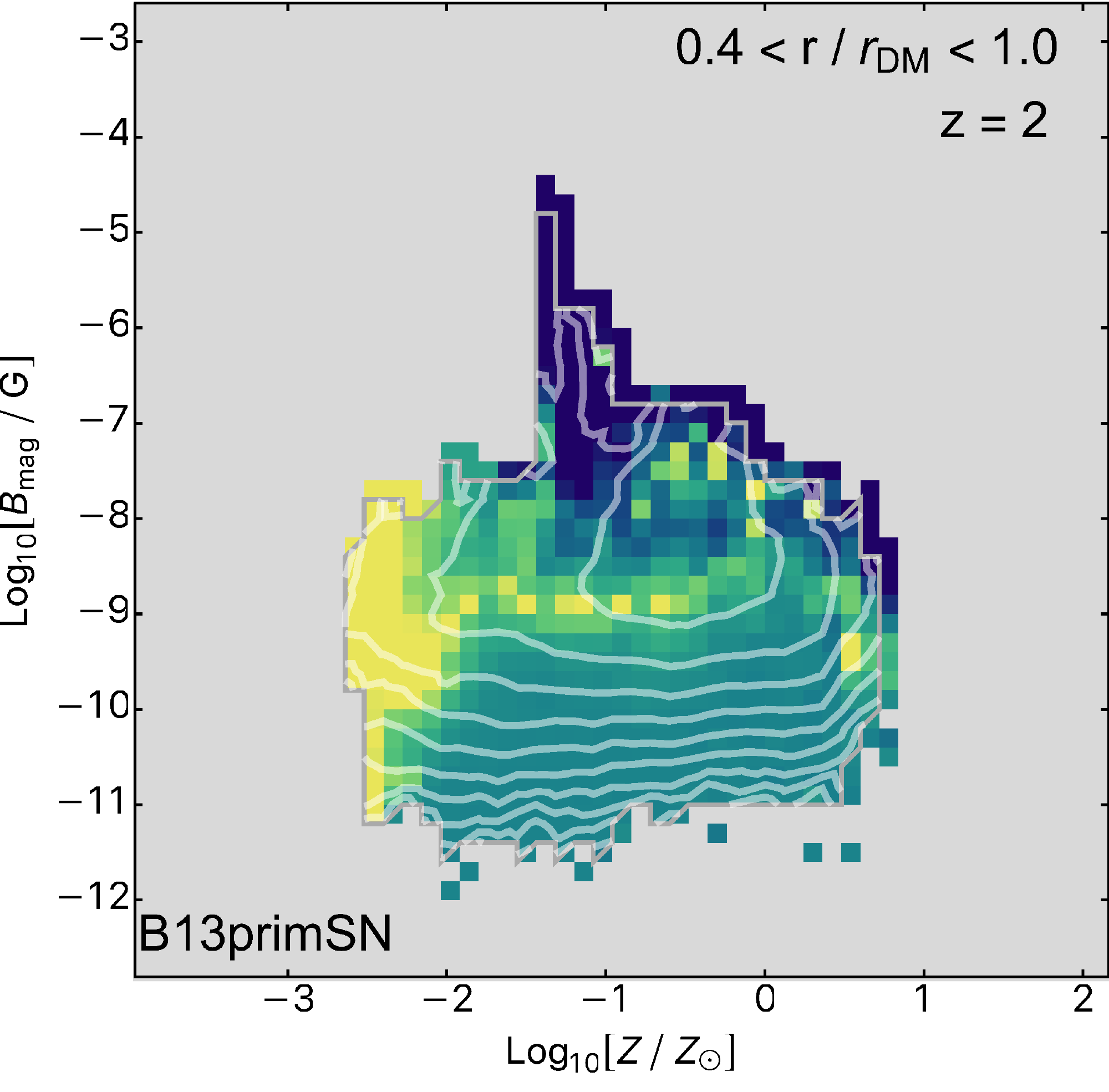}\\
    \caption{\TrMBUp~(top row) and \TrMBDown~(bottom row) magnetic field-metallicity phase space coloured by the ratio of SN-injected to primordial magnetic energy in the galactic region {\bf(left column)}, SN-injected to primordial magnetic energy in the outer halo ($0.4 r_\text{DM} < r < 1.0 r_\text{DM}$; {\bf central column}), and primordial to cross term magnetic energy in the outer halo {\bf(right column)}. White contours show the total magnetic energy distribution. Additional contours for the total magnetic energy in the \MBDoce~(cyan) and \MBInj~(orange) are included for comparison. Strong magnetic fields ($B > 10^{-8}$ G) in halo regions with low gas metallicity are a potential signature of important primordial magnetic fields.}
    \label{fig:MetalPhases}
\end{figure*}

If magnetic fields are generated by stars, or primarily inside galaxies, we expect a correlation of the magnetic field associated with a parcel of gas and its metallicity. This correlation has been shown to hold in isolated galaxies \citep{Butsky2017} and in their halo \citep{Pakmor2020}. These two studies find that this correlation holds for magnetisation of astrophysical (inside the galaxy) nature, either by SN-injection or dynamical amplification, respectively. In this Section we review whether this correlation also holds for primordial magnetic fields, and taking advantage of our tracer algorithm, we also show whether strong primordial magnetic fields can be identified and discerned from astrophysical ones by means of studying the local gas metallicity. 

In Fig.~\ref{fig:MetalPhases} we colour the phase space as a function of the ratio between two traced magnetic energies at $z = 2$ for the \TrMBUp~(top row) and \TrMBDown~(bottom row) simulations. As before, white contours show the distribution of the total magnetic energy and whenever present, cyan and orange show the total magnetic energy in the \MBDoce~(exclusively primordial) and \MBInj~(exclusively astrophysical) runs, respectively. The first two columns show the ratio of SN-injected to primordial magnetic energy in the galactic region (left column) and the external part of the halo (right column; $0.4\;r_\text{DM} < r < 1.0\;r_\text{DM}$). Both in the galaxies and their haloes we find a clear separation of SN and primordial magnetic fields, with the former concentrating at higher metallicities and the second at pristine gas. When comparing the distribution of the total magnetic energies, the one generated inside the galaxy accumulates more uniformly across metallicities $Z \gg 10^{-1} \mathrm{Z_\odot}$, whereas the primordial energy has a flatter distribution towards the lowest metallicities in the simulation. The low metallicity but high magnetic field region of the phase diagram is particularly interesting, as SN-produced magnetic fields do not seem capable to penetrate it. Attending to our tracer simulations, we predict the presence of strong magnetic fields ($B > 10^{-8}$~G) in regions of extremely low metal content to be a signature of primordial magnetic fields. These could be measured in the CGM of galaxies even when accounting for the pollution of galactic magnetic fields. Additional physics or the calibration of our employed feedback prescription might drive more efficient galactic outflows that may introduce some pollution of this low metallicity / high magnetic field regime dominated by primordial magnetic fields. When reviewing the correlation of the magnetic field with metallicity $B \propto Z
^\alpha$ in the galactic region. We fit $\alpha$ assuming magnetic energy-weighting. Low correlations are found for simulations dominated by primordial magnetic fields \MBDoce~($\alpha=0.08$) and \TrMBUp~($\alpha=0.20$). However, when the runs are dominated by SN injection the values found for $\alpha$ are higher: $\alpha=0.40$ for \MBInj~and $\alpha=0.56$ for  \TrMBDown\footnote{Mass-weighted values for \MBDoce, \TrMBUp, \MBInj, and \TrMBDown~are respectively $\alpha=0.04,0.08,0.17,0.2$.}. These two regimes of $B-Z$ correlation are also found when separating the magnetic energy in \TrMBUp~into the two tracer fields, with $\alpha_\text{prim}=0.16$ and $\alpha_\text{SN}=0.22$. This indicates that even in the presence of various magnetic sources, the division based on metallicity is maintained. In the rightmost column of Fig.~\ref{fig:MetalPhases} we show the ratio of primordial to cross-term magnetic energy in the halo of both galaxies (\TrMBUp, top; \TrMBDown, bottom). Altogether with the central column, this confirms that any potential pollution by astrophysically-generated magnetic fields in the low-metallicity region is unimportant both for SN-injected magnetic fields and for the energy produced by the interaction between the two magnetic fields.

\section{Discussion and caveats}
\label{s:Caveats}
We have shown that strong primordial magnetic fields (i.e. $B_0 \gtrsim 10^{-12}$~G) produce realistic galactic magnetisation, compatible with observations and comparable with the one attained when magnetic fields are instead the product of SN feedback. Furthermore, we have found that when primordial and SN-generated magnetic fields are considered in combination, the attained magnetisation is not distinguishable from any of the two former cases. In a similar study, \citet{Garaldi2020} found a comparable result when evolving a zoom simulation of a galaxy to $z = 0$, exploring separately magnetisation from either a Biermann battery \citep{Biermann1950}, the \citet{Durrive2015} mechanism, a primordial seed, or SN injection. They find their SN and primordial channels provide a galaxy magnetisation roughly indistinguishable by $z \sim 2 - 3$. We note that they probe a weaker primordial seed ($B_0 = 10^{-14}$ G) and a lower magnetic energy injection by SN ($\Einj = 10^{-4} E_\text{SN}$) than those studied here. As a result, their considered channels of galaxy magnetisation rely more significantly on dynamical amplification than those in our work. We also find that the origin of galactic magnetic fields cannot be discriminated based on magnetisation once they have reached comparable levels, shortly after the formation of the studied galaxy in our work. However, we have found using our magnetic field tracing algorithm that in our simulations strong primordial magnetic fields and those produced inside the galaxy have differentiated properties and can be distinguished from high redshift ($z \sim 4$) onwards. In particular, we find changes in their energy distribution across spatial scales, as well as in terms of correlations with ISM temperature and gas metallicity. This differentiation can be taken advantage of by future observations, using galaxies as probes of the cosmic magnetic field and potentially improving our constraint of its upper limit ($B_0 < 10^{-9}$~G; \citealt{PlanckCollaboration2015}) by up to three orders of magnitude. 

However, a series of important considerations must be taken into account when interpreting the results presented in this work. Firstly, we only considered two main channels of magnetic field generation, namely primordial (in the form of an ab-initio uniform field) and SN-injected magnetic fields. A third suitable channel of magnetic field generation in galaxies is through the action of a small-scale turbulent dynamo, which would amplify weak magnetic fields to $\sim\muG$ values on short timescales. In the scenarios considered here, one can interpret the SN-injected magnetic fields as a proxy for magnetic fields amplified by a turbulent dynamo, as both are generated at the smallest scales in the simulated galaxy. We remark that no important turbulent dynamo activity is expected in our simulations: due to limited resolution, our CT method has modest magnetic amplification growth rates. This is even true when the magnetic field is weak and only in the kinematic regime \citep{Martin-Alvarez2018}. Furthermore, as the magnetic fields in our simulations rapidly reach $\gtrsim\muG$, we expect their back-reaction on small-scales to inhibit the activity of the turbulent dynamo. As a result, small-scale dynamo amplification would only play a secondary role in our study, and would be more relevant for the SN-injected component of the field that prevails at small scales (Section \ref{ss:Spectra}). Our separation into two tracer fields can safely be interpreted as separating magnetic fields produced either inside the galaxy or in the early Universe. Apart from the turbulent dynamo, other dynamo processes are expected to be active in our galaxy. These will correspondingly modify dynamically the evolution of the total and traced magnetic fields through the induction equation (Equation~\ref{eq:Induction}). An example is shear and differential rotation of the galactic disk after its formation at $z \sim 4$ \citep{Chamandy2013,Bendre2015}.

Secondly, as shown by Fig.~\ref{fig:HMSM}, our galaxies have an excess of stellar mass for their given halo mass compared with expected relations \citep[e.g.][]{Behroozi2013,Moster2018}. A frequent solution employed by simulations to avoid this is to increase feedback efficiency by e.g., increasing the specific energy of stellar feedback, which thwarts the process of star formation. The excess of stellar mass in our simulations implies that more SN-events than expected for a Milky-Way mass galaxy are likely to occur. This in turn could overestimate the amount of SN-injected magnetic energy. As the main conclusion presented in this work is that strong primordial magnetic fields ($B_0 \gtrsim 10^{-12}$ G) remain important in the studied galaxy down to $z \sim 1$, well after the peak of star formation of the Universe, the presence of additional SN-injected energy only reinforces this conclusion.

As our simulations are only evolved until $z = 2$ ($z = 1$ for the \TrMBUp~case), our comparison with observations at $z \sim 0$ (e.g. as those presented in Fig. \ref{fig:Profiles}) relies on the assumption that no significant redshift evolution takes places between $z \sim 1$ and $z = 0$. Evidence for strong magnetic fields has been observed at redshifts as high as $z \sim 4$ \citep{Bernet2008}, but other field properties may differ. Similarly, the behaviour observed in our simulations remains approximately unchanged between $z \sim 4$ and $z \sim 1$. However, as our simulations do not reach $z = 0$, some of their properties, including their deviation from the stellar mass vs halo mass may evolve if they were to reach this redshift.

When studying magnetic fields in the halo of galaxies, we find no significant evidence for dynamo activity in these regions. While halo dynamo amplification is often deemed unimportant \citep{Beck2012,Marinacci2015,Butsky2017}, \citet{Pakmor2020} suggested that some degree of amplification might occur once simulations reach high enough spatial resolution. If halo dynamo activity in nature causes an important amplification of weak primordial magnetic fields to values $> 10^{-8}$ G, this would complicate a clear detection of primordial magnetic fields in the CGM of galaxies. 

While the primordial magnetic field seeded in this work is uniform and homogeneous for the sake of simplicity, the primordial magnetic field of our Universe will be inhomogeneous and likely to approximately follow a power spectrum \citep{PlanckCollaboration2015,Hutschenreuter2018}. One of the potential consequences is that employing a uniform seed instead of an spectrum could imprint large-scale features on the simulated primordial field, particularly outside of virialised structures \citep{Marinacci2015}. Therefore, it is important to consider that, regardless of whether primordial magnetic fields are strong or weak in nature, different regions of the Universe will present local variations of the strength and importance of the primordial magnetic field.

Finally, we note that current galaxy formation simulations often lack additional physics that are important to capture the correct properties of the ISM. While we account for magnetic fields, radiative transfer \citep{Rosdahl2015b,Emerick2018} and cosmic rays \citep{Pfrommer2017b,Hopkins2019,Dashyan2020} influence the ISM of galaxies and the evolution of their magnetic fields \citep{Hanasz2009}. Equally, while AGN activity could be relatively unimportant in the explored mass regime \citep{Crain2015,Martin-Navarro2018}, this mode of feedback could serve as a considerable source of magnetic fields as well as also producing magnetised outflows \citep{Dashyan2018,Koudmani2019} such as those observed in the MW \citep{Heywood2019}. Indeed, \citet{Vazza2017} show that AGN can eject magnetic fields to intergalactic distances. The importance of AGN in magnetising galaxies and their surroundings will be revisited with our magnetic tracers method in future work.

\section{Conclusions}
\label{s:Conclusions}
In this work we study four new magneto-hydrodynamical cosmological zoom-in simulations of a Milky Way-like galaxy. The simulations are generated using our modified version of the constrained transport MHD {\sc ramses} code \citep{Teyssier2002,Fromang2006,Teyssier2006} capable of tracing the separate evolution of different types of magnetic fields \citep{KMA2019}. In particular, we trace magnetic fields of primordial nature as well as astrophysical magnetic fields generated in the galaxy during SN events. Our simulations span the cases of exclusively primordial (\MBDoce), primordially dominated (\TrMBUp), SN-injection dominated (\TrMBDown), and purely SN-injected (\MBInj) magnetisation. With the focus on disentangling the primordially generated magnetic fields from the SN-injected ones in our simulations (\TrMBUp~and \TrMBDown), we find the following results:
\begin{enumerate}
    \item Both primordial magnetic fields (with $B_0 \sim 10^{-12}$ G) and SN-injected magnetic fields (with $\Einj \sim 0.01 E_\text{SN}$) produce realistic magnetisations in simulated galaxies, and compare well with observations.
    \item Specifically, the profiles of the magnetic field in all our simulations compare well with observations \citep[e.g.][]{Basu2013,Beck2005,Berkhuijsen2016}, and are similar to other MHD simulations \citep{Pakmor2017}. While the profile of primordial magnetic fields appear a better match to observations at the core of the galactic disk, this is likely due to the lack to resolution in observations as magnetic fields on the order of $m$G are observed in the centre of our own galaxy \citep{Aitken1998}.
    \item The studied channels of magnetisation and their combinations produce galaxies with a similar appearance and almost identical stellar masses and star formation rates. This is in agreement with previous studies, which have suggested magnetic fields to have a minor impact on such properties \citep{Pakmor2017,Su2017,Martin-Alvarez2020}.
    \item Once strong enough primordial magnetic fields are considered with $B_0 \sim 10^{-12}$~G, their resulting magnetic energy budget in galaxies is roughly equal to that attained by magnetic fields generated through SN feedback. $B_0 \sim 10^{-12}$ G is the approximate value found in \citet{Martin-Alvarez2020} for which the first noticeable effects in galaxy properties due to magnetic fields appear. For even higher values of $B_0$, primordial magnetic fields will have a significant impact on the global properties of galaxies.
    \item In the inner regions of the galaxy, bipolar outflows from the galaxy pollute the halo with SN-injected magnetic fields, which mostly concentrate at small radii ($r < 3$ kpc). The strong primordial fields are distributed approximately homogeneously and become dominant at larger distances ($r > 4$ kpc) as also evidenced by the analysis of magnetic energy spectra. This indicates that observations of magnetic fields in the CGM and IGM are potential probes of the primordial magnetic field of our Universe.
    \item Moreover, while SN-injected fields are dominant in hot SN-driven bubbles and outflows, strong primordial fields dominate in the warm phase of the ISM. While both type of fields correlate well with gas metallicity they scale in different way, with only the strong primordial fields found in pristine gas, which may help to unravel their origin.
\end{enumerate}

Anticipating facilities that will be capable of unveiling the magnetic Universe, such as SKA, we showed that galaxies and their CGM are a potential window to the primordial magnetic field of our Universe. As emerging numerical work highlights the importance of magnetic fields, MHD simulations of galaxy formation will become an important tool to understand galaxy formation. In future work, we aim to study detailed mock observational signatures associated with our magnetic tracers which will allow us to make a direct link with observations to ultimately unravel the origin of cosmic magnetic fields.

\section*{Acknowledgements}
We kindly thank the referee for insightful comments and suggestions that contributed to improve the quality of this manuscript. We would like to thank Jim Pringle and Franco Vazza for useful comments and suggestions. DS would like to thank Prof. D'Anchise for all the help and support in finalizing this manuscript. This work was supported by the ERC Starting Grant 638707 ``Black holes and their host galaxies: co-evolution across cosmic time" and by STFC. This work is part of the Horizon-UK project, which used the DiRAC Complexity system, operated by the University of Leicester IT Services, which forms part of the STFC DiRAC HPC Facility (\href{www.dirac.ac.uk}{www.dirac.ac.uk}). This equipment is funded by BIS National E-Infrastructure capital grant ST/K000373/1 and STFC DiRAC Operations grant ST/K0003259/1. The equipment was funded by BEIS capital funding via STFC capital grants ST/K000373/1 and ST/R002363/1 and STFC DiRAC Operations grant ST/R001014/1. DiRAC is part of the National e-Infrastructure. The authors would like to acknowledge the use of the University of Oxford Advanced Research Computing (ARC) facility in carrying out this work. \href{http://dx.doi.org/10.5281/zenodo.22558}{http://dx.doi.org/10.5281/zenodo.22558}. The authors also acknowledge the usage of the FFTW library: \href{http://www.fftw.org/}{http://www.fftw.org/}. Some of the results in this paper have been derived using the healpy and HEALPix package (\href{http://healpix.sourceforge.net}{http://healpix.sourceforge.net}; \citealt{Zonca2019}).

\section*{Data availability}
The data employed in this manuscript is to be shared upon reasonable request contacting the corresponding author.

\bibliographystyle{mnras}
\bibliography{references.bib}

\appendix
\section{Magnetised stellar feedback}
\label{ap:MagInjection}

\begin{figure}
    \centering
    \includegraphics[width=\columnwidth]{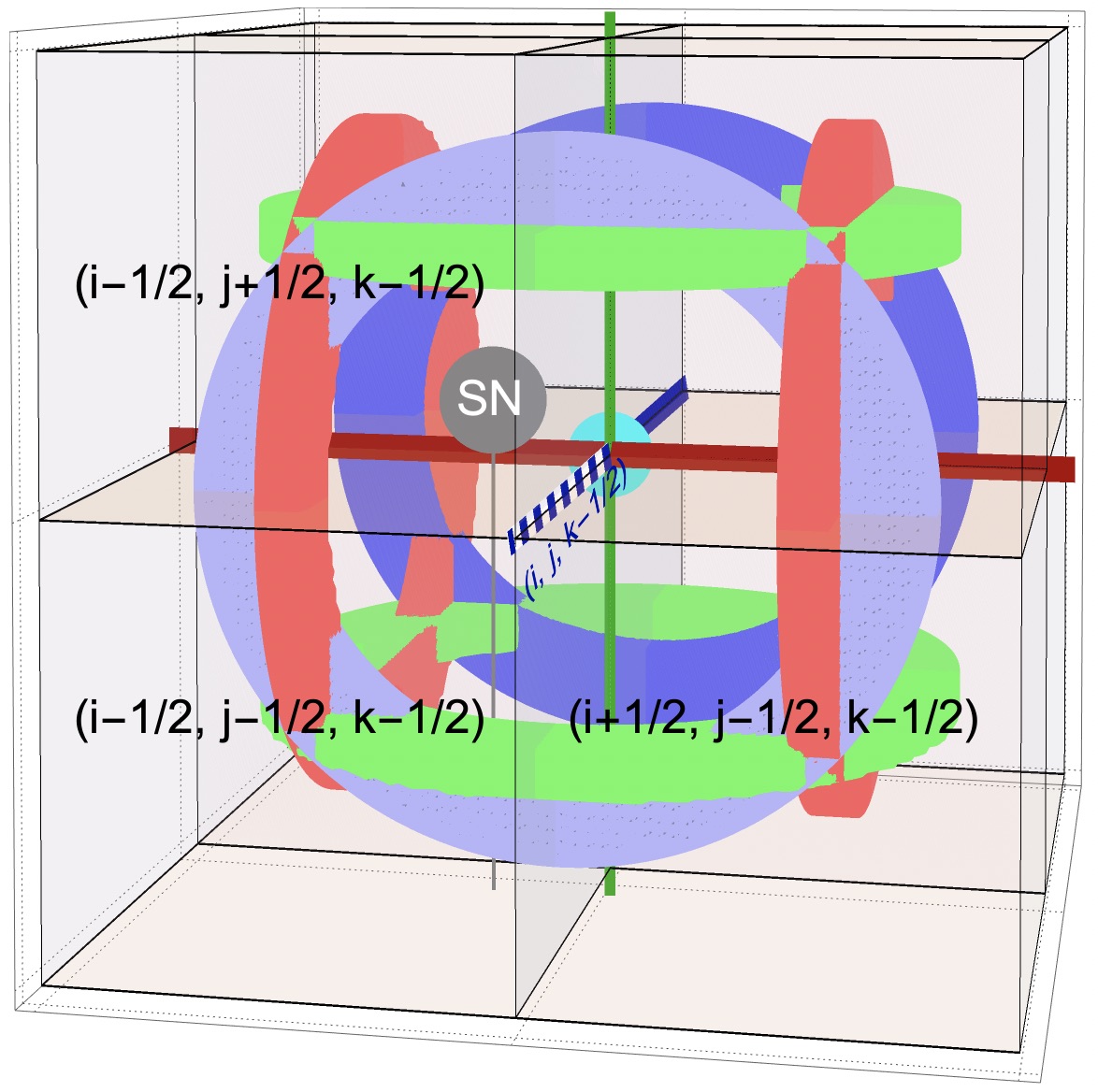}\\
    \caption{Schematic of the magnetic field injection by SN. Each events modifies the magnetic field of the closest 8 cells. Coordinates in black text indicate the positions of cell centres. A SN event is assumed to occur in cell $(i-1/2, j+1/2, k-1/2)$ at the position of the gray circle. The cyan circle represents the point of intersection of the 8 neighbouring cells. We depict the six magnetic loops injected by the SN as rings, which modify the magnetic field of each cell face they traverse by $B_\text{inj}$. Each ring is centred on a unique cell edge. All six injection cell edges are depicted by thick lines with the same colour of their corresponding ring. The text describes in more detail the injection around edge $(i, j, k-1/2)$, in dashed blue/white.}
    \label{fig:InjectionCube}
\end{figure}

Our injection runs (i.e. \MBInj, \TrMBUp, and \TrMBDown) return magnetised gas to the ISM of the galaxy during each SN event. When a SN event takes place, the algorithm injects 6 closed loops or rings of magnetic field into the 8 cells surrounding the stellar particle. Each of these loops will traverse 4 faces, crossing the interfaces of each cell twice, once with the injected magnetic field $\Binj$ pointing inwards and a second time directed outwards. As a result, each injection fulfils $\vec{\nabla} \cdot \vec{B} = 0$. A useful system to express this injection is portrayed in Fig.~\ref{fig:InjectionCube}, using a set of coordinates centred at the closest vertex to the stellar particle, with position $(i,j,k)$. The 6 injected loops are placed around the 6 cell edges connected to the central vertex, with each vertex having coordinates $(i \pm 1/2, j, k)$, where we allow $(i,j,k)$ to cycle with positive permutations across $(x,y,z)$. The magnetic loop injected around the edge $(i,j,k-1/2)$ will modify the magnetic fields at the cell faces (which are defined at positions $(i\pm1/2,j,k-1/2)$, $(i,j\pm1/2,k-1/2)$ for this particular edge) will have components $B_x$ and $B_y$:
\begin{equation}
\begin{split}
B_x^{i,j+1/2,k-1/2} &= B_x^{i,j+1/2,k-1/2} \pm \Binj \, , \\
B_y^{i+1/2,j,k-1/2} &= B_y^{i+1/2,j,k-1/2} \mp \Binj \, , \\
B_x^{i,j-1/2,k-1/2} &= B_x^{i,j-1/2,k-1/2} \mp \Binj \, , \\
B_y^{i-1/2,j,k-1/2} &= B_y^{i-1/2,j,k-1/2} \pm \Binj \, . \\
\end{split}
\label{eq:InjB}
\end{equation}
As indicated in the main text, we select $\Einj = 0.01 E_\text{SN}$, comparable to the magnetic energy per SN injected in other studies \citep[e.g.,][]{Beck2013a,Butsky2017,Vazza2017}. As a result, we choose in code units\footnote{{\sc ramses} magnetic code units are rational units (i.e. Lorentz-Heaviside units).} $\Binj = \sqrt{0.02 \epsilon_\text{SN}}$, where $\epsilon_\text{SN} = E_\text{SN} / V_\text{cell}$ and $V_\text{cell}$ is the volume of the cell. This is different to what was used in \citet{KMA2019}, where the injected magnetic field had a fixed physical strength instead. As our simulations do not display turbulent magnetic amplification capable of reaching saturation in short timescales \citep{Martin-Alvarez2018}, our results will not depend significantly on the fraction of the magnetic injected, so long as it is high enough to reproduced observed magnetic fields but too high as to represent a large fraction of the total energy injected per SN. In order to source the SN-injected magnetic tracer when the simulations have two magnetic sources, the same injection operation performed for the total field is also done in the traced magnetic field $\vec{B}_\text{ast}$. 

\begin{figure}
    \centering
    \includegraphics[width=\columnwidth]{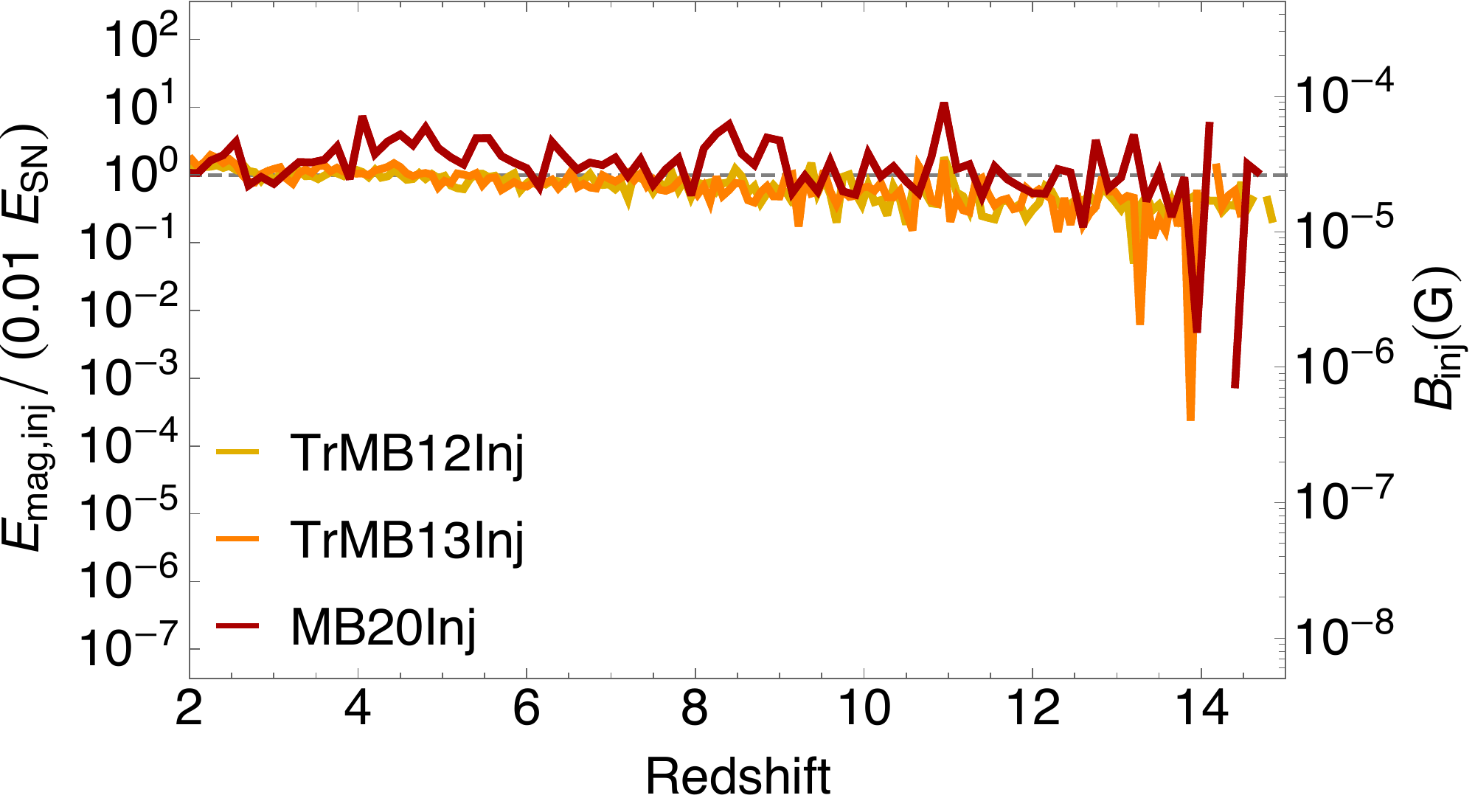}\\
    \caption{Average SN-injected magnetic energy in the three simulations including injection. The horizontal gray dashed line is the target magnetic energy, to which simulations adjust reasonably well. We note that \MBInj~has a larger injection efficiency than the other runs due to the absence of dynamically important large-scale ab-initio magnetic fields.}
    \label{fig:InjectionCheck}
\end{figure}

We select the orientation of the loop (and therefore, the respective set of signs in Equation~\ref{eq:InjB} as a function of the local magnetic field in such a way that the injection maximises the amount of magnetic energy generated. This selection is made so that the injected energy matches the target injection energy. When a simple injection configuration is selected, there is no guarantee that at least a part of the injected magnetic field will not oppose the pre-existing magnetic field. As a result, the amount of injected energy in each event is not fixed, but oscillates around the desired value. In our simulations, we aim to inject a magnetic energy $\sim 0.01 E_\text{SN}$. Fig.~\ref{fig:InjectionCheck} shows the average injected magnetic energy $E_\text{mag,inj}$ in the three runs with injection as a function of redshift. The desired energy is represented by the horizontal dashed gray line. Overall, all our runs tend to inject a magnetic energy on the desired order of magnitude. We note that the \MBInj~simulation has a slightly higher injection efficiency, which is due to the initial absence of magnetic fields, and therefore the selected injection geometry opposing a smaller fraction of the local field. Finally, we note that our selection for the direction of the injection will have a slight preference to align the injected field with any existing field, as this maximises the injected energy. As a result, injections will tend to positive cross-term instead of negative cross-term magnetic energy when separated by the magnetic tracers.

\bsp	
\label{lastpage}
\end{document}